# First M87 Event Horizon Telescope Results. VII. Polarization of the Ring

The Event Horizon Telescope Collaboration

(See the end matter for the full list of authors.)

## Abstract

In 2017 April, the Event Horizon Telescope (EHT) observed the near-horizon region around the supermassive black hole at the core of the M87 galaxy. These 1.3 mm wavelength observations revealed a compact asymmetric ring-like source morphology. This structure originates from synchrotron emission produced by relativistic plasma located in the immediate vicinity of the black hole. Here we present the corresponding linear-polarimetric EHT images of the center of M87. We find that only a part of the ring is significantly polarized. The resolved fractional linear polarization has a maximum located in the southwest part of the ring, where it rises to the level of ~15%. The polarization position angles are arranged in a nearly azimuthal pattern. We perform quantitative measurements of relevant polarimetric properties of the compact emission and find evidence for the temporal evolution of the polarized source structure over one week of EHT observations. The details of the polarimetric data reduction and calibration methodology are provided. We carry out the data analysis using multiple independent imaging and modeling techniques, each of which is validated against a suite of synthetic data sets. The gross polarimetric structure and its apparent evolution with time are insensitive to the method used to reconstruct the image. These polarimetric images carry information about the structure of the magnetic fields responsible for the synchrotron emission. Their physical interpretation is discussed in an accompanying publication.

*Unified Astronomy Thesaurus concepts:* Polarimetry (1278); Radio interferometry (1346); Very long baseline interferometry (1769); Supermassive black holes (1663); Active galactic nuclei (16); Low-luminosity active galactic nuclei (2033); Astronomy data modeling (1859); Galaxy accretion disks (562); Galaxies: individual: M87

## 1. Introduction

The Event Horizon Telescope (EHT) Collaboration has recently reported the first images of the event-horizon-scale structure around the supermassive black hole in the core of the massive elliptical galaxy M87, one of its two main targets.[130] The EHT images of M87's core at 230 GHz (1.3 mm wavelength) revealed a ring-like structure whose diameter of 42 $\mu$as, brightness temperature, shape, and asymmetry are interpreted as synchrotron emission from relativistic electrons gyrating around magnetic field lines in close vicinity to the event horizon. We have described the details of the EHT's instrumentation, data calibration pipelines, data analyses and imaging procedures, and the theoretical interpretation of these first images in a series of publications (Event Horizon Telescope Collaboration et al. 2019a, 2019b, 2019c, 2019d, 2019e, 2019f, hereafter Papers I, II, III, IV, V, VI, respectively).

In this Letter, we present the first *polarimetric* analysis of the 2017 EHT observations of M87 and the first images of the linearly polarized radiation surrounding the M87 black hole shadow. These polarimetric images provide essential new information about the structure of magnetic field lines near the event horizon of M87's central supermassive black hole, and they put tight constraints on the theoretical interpretations of the nature of the ring and of relativistic jet-launching theories. The theoretical implications of these images and the constraints that they place on the magnetic field structure and accretion state of the black hole are discussed in an accompanying work (Event Horizon Telescope Collaboration et al. 2021, hereafter Paper VIII). Readers interested in the details of the data reduction, methodology, and validation can find a detailed index of this Letter in Section 1.2. Readers primarily interested in the results may skip directly to Section 5 and to subsequent discussion and conclusions in Section 6.

### 1.1. Previous Polarimetric Observations of the M87 Jet

The giant elliptical galaxy Messier 87 (M87, NGC 4486) is the central member of the Virgo cluster of galaxies and hosts a low-luminosity radio source (Virgo A, 3C 274, B1228+126). M87 is nearby and bright, and at its center is one of the best-studied active galactic nuclei (AGNs). M87 was the first galaxy in which an extragalactic jet (first described as a "narrow ray") extending from the nucleus was discovered (Curtis 1918). This kiloparsec-scale jet is visible, with remarkably similar morphology, at all wavelengths from radio to X-ray. The optical radiation from the jet on kpc scales was found to be linearly polarized by Baade (1956), which was confirmed by Hiltner (1959), suggesting that the emission mechanism is synchrotron radiation.

The central engine that powers the jet contains one of the most massive black holes known, measured from the central stellar velocity dispersion (Gebhardt et al. 2011; $M = (6.6 \pm 0.4) \times 10^9 M_\odot$) and directly from the size of the observed emitting

---

region surrounding the black hole shadow (Paper VI; $M = (6.5 \pm 0.7) \times 10^9\,M_\odot$). For this mass, the Schwarzschild radius is $R_s = 2GM/c^2 = 1.8 \times 10^{15}$ cm. At the distance of M87, $16.8^{+0.8}_{-0.7}$ Mpc (Blakeslee et al. 2009; Bird et al. 2010; Cantiello et al. 2018, Paper VI), the EHT resolution of about 20 micro-arcseconds ($\mu$as) translates into a linear scale of 0.0016 pc = 2.5 $R_s$.

The M87 jet has been imaged at subarcsecond resolution in both total intensity and linear polarization at optical wavelengths with the Hubble Space Telescope (Thomson et al. 1995; Capetti et al. 1997), and at radio wavelengths with the Very Large Array (e.g., Owen et al. 1989). Observing the launching region of the jet closer to the black hole and the region surrounding the black hole requires milliarcsecond (mas) resolution or better, and hence very-long-baseline interferometry (VLBI) techniques used at the highest frequencies (e.g., Boccardi et al. 2017 and references therein).

Milliarcsecond-scale VLBI observations show that the core itself is unpolarized even at millimeter wavelengths. Zavala & Taylor (2002), observing at 8, 12, and 15 GHz, set upper limits on the fractional polarization of the compact core of $m < 0.1\%$. About 20 mas downstream from the core, patchy linear polarization starts to become visible in the jet at the level of 5%–10%, although no large-scale coherent pattern to the electric-vector position angles (EVPAs) $\chi$ is apparent. However, at each patch in the downstream jet, the EVPAs exhibit a linear change with $\lambda^2$, allowing the rotation measures (RMs) to be estimated. These RMs range from $-4000\,\mathrm{rad\,m^{-2}}$ to $9000\,\mathrm{rad\,m^{-2}}$ (Zavala & Taylor 2002). The linear dependence of EVPA on $\lambda^2$ over several radians is important, as it shows that the Faraday-rotating plasma in the jet cannot be mixed in with the relativistic emitting particles (Burn 1966) but must be in a cooler (sub-relativistic) foreground screen.

On kiloparsec scales, Owen et al. (1990) found a complex distribution of RM. Over most of the source the RM is typically of order $1000\,\mathrm{rad\,m^{-2}}$, but there are patches where values are as high as $8000\,\mathrm{rad\,m^{-2}}$ are found.

More recently, Park et al. (2019) studied Faraday RMs in the jet using multifrequency Very Long Baseline Array (VLBA) data at $\lesssim 8$ GHz. They found that the RM magnitude systematically decreases with increasing distance from 5,000 to 200,000 $R_s$. The observed large ($\gtrsim 45°$) EVPA rotations at various locations of the jet suggest that the dominant Faraday screen in this distance range would be external to the jet, similar to the conclusion of Zavala & Taylor (2002). Homan & Lister (2006), also observing at 15 GHz with the VLBA (as part of the MOJAVE project) found a tight upper limit on the fractional linear polarization of the core of $< 0.07\%$. They also detect circular polarization of $(-0.49 \pm 0.10)\%$.

At 43 GHz, Walker et al. (2018) presented results from 17 years of VLBA observations of M87, with polarimetric images presented at two epochs. These show significant polarization (up to 4%) in the jet near the 43 GHz core, but at the position of the total intensity peaks the fractional polarizations are only 1.5% and 1.1%. They interpret these fractions as coming from a mix of emission from the unresolved, unpolarized core and a more polarized inner jet.

Hada et al. (2016) showed images at four epochs at 86 GHz made with the VLBA and the Green Bank Telescope. At this frequency, the resolution is about $(0.4 \times 0.1)$ mas, corresponding to $(56 \times 14)$ $R_s$. Again, the core is unpolarized with no linear polarization detected at the position of the total intensity emission's peak, while there is a small patch of significant (3.5%) polarization located 0.1 mas downstream. At 0.4 mas downstream from the peak, there is another patch of significant polarization (20%). These results indicate that there are regions of significantly ordered magnetic field very close to the central engine.

Very recently, new observations by Kravchenko et al. (2020) using the VLBA at 22 and 43 GHz show two components of linear polarization and a smooth rotation of EVPA around the 43 GHz core. Comparison with earlier observations show that the global polarization pattern in the jet is largely stable over an 11 year timescale. They suggest that the polarization pattern is associated with the magnetic structure in a confining magnetohydrodynamic wind, which is also the source of the observed Faraday rotation.

The EHT presently observes at $\sim$230 GHz and has previously reported polarimetric measurements only for Sagittarius A* (Sgr A*; Johnson et al. 2015). The only previous polarimetric measurements of M87 at this frequency were done by Kuo et al. (2014) using the Submillimeter Array (SMA) on Maunakea, Hawai'i, USA. The SMA is a compact array with a $(1.2 \times 0.8)$ arcsec beam, 10000 times larger than the EHT beam. Li et al. (2016) used the value from this work to calculate a limit on the accretion rate onto the M87 black hole. Most recently, Goddi et al. (2021) reported results on M87 around 230 GHz as part of the Atacama Large Millimeter/submillimeter Array (ALMA) interferometric connected-element array portion of the EHT observations in 2017. The ALMA-only 230 GHz observations (with a FWHM synthesized beam in the range $1''–2''$, depending on the day) resolve the M87 inner region into a compact central core and a kpc-scale jet across approximately $25''$. It has been found that the 230 GHz core at these scales has a total flux density of $\sim$1.3 Jy, a low linear polarization fraction $|m| \sim 2.7\%$, and even less circular polarization, $|v| < 0.3\%$. Notably, ALMA-only observations show strong variability in the RM estimated based on four frequencies within ALMA Band 6 (four spectral windows centered at 213, 215, 227, and 229 GHz; Matthews et al. 2018). The RM difference is clear between the start of the EHT observing campaign on 2017 April 5 (RM $\approx 0.6 \times 10^5\,\mathrm{rad\ m^{-2}}$) and the end on 2017 April 11 (RM $\approx -0.4 \times 10^5\,\mathrm{rad\ m^{-2}}$). Because these measurements were taken simultaneously with the EHT VLBI observations presented here, the ALMA-only linear polarization fraction measurements can be used as a point of reference, and we discuss possible implications of the strong RM evolution on the EHT polarimetric images of M87.

### 1.2. This Work

This Letter presents the details of the polarimetric data calibration, the procedures for polarimetric imaging, and the resulting images of the M87 core. In Section 2, we briefly overview the basics of polarimetric VLBI. In Section 3, we summarize the EHT 2017 observations, describe the initial data calibration procedure and validation tests, and describe the basic properties of the polarimetric data. In Section 4, we describe our methods, strategy, and test suite for our polarimetric calibration and imaging. In Section 5, we present and analyze the polarimetric images of the M87 ring and examine the calibration's impact on the polarimetric image. We discuss the results and summarize the work in Sections 6 and 7.

This Letter is supplemented with a number of appendices supporting our analysis and results. The appendices summarize: polarimetric data issues (Appendix A); novel VLBI closure data products (Appendix B); details of calibration and imaging methods (Appendix C); validation of polarimetric calibration for telescopes with an intra-site partner (Appendix D); fiducial leakage D-terms from M87 imaging (Appendix E); preliminary results of polarimetric imaging of M87 (Appendix F); polarimetric imaging scoring procedures (Appendix G); details of Monte Carlo D-term simulations (Appendix H); consistency of low- and high-band results for M87 (Appendix I); comparison to polarimetric properties of calibrator sources (Appendix J); and validations of assumptions made in polarimetric imaging of the main target and the calibrators (Appendix K).

## 2. Basic Definitions

A detailed introduction to polarimetric VLBI can be found in Thompson et al. (2017, their Chapter 4). Here we briefly introduce the basic concepts and notation necessary to understand the analysis presented throughout this Letter. The polarized state of the electromagnetic radiation at a given spatial coordinate $x = (x, y)$ is described in terms of four Stokes parameters, $\mathcal{I}(x)$ (total intensity), $\mathcal{Q}(x)$ (difference in horizontal and vertical linear polarization), $\mathcal{U}(x)$ (difference in linear polarization at $45°$ and $-45°$ position angle), and $\mathcal{V}(x)$ (circular polarization). We define the complex linear polarization $\mathcal{P}$ as

$$\mathcal{P} = \mathcal{Q} + i\mathcal{U} = \mathcal{I}|m|e^{2i\chi}, \tag{1}$$

where $m = (\mathcal{Q} + i\mathcal{U})/\mathcal{I}$ represents the (complex) fractional polarization, and $\chi = 0.5 \arg(\mathcal{P})$ is the EVPA, measured from north to east. Total-intensity VLBI observations directly sample the Fourier transform $\tilde{\mathcal{I}}$ as a function of the spatial frequency $u = (u, v)$ of the total-intensity image; similarly, polarimetric VLBI observations also sample the Fourier transform of the other Stokes parameters $\tilde{\mathcal{Q}}, \tilde{\mathcal{U}}, \tilde{\mathcal{V}}$.

EHT data are represented in a circular basis, related to the Stokes visibility components by the following coordinate system transformation:

$$\rho_{jk} \equiv \begin{pmatrix} R_j R_k^* & R_j L_k^* \\ L_j R_k^* & L_j L_k^* \end{pmatrix} = \begin{pmatrix} \tilde{\mathcal{I}}_{jk} + \tilde{\mathcal{V}}_{jk} & \tilde{\mathcal{Q}}_{jk} + i\tilde{\mathcal{U}}_{jk} \\ \tilde{\mathcal{Q}}_{jk} - i\tilde{\mathcal{U}}_{jk} & \tilde{\mathcal{I}}_{jk} - \tilde{\mathcal{V}}_{jk} \end{pmatrix} \tag{2}$$

for a baseline between two stations $j$ and $k$. The notation $R_j R_k^*$ indicates the complex correlation (where the asterisk denotes conjugation) of the electric field components measured by the telescopes; in this example, the right-hand circularly polarized component $R_j$ measured by the telescope $j$ and the left-hand circularly polarized component $L_k$ measured by the telescope $k$. Equation (2) defines the coherency matrix $\rho_{jk}$. Following Johnson et al. (2015), we also define the fractional polarization in the visibility domain,

$$\breve{m} \equiv \frac{\tilde{\mathcal{Q}} + i\tilde{\mathcal{U}}}{\tilde{\mathcal{I}}} = \frac{\mathcal{P}}{\tilde{\mathcal{I}}} = \frac{2RL^*}{RR^* + LL^*}. \tag{3}$$

Note that Equation (3) implies that $\breve{m}(u)$ and $\breve{m}(-u)$ constitute independent measurements for $u \neq 0$. Moreover, $\breve{m}(u)$ and $m(x)$ are *not* a Fourier pair. While the image-domain fractional polarization magnitude is restricted to values between 0 (unpolarized radiation) and 1 (full linear polarization), there is no such restriction on the absolute value of $\breve{m}$. Useful relationships between $\breve{m}$ and $m$ are discussed in Johnson et al. (2015).

Imperfections in the instrumental response distort the relationship between the measured polarimetric visibilities and the source's intrinsic polarization. These imperfections can be conveniently described by a Jones matrix formalism (Jones 1941), and estimates of the Jones matrix coefficients can then be used to correct the distortions. The Jones matrix characterizing a particular station can be decomposed into a series of complex matrices $\boldsymbol{G}$, $\boldsymbol{D}$, and $\Phi$ (Thompson et al. 2017):

$$\boldsymbol{J} = \boldsymbol{G}\boldsymbol{D}\Phi = \begin{pmatrix} G_R & 0 \\ 0 & G_L \end{pmatrix} \begin{pmatrix} 1 & D_R \\ D_L & 1 \end{pmatrix} \begin{pmatrix} e^{-i\phi} & 0 \\ 0 & e^{i\phi} \end{pmatrix}. \tag{4}$$

Time-dependent field rotation matrices $\Phi \equiv \Phi(t)$ are known a priori, with the field rotation angle $\phi(t)$ dependent on the source's elevation $\theta_{el}(t)$ and parallactic angle $\psi_{par}(t)$. The angle $\phi$ takes the form

$$\phi = f_{el}\,\theta_{el} + f_{par}\,\psi_{par} + \phi_{off}, \tag{5}$$

where $\phi_{off}$ is a constant offset, and the coefficients $f_{el}$ and $f_{par}$ are specific to the receiver position. The gain matrices $\boldsymbol{G}$, containing complex station gains $G_R$ and $G_L$, are estimated within the EHT's upstream calibration and total-intensity imaging pipeline; see Section 3.2. Estimation of the D-terms, the complex coefficients $D_R$ and $D_L$ of the leakage matrix $\boldsymbol{D}$, generally requires simultaneous modeling of the resolved calibration source, and hence cannot be easily applied at the upstream data calibration stage. The details of the leakage calibration procedures adopted for the EHT polarimetric data sets analysis are described in Section 4.

For a pair of VLBI stations $j$ and $k$ the measured coherency matrix $\rho'_{jk}$ is related to the true-source coherency matrix $\rho_{jk}$ via the Radio Interferometer Measurement Equation (RIME; Hamaker et al. 1996; Smirnov 2011),

$$\rho'_{jk} = \boldsymbol{J}_j \rho_{jk} \boldsymbol{J}_k^\dagger, \tag{6}$$

where the dagger † symbol denotes conjugate transposition. Once the Jones matrices for the stations $j$ and $k$ are well characterized, Equation (6) can be inverted to give the source coherency matrix $\rho_{jk}$. From $\rho_{jk}$, Stokes visibilities can be obtained by inverting Equation (2):

$$\begin{pmatrix} \tilde{\mathcal{I}}_{jk} \\ \tilde{\mathcal{Q}}_{jk} \\ \tilde{\mathcal{U}}_{jk} \\ \tilde{\mathcal{V}}_{jk} \end{pmatrix} = \frac{1}{2} \begin{pmatrix} R_j R_k^* + L_j L_k^* \\ R_j L_k^* + L_j R_k^* \\ -i(R_j L_k^* - L_j R_k^*) \\ R_j R_k^* - L_j L_k^* \end{pmatrix}. \tag{7}$$

The collection of Stokes visibilities sampled in $(u, v)$ space by the VLBI array can finally be used to reconstruct the polarimetric images $\mathcal{I}(x)$, $\mathcal{Q}(x)$, $\mathcal{U}(x)$, and $\mathcal{V}(x)$.

The coherency matrices on a quadrangle of baselines can be combined to form "closure traces," data products that are insensitive to any calibration effects that can be described using Jones matrices. Appendix B defines these closure traces and outlines their utility for describing the EHT data.

## 3. EHT 2017 Polarimetric Data

### 3.1. Observations and Initial Processing

Eight observatories at six geographical locations participated in the 2017 EHT observing campaign: ALMA and the Atacama Pathfinder Experiment (APEX) in the Atacama Desert in Chile;

the Large Millimeter Telescope Alfonso Serrano (LMT) on the Volcán Sierra Negra in Mexico; the South Pole Telescope (SPT) at the geographic south pole; the IRAM 30 m telescope (PV) on Pico Veleta in Spain; the Submillimeter Telescope (SMT) on Mt. Graham in Arizona, USA; SMA and the James Clerk Maxwell Telescope (JCMT) on Maunakea in Hawai'i, USA.[131]

The EHT observations were carried out on five nights between 2017 April 5 and 11. M87 was observed on April 5, 6, 10, and 11. Along with the main EHT targets M87 and Sgr A*, several other AGN sources were observed as science targets and calibrators.

Observations were conducted using two contiguous frequency bands of 2 GHz bandwidth each, centered at frequencies of 227.1 and 229.1 GHz, hereby referred to as low and high band, respectively. The observations were arranged in scans alternating different sources, with durations lasting between 3 and 7 minutes. Apart from the JCMT, which observed only a single polarization (right-circular polarization on 2017 April 5–7 and left-circular polarization on 2017 April 10–11), all stations observed in full polarization mode. ALMA is the only station to natively record data in a linear polarization basis. Visibilities measured on baselines to ALMA were converted from a mixed linear-circular basis to circular polarization after correlation using the `PolConvert` software (Martí-Vidal et al. 2016; Matthews et al. 2018; Goddi et al. 2019). A technical description of the EHT array is presented in Paper II and a summary of the 2017 observations and data reduction is presented in Paper III.

### 3.2. Correlation and Data Calibration

After the sky signal received at each telescope was mixed to baseband, digitized, and recorded directly to hard disk, the data from each station were sent to MIT Haystack Observatory and the Max-Planck-Institut für Radioastronomie (MPIfR) for correlation using the DiFX software correlators (Deller et al. 2011). The accumulation period adopted at correlation is 0.4 s, with a frequency resolution of 0.5 MHz. The clock model used during correlation to align the wavefronts arriving at different telescopes is imperfect, owing to an approximate a priori model for Earth's geometry as well as rapid stochastic variations in path length due to local atmospheric turbulence (Paper III). Before the data can be averaged coherently to build up signal-to-noise ratio (S/N), these effects must be accurately measured and corrected. This process, referred to as fringe fitting, was conducted using three independent software packages: the Haystack Observatory Processing System (HOPS; Whitney et al. 2004; Blackburn et al. 2019); the Common Astronomy Software Applications package (CASA; McMullin et al. 2007; Janssen et al. 2019a); and the NRAO Astronomical Image Processing System (AIPS; Greisen 2003, Paper III). Automated reduction pipelines were designed specifically to address the unique challenges related to the heterogeneity, wide bandwidth, and high observing frequency of EHT data. The field rotation angle is corrected with Equations (4)–(5), using coefficients given in Table 1. Flux density (amplitude) calibration is applied via a common post-processing framework for all pipelines (Blackburn et al. 2019; Paper III), taking into account estimated station sensitivities (Issaoun et al. 2017; Janssen et al. 2019b). Under the assumption of zero circular polarization of the primary (solar system) calibrator sources, elevation-independent station gains possess independent statistical uncertainties for the right-hand-circular polarization (RCP) and left-hand-circular polarization (LCP) signal paths, estimated to be ∼20% for the LMT and ∼10% for all other stations (Janssen et al. 2019b).

To remove the instrumental amplitude mismatch between the $LL^*$ and $RR^*$ visibility components (the $R$–$L$ phases are correctly calibrated in all scans by using ALMA as the reference station), calibration of the complex polarimetric gain ratios (the ratios of the $G_R$ and $G_L$ terms in the $G$ matrices) is performed. This is done by fitting global (multi-source, multi-days) piecewise polynomial gain ratios as functions of time. The aim of this approach is to preserve differences in $LL^*$ and $RR^*$ visibilities intrinsic to the source (Steel et al. 2019). After this step, preliminary polarimetric Stokes visibilities $\tilde{\mathcal{I}}$, $\tilde{\mathcal{Q}}$, $\tilde{\mathcal{U}}$, $\tilde{\mathcal{V}}$ can be constructed. However, the gain calibration requires significant additional improvements. The final calibration of the station phase and amplitude gains takes place in a self-calibration step as part of imaging or modeling the Stokes $\mathcal{I}$ brightness distribution, preserving the complex polarimetric gain ratios (e.g., Papers IV, VI). Fully calibrating the D-terms requires modeling the polarized emission.

The Stokes $\mathcal{I}$ (total intensity) analysis of a subset of the 2017 observations (Science Release 1 (SR1)), including M87, was the subject of Papers I–VI. The quality of these Stokes $\mathcal{I}$ data was assured by a series of tests covering self-consistency over bands and parallel hand polarizations, and consistency of trivial closure quantities (Wielgus et al. 2019). Constraints on the residual non-closing errors were found to be at a 2% level.

For additional information on the calibration, data reduction, and validation procedures for EHT, see Paper III. Information about accessing SR1 data and the software used for analysis can be found on the EHT website's data portal.[132] In this Letter, we utilize the HOPS pipeline full-polarization band-averaged (i.e., averaged over frequency within each band) and 10-second averaged data set from the same reduction path as SR1, but containing a larger sample of calibrator sources for polarimetric leakage studies. In addition, the ALMA linear-polarization observing mode allows us to measure and recover the absolute EVPA in the calibrated VLBI visibilities (Martí-Vidal et al. 2016; Goddi et al. 2019). Other minor subtleties in the handling of polarimetric data are presented in Appendix A.

**Table 1**
Field Rotation Parameters for the EHT Stations

| Station | Receiver Location | $f_{par}$ | $f_{el}$ | $\phi_{off}^\circ$ |
|---|---|---|---|---|
| ALMA | Cassegrain | 1 | 0 | 0 |
| APEX | Nasmyth-Right | 1 | 1 | 0 |
| JCMT | Cassegrain | 1 | 0 | 0 |
| SMA | Nasmyth-Left | 1 | −1 | 45 |
| LMT | Nasmyth-Left | 1 | −1 | 0 |
| SMT | Nasmyth-Right | 1 | 1 | 0 |
| PV | Nasmyth-Left | 1 | −1 | 0 |
| SPT | Cassegrain | 1 | 0 | 0 |

---

[131] In the EHT array, there are stations with a co-located element of the array: ALMA and APEX (with ∼2 km baseline) and JCMT and SMA (with ∼0.2 km baseline). We further refer to these two baselines as *zero* baselines or *intra-site* baselines.

[132] https://eventhorizontelescope.org/for-astronomers/data

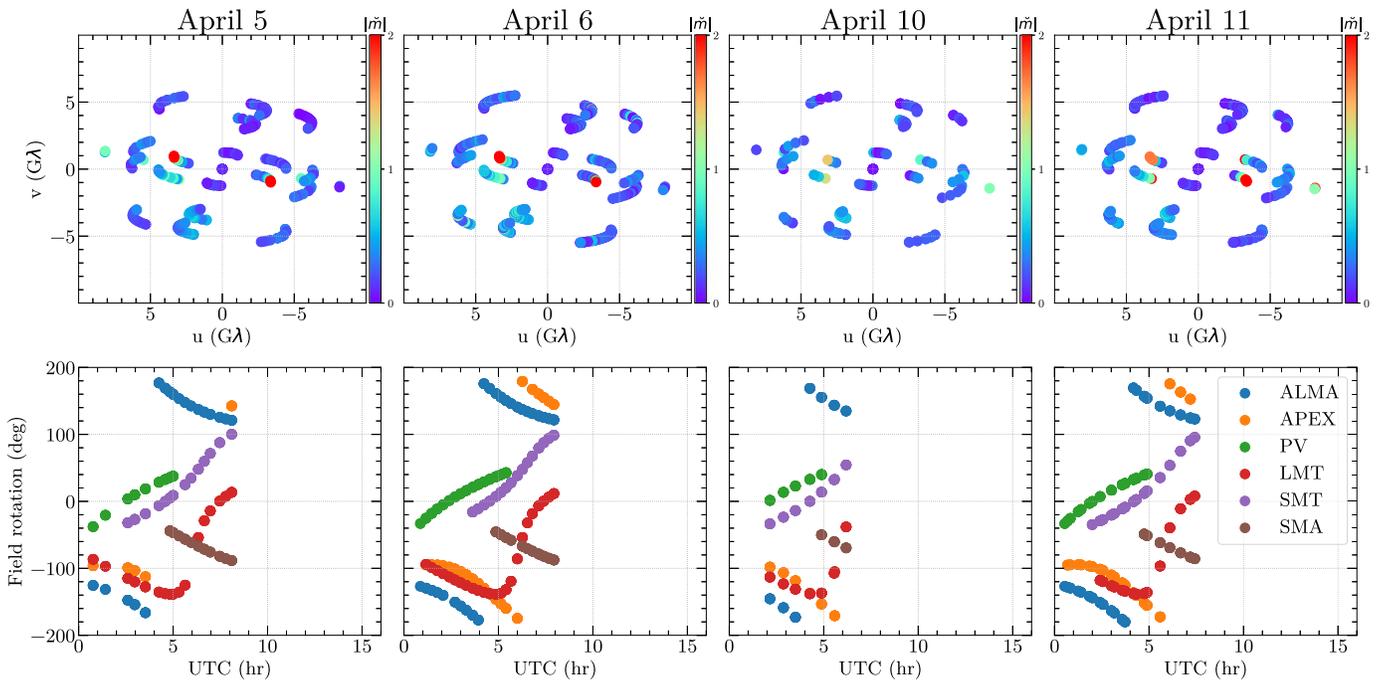

### 3.3. Polarimetric Data Properties

In Figure 1 (top row), we show the $(u, v)$ coverage and low-band interferometric polarization of our main target M87 as a function of the baseline $(u, v)$ after the initial calibration stage but before D-term calibration. The colors code the scan-averaged amplitude of the complex fractional polarization $\breve{m}$ (i.e., the fractional polarization in visibility space; for analysis of $\breve{m}$ in another source, Sgr A*, see Johnson et al. 2015). M87 is weakly polarized on most baselines, $|\breve{m}| \lesssim 0.5$. Several data points on SMA–SMT baselines have very high fractional polarization $|\breve{m}(u, v)| \sim 2$ that occur at $(u, v)$ spacings where the Stokes $\mathcal{I}$ visibility amplitude enters a deep minimum. The fractional polarization $\breve{m}$ of the M87 core is broadly consistent across the four days of observations and between low- and high-frequency bands, therefore high-band results are omitted in the display.

In Figure 1 (bottom row) we show the field rotation angles $\phi$ for each station observing M87 on the four observing days. The data are corrected for this angle during the initial calibration stage, but the precision of the leakage calibration depends on how well this angle is covered and on the difference in the field angles at the two stations forming a baseline. In the M87 data the field rotation for stations forming long baselines (LMT, SMT, and PV) is frequently larger than $100°$ except for April 10, for which the $(u, v)$ tracks are shorter.

In addition to the M87 data, a number of calibrators are utilized in this Letter for leakage calibration studies. To estimate D-terms for each of the EHT stations we use several EHT targets observed near in time to M87. In VLBI, weakly polarized sources are more sensitive to polarimetric calibration errors so they are preferred calibrators. For full-array leakage calibration, we focus on two additional sources: J1924–2914 and NRAO 530 (calibrators for the second EHT primary target, Sgr A*), which are compact and relatively weakly polarized. The main calibrator for M87 in total intensity, 3C 279

(Kim et al. 2020; Paper IV), is bright and strongly polarized on longer baselines and is not used in this work. The properties and analysis of the calibrators are discussed in more detail in Appendices J and K.

The closure traces for M87 and the calibrators can be used both to probe the data for uncalibrated systematic effects (see Appendix B.2) and to ascertain the presence of polarized flux density in a calibration-insensitive manner (see Appendix B.3).

Unless otherwise stated, the following analysis is focused on the low-band half of the data sets.

## 4. Methods for Polarimetric Imaging and Leakage Calibration

### 4.1. Methods

Producing an image of the linearly polarized emission requires both solving for the sky distribution of Stokes parameters $\mathcal{Q}$ and $\mathcal{U}$ and for the instrumental polarization of the antennas in the EHT array. In this work, we use several distinct methods to accomplish these tasks. Our approaches can be classified into three main categories: imaging via sub-component fitting; imaging via regularized maximum likelihood; and imaging as posterior exploration. In this section we only briefly describe each method; fuller descriptions are presented in Appendix C.

The calibration of the instrumental polarization by sub-component fitting was performed using three different codes (LPCAL, GPCAL, and polsolve) that depend on two standard software packages for interferometric data analysis: AIPS[133] and CASA.[134] In all of these methods, the Stokes $\mathcal{I}$ imaging step is performed using the CLEAN algorithm (Högbom 1974), and sub-components with constant complex fractional polarization

---

are then constructed from collections of the total intensity CLEAN components and fit to the data. In AIPS, two algorithms for D-term calibration are available: LPCAL (extensively used in VLBI polarimetry for more than 20 years; Leppänen et al. 1995) and GPCAL[135] (Park et al. 2021). In CASA, we use the polsolve algorithm (Martí-Vidal et al. 2021), which uses data from multiple calibrator sources simultaneously to fit polarimetric sub-components and allows for D-terms to be frequency dependent (see Appendix D). In all sub-component fitting and imaging methods, we assume that Stokes $\mathcal{V} = 0$. Further details on LPCAL, GPCAL, and polsolve can be found in Appendix C.1.

Image reconstruction via the Regularized Maximum Likelihood (RML) method was used in Paper IV along with CLEAN to produce the first total intensity images of the 230 GHz core in M87. RML algorithms find an image that maximizes an objective function composed of a likelihood term and regularizer terms that penalize or favor certain image features. In this work, we use the RML method implemented in the eht-imaging[136] software library (Chael et al. 2016, 2018) to solve for images in both total intensity and linear polarization. Like the CLEAN-based methods, eht-imaging does not solve for Stokes $\mathcal{V}$. Details on the specific imaging methods in eht-imaging used in the reconstructions presented in this work can be found in Appendix C.2.

Imaging as posterior exploration is carried out using two independent Markov chain Monte Carlo (MCMC) schemes: D-term Modeling Code (DMC) and THEMIS. Both codes simultaneously explore the posterior space of the full Stokes image (including Stokes $\mathcal{V}$) alongside the complex gains and leakages at every station; station gains are permitted to vary independently on every scan, while leakage parameters are modeled as constant in time throughout an observation. We provide more detailed model specifications for both codes in Appendix C.3 and in separate publications (Pesce 2021; A. E. Broderick et al. 2021, in preparation).

Hereafter, we often refer to eht-imaging, polsolve, and LPCAL methods as *imaging* methods/pipelines and to DMC and THEMIS methods as *posterior exploration* methods/ pipelines.

### 4.2. Leakage and Gain Calibration Strategy

In the imaging methods we divide the polarimetric calibration procedure for EHT data into two steps. In the first step, we calibrate the stations with an intra-site partner (ALMA–APEX, SMA–JCMT) using the assumption that sources are unresolved on intra-site baselines, where the brightness distribution can be approximated with a simple point source model. In the imaging pipelines we apply the D-terms for ALMA, APEX, and SMA to the data before polarimetric imaging and D-term calibration of the remaining stations. Baselines to the JCMT (which are redundant with SMA baselines) are removed from the data sets, to reduce complications from handling single-polarization data. The

ALMA, APEX, and SMA D-terms are fixed in imaging with eht-imaging and polsolve; because LPCAL is unable to fix D-terms of specific stations to zero, it derives a residual leakage for these stations, which remains small.[137] In the second step, we perform simultaneous imaging of the source brightness distribution and D-term calibration of stations for which only long, source-resolving baselines are available. In contrast, the posterior exploration pipelines do not use the D-terms derived using the intra-site baseline approach and instead solve for all D-terms (and station gains) starting with the base data product described in Section 3.2.

The point source assumption adopted in the imaging method intra-site baseline D-term calibration step is an extension to the intra-site redundancies already exploited in the EHT network calibration (Paper III), allowing us to obtain a model-independent gain calibration for ALMA, APEX, SMA, and JCMT. For an unresolved, slowly evolving source we can assume the true parameters of the coherency matrix $\rho_{jk}$ in Equation (6) to be constant during a day of observations, as very low spatial frequencies $\boldsymbol{u}$ are sampled, $\rho_{jk} \approx \rho_{jk}(\boldsymbol{u} = 0)$. Hence, only four intrinsic visibility components of $\rho_{jk}$ per source and four complex D-terms (two for each station) need to be determined from all the data on an available baseline.

We fit the D-terms of ALMA, APEX, JCMT, and SMA for each day using the multi-source feature of polsolve, combining band-averaged observations of multiple sources (3C 279, M87, J1924–2914, NRAO 530, 3C 273, 1055+018, OJ287, and Cen A as shown in Appendix D) on each day in one single fit per band. The results of these fits per station, polarization, day, and band are presented in Figure 2 (left panel), where we also plot the mean and standard deviation of the D-terms across all days and both bands for each station and polarization. In Appendix D, we provide tables with D-term values and further discuss the time and frequency dependence of D-terms and JCMT single polarization handling. In Appendix D we also present several validation tests of our intra-site baseline D-term estimation method carried out to motivate the use of band-averaged data products, comparisons to independent polarimetric source properties measured from simultaneous interferometric-ALMA observations (Goddi et al. 2021) near-in-time interferometric-SMA leakage estimates, and comparisons to results from a model fitting approach.

In addition to intra-site baseline D-term calibration in the imaging pipelines, we also account for residual station-based amplitude gain errors by calibrating the data to pre-determined fiducial Stokes $\mathcal{I}$ images of chosen calibrator sources. Given the extreme resolving power of the EHT array, all available calibrators are resolved on long baselines. Therefore, we must select sources that are best imaged by the EHT array; these are compact non-variable sources with sufficient $(u, v)$ coverage. Four targets in the EHT 2017 observations fit these criteria: M87, 3C 279, J1924–2914, and NRAO 530. Stokes $\mathcal{I}$ images of M87 and 3C 279 have been published (Papers I–VI; Kim et al. 2020). Final Stokes $\mathcal{I}$ images for the Sgr A* calibrators NRAO 530 and J1924–2914 will be presented in upcoming publications (S. Issaoun et al. 2021, in preparation; S. Jorstad et al. 2021, in preparation) but the best available preliminary images are used to self-calibrate our visibility data for D-term comparisons in this Letter (Appendix J).

---

[135] GPCAL is a new automated pipeline written in Python and based on AIPS and the CLEAN imaging software Difmap. GPCAL adopts a similar calibration scheme to LPCAL but allows users to (i) fit the D-term model to multiple calibrators simultaneously and (ii) use more accurate linear polarization models of the calibrators for D-term estimation. In this Letter, we use GPCAL to complement the LPCAL analysis of the M87 data (Appendices G.3 and K) and the D-term estimation using calibrators (Appendix J). We do not show GPCAL results in the main text.

[136] https://github.com/achael/eht-imaging

[137] The non-zero LPCAL D-terms for ALMA, APEX, and SMA indicate that there may either be possible residual leakage after intra-site baseline fitting or that uncertainties in the LPCAL estimates originate from e.g., a breakdown of the similarity approximation.

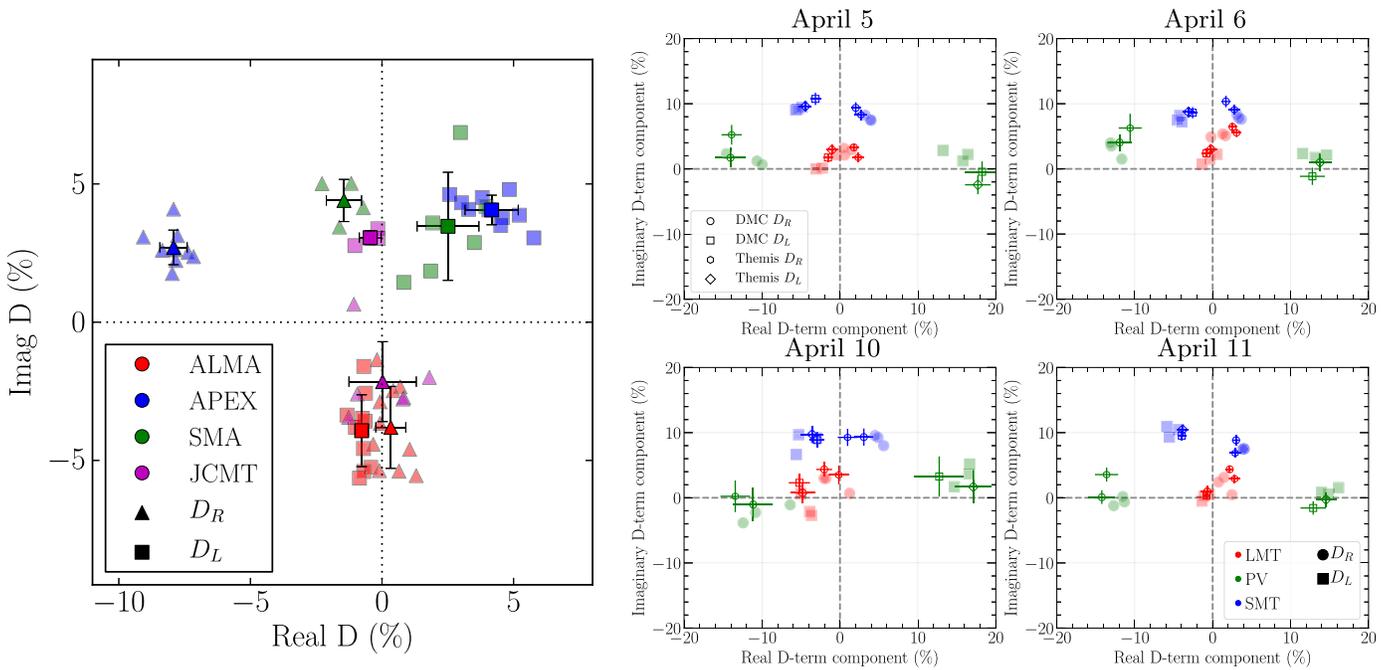

**Figure 2.** Left panel: D-term estimates for ALMA, APEX, JCMT, and SMA from `polsolve` multi-source intra-site baseline fitting; one point per day and band (low and high) for each station across the EHT 2017 campaign. Both polarizations are shown for ALMA and APEX per day, but only one polarization is shown for JCMT and SMA per day due to JCMT polarization setup limitations. Station averages across days and high/low bands are shown as solid points with error bars. The depicted D-terms are provided in tabulated form in Appendix D. Right panels: fiducial D-terms for LMT, PV, and SMT derived from the low-band data via leakage calibration in tandem with polarimetric imaging methods and posterior modeling of M87 observations. We depict fiducial D-terms per day, where each point corresponds to one station, polarization, and method. Filled symbols depict D-terms from imaging methods and symbols for posterior exploration methods have error bars corresponding to the 1σ standard deviations estimated from the posterior distributions of the resulting D-terms.

For M87, although multiple imaging packages and pipelines were utilized in the Stokes $\mathcal{I}$ imaging process, the resulting final "fiducial" images from each method are highly consistent at the EHT instrumental resolution (e.g., Paper IV, Figure 15). Therefore, we selected a set of Stokes $\mathcal{I}$ images for self-calibration from the RML-based `SMILI` imaging software pipeline (Akiyama et al. 2017a, 2017b; Paper IV). The images we use for self-calibration are at `SMILI`'s native imaging resolution (~10 μas), which provide the best fits to the data and are not convolved with any restoring beam. We self-calibrate our visibility data to these images, thereby accounting for residual station gain variations in the data that make imaging challenging. Using these self-calibrated data sets allows the imaging methods to focus on accurate reconstructions of the polarimetric Stokes $\mathcal{Q}$ and $\mathcal{U}$ brightness distributions and D-term estimation.

Preliminary D-terms estimated by the three imaging methods before testing and optimizing imaging parameters on synthetic data are reported in Appendix F. The right panels of Figure 2 show the final D-terms for LMT, PV, and SMT derived from the imaging and posterior modeling methods after optimization on synthetic data (see Section 4.3). To quantify the agreement (or distance in the complex plane) between D-term estimates from different methods we calculate $L_1$ norms. The $L_1$ norms averaged over left and right (also real and imaginary) D-term components, over all stations and over the four observing days, are all less than 1% for each pair of imaging methods (see Figure 20 in Appendix E). The mean values of the D-terms from the posterior exploration methods correlate well with the D-terms estimated by the imaging methods. For each combination of imaging and posterior exploration method the station averaged $L_1$ norms range from 1.5% to 1.89%.

DMC is the only method that solves for independent left-and right-circular station gains. The ratio of the gains derived by DMC on April 11 is shown in Figure 3. As expected, the assumption made by all of the imaging pipelines and one of the posterior exploration pipelines (THEMIS), that right- and left-hand gains are equal for all stations at all times, holds. For verification purposes, we also estimate D-terms using data of several calibrator sources. We find that the D-terms derived by polarimetric imaging of these other sources are consistent with those of M87 (Appendix J). Finally, we note our estimated SMT D-terms are similar to those computed previously using early EHT observations of Sgr A* (Johnson et al. 2015).

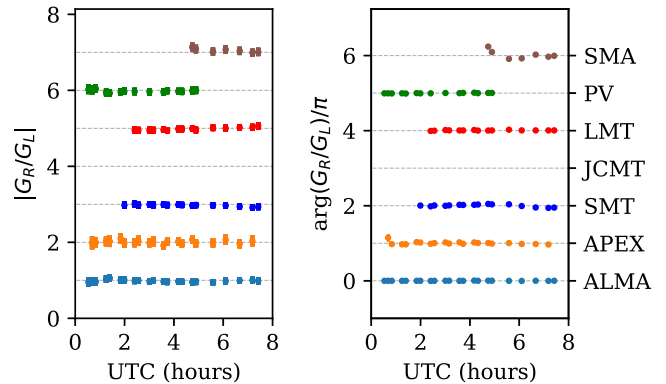

**Figure 3.** Amplitudes (left panel) and phases (right panel) of the ratio of $R$ to $L$ station gains from the DMC fit to M87 2017 April 11 low-band data. Individual station gain ratios are offset vertically for clarity, with the dashed horizontal lines indicating a unit ratio for each station (i.e., unity for amplitudes and zero for phases). Note that JCMT only observes one polarization at a time, and so provides no constraints on gain ratios. We see that the assumption made by the three imaging pipelines and one posterior exploration pipeline (THEMIS)—namely, that the right- and left-hand gains are equal for all stations at all times—largely holds. The behavior in this plot is representative of that seen across days and bands.

### 4.3. Parameter Surveys and Validation on Synthetic Data

Each imaging and leakage calibration method has free parameters that must be set by the user before the optimization or posterior exploration takes place. Some of these parameters (e.g., field of view, number of pixels) are common to all methods, but many are unique to each method (e.g., the sub-component definitions in `LPCAL` or `polsolve`, or the regularizer weights in `eht-imaging`). In VLBI imaging, these parameters are often simply set by the user given their experience on similar data sets, or based on what appears to produce an image that is a good fit to the data and free of noticeable imaging artifacts. In this work, we follow Paper IV in choosing the method parameters that we use in our final image reconstructions more objectively by surveying a portion of the parameter space available to each method.

We perform surveys over the different free parameters available to each method and attempt to choose an optimal set of parameters based on their performance in recovering the source structure and input D-terms from several synthetic data models. Appendix G provides more detail on the individual parameter surveys performed by each method. The parameter set that performs best on the synthetic data for each method is considered the "fiducial" parameter set for imaging M87 with that method.[138] The corresponding images reconstructed from various data sets using these parameters are the method's "fiducial images."

The synthetic data sets that we used for scoring the imaging parameter combinations consist of six synthetic EHT observations using the M87 2017 April 11 equivalent low-band $(u, v)$ coverage. The source structure models used in the six sets vary from complex images generated using general relativistic magnetohydrodynamic (GRMHD) simulations of M87's core and jet base (Models 1 and 2 from Chael et al. 2019) to simple geometrical models (a filled disk, Model 3, and simple rings with differing EVPA patterns, Models 4–6). The synthetic source models have varying degrees of fractional polarization and diverse EVPA structures. The synthetic source models blurred to the EHT nominal resolution are displayed in the first column of Figure 4.

All M87 synthetic data sets were generated using the synthetic data generation routines in `eht-imaging`. We followed the synthetic data generation procedure in Appendix C.2 of Paper IV, but with models featuring complex polarization structure. The synthetic visibilities sampled on EHT baselines are corrupted with thermal noise, phase and gain offsets, and polarimetric leakage terms. Mock D-terms for the SMT, LMT, and PV stations were chosen to be similar to those found by the initial exploration of the M87 EHT 2017 data reported in Appendix F. Random residual D-terms for ALMA, APEX, JCMT, and SMA (reflecting possible errors in the intra-site baseline calibration procedure) were drawn from normal distributions with 1% standard deviation. After generation, the phase and amplitude gains in the synthetic data were calibrated for use in imaging pipelines in the same way as the real M87 data; that is, they were self-calibrated to a Stokes $\mathcal{I}$ image reconstructed via the `SMILI` fiducial script for M87 developed in Paper IV.

In Figure 4, we present our fiducial set of images (in a uniform scale) from synthetic data surveys carried within each method. In each panel we report a correlation coefficient $\langle I \cdot I_0 \rangle$ between recovered Stokes $\mathcal{I}$ and the ground-truth $\mathcal{I}_0$ images,

$$\langle I \cdot I_0 \rangle = \frac{\langle (\mathcal{I} - \overline{\mathcal{I}})(\mathcal{I}_0 - \overline{\mathcal{I}}_0) \rangle}{\sqrt{\langle (\mathcal{I} - \overline{\mathcal{I}})^2 \rangle} \sqrt{\langle (\mathcal{I}_0 - \overline{\mathcal{I}}_0)^2 \rangle}}. \quad (8)$$

This reflects the dot product of the two mean-subtracted images when treated as unit vectors. We also calculate a correlation coefficient for the reconstructed linear polarization image $\mathcal{P} \equiv \mathcal{Q} + i\mathcal{U}$,

$$\langle \vec{P} \cdot \vec{P}_0 \rangle = \frac{\mathrm{Re}[\langle \mathcal{P}\mathcal{P}_0^* \rangle]}{\sqrt{\langle \mathcal{P}\mathcal{P}^* \rangle} \sqrt{\langle \mathcal{P}_0\mathcal{P}_0^* \rangle}}. \quad (9)$$

The real part is chosen to measure the degree of alignment of the polarization vectors $(\mathcal{Q}, \mathcal{U})$. In both cases, images are first shifted to give the maximum correlation coefficient for Stokes $\mathcal{I}$. Because Stokes $\mathcal{I}$ image reconstructions are tightly constrained by an a priori known total image flux density, the Stokes $\mathcal{I}$ correlation coefficients are mean subtracted to increase the dynamic range of the comparison. This introduces a field-of-view dependence to the metric, as only spatial frequencies above (field of view)$^{-1}$ are considered; up to the beam resolution. There is no such dependence in the linear polarization coefficient, which is not mean subtracted.

The correlation is equally strong independently of the employed method. The polarization structure is more difficult to recover for models with high or complex extended polarization (Models 1 and 2) for which correlation of the recovered polarization vectors is strong to moderate. In Figure 5 we present a uniform comparison of the recovered D-terms and the ground-truth D-terms for all synthetic data sets and methods. For all methods the recovered D-terms show a strong correlation with the model D-terms. To quantify the agreement (or distance in the complex plane) between D-term estimates and the ground-truth values $D_{\mathrm{Truth}}$ in each approach, we calculate the $L_1 \equiv |D_i - D_{\mathrm{Truth}}|$ norm, where $D_i$ is a D-term component derived within a method $i$. Overall, for the fiducial set of parameters the agreement between the ground truth and the recovered D-terms in synthetic data measured using the $L_1$ norm is $\leqslant 1.3\%$ on average (when averaging is done over stations, D-term components, and models). The reported averaged $L_1$ norms give us a sense of the expected discrepancies in D-terms between employed methods for their fiducial set of parameters. However, we notice again that the discrepancies do depend on source structure. For example, in models with no polarization substructure (e.g., Model 3) all methods had difficulty in recovering D-terms for PV (visible as large error bars for the station), a station forming only very long baselines on a short $(u, v)$ track. If we exclude PV from the $L_1$ metrics the expected $L_1$ norms for LMT and SMT alone for all methods are $L_1 \sim 0.6\%$–$0.8\%$ when averaged over models.

---

[138] In the case of `LPCAL`, the parameter survey on synthetic data only constrained one parameter (the number of sub-components). For M87, the `LPCAL` pipeline explored choices in the other imaging and calibration parameters by a number of users, whose final D-terms were then synthesized to obtain the fiducial results. See Appendix G.3 for details.

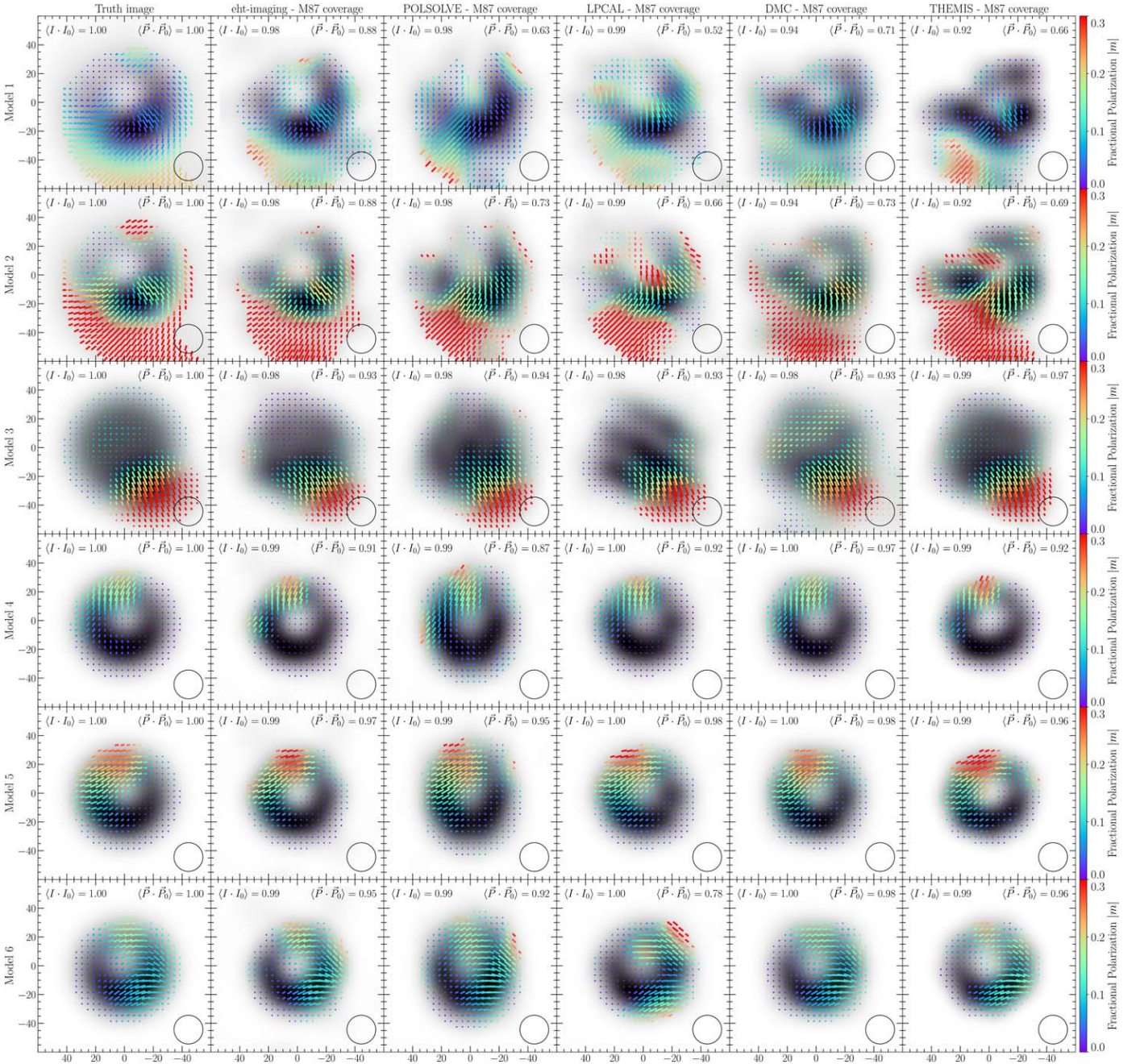

**Figure 4.** Fiducial images from synthetic data model reconstructions using M87 2017 April 11 low-band $(u, v)$ coverage. Rows from top to bottom correspond to six different synthetic data sets. Columns from left to right show ground-truth synthetic image (column 1) and the best image reconstructions by each method (columns 2–6). The polarization tick length reflects total linear polarization, while the color reflects fractional polarization from 0 to 0.3. The normalized overlap is calculated against the respective ground-truth image, and in the case of the total intensity it is mean-subtracted.

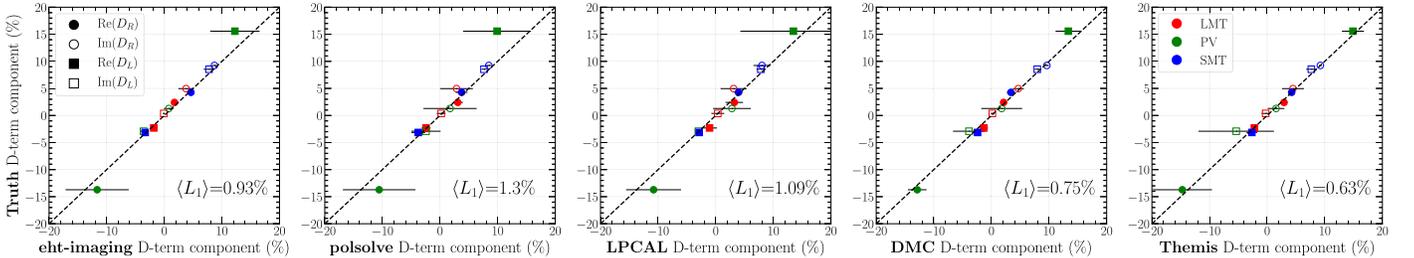

**Figure 5.** A comparison of LMT, SMT, and PV D-term estimates to ground-truth values in the synthetic data sets 1 through 6 (shown in Figure 4). Each panel shows correlation of the estimated and the truth D-terms for a single method. Each data point in each panel depicts an average and standard deviation for each D-term estimate derived from the six synthetic data sets. The norm $L_1 \equiv |D - D_{\mathrm{Truth}}|$ is averaged over left, right, real, and imaginary components of the D-terms and over all shown EHT stations. Notice that each method recovers the ground-truth D-terms to within ~1%, on average.

## 5. Results

### 5.1. Fiducial Polarimetric Images of M87

In Figure 6, we present the fiducial M87 linear-polarimetric images produced by each method from the low-band data on all four observing days. The fiducial images from each method are broadly consistent with those from the preliminary imaging stage shown in Figure 21 of Appendix F.

Unless otherwise explicitly indicated, we display low-band results in the main text. The high-band results are given in Appendix I. We decided to keep the analysis of the high- and low-band data separate for several reasons. First, the main limitations in the dynamic range and image fidelity in EHT reconstructions arise from the sparse sampling of spatial frequencies, not the data S/N. Increasing the S/N by performing band averaging does not improve the dynamic range of the reconstructed images. Second, treating each band separately minimizes any potential chromatic effects that might add extra limitations to the dynamic range, such as intra-field differential Faraday rotation. Finally, separating the bands in the analysis allowed us to use the high-band results as a consistency check on the calibration of the instrumental polarization and image reconstruction for the low-band data. We perform this comparison of the results obtained at both bands in Appendix I. We conclude that both the recovered D-terms and main image structures are broadly consistent between the low and high bands.

The different reconstruction methods have different intrinsic resolution scales; for instance, the CLEAN reconstruction methods model the data as an array of point sources, while the RML and MCMC methods have a resolution scale set by the pixel size. In Figure 6, we display the fiducial images from each method at the same resolution scale by convolving each with a circular Gaussian kernel with a different FWHM. The FWHM for each method is set by maximizing the normalized cross-correlation of the blurred Stokes $\mathcal{I}$ image with the April 11 "consensus" image presented in Figure 15 of Paper IV. The blurring kernel FWHMs selected by this method are 19 $\mu$as for eht-imaging, DMC, and THEMIS, 20 $\mu$as for LPCAL, and 23 $\mu$as for polsolve.

The M87 emission ring is polarized only in its southwest region and the peak fractional polarization at $\approx 20\ \mu$as resolution is at the level of about 15%. The residual rms in linear polarization (as estimated from the CLEAN images) is between 1.10–1.30 mJy/beam in all epochs, which implies a polarization dynamic range of $\sim 10$. The nearly azimuthal EVPA pattern is a robust feature evident in all our reconstructions across time, frequency, and imaging method. The images show slight differences in the polarization structure between the first two days, 2017 April 5/6 and the last two, 2017 April 10/11. Notably, the southern part of the ring appears less polarized on the later days. This evolution in the polarized brightness is consistent with the evolution in the Stokes $\mathcal{I}$ image apparent in the underlying closure phase data (Paper III, Figure 14; Paper IV, Figure 23). However, as with the Stokes $\mathcal{I}$ image, the structural changes in the polarization images with time over this short timescale (6 days $\approx 16\ GM/c^3$) are relatively small, and it is difficult to disentangle which differences in the polarized images are robust and which are influenced by differences in the interferometric ($u$, $v$) coverage between April 5 and 11 (Paper IV, Section 8.3).

In Figure 7, we show the simple average of the five equivalently blurred fiducial images (one per method) for each of the four observed days. The averaging is done independently for each Stokes intensity distribution. These method-averaged images are consistent with the EHT closure traces, as shown in Figure 13 in Appendix B. We adopt the images in Figure 7 as a conservative representation of our final M87 polarimetric imaging results.

### 5.2. Azimuthal Distribution of the Polarization Brightness

While the overall pattern of the linearly polarized emission from M87 is consistent from method to method, the details of the emission pattern can depend sensitively on the remaining statistical uncertainties in our leakage calibration. In addition, the different assumptions and parameters used in each reconstruction method affect the recovered polarized intensity pattern, introducing an additional source of systematic uncertainty in our recovered images. In this section, we assess the consistency of the recovered polarized images across different D-term calibration solutions within and between methods.

We explore the consistency of our image reconstructions against the uncertainties in the calibrated D-terms by generating a sample of 1000 images for each method, each generated with a different D-term solution. For the imaging methods, we define complex normal distributions for each D-term based on the scatter in recovered D-terms in Figure 2 and reconstruct images after calibrating to each set of random D-terms without additional calibration. This procedure is explained in detail in Appendix H. For the posterior exploration methods we simply draw 1000 images from the posterior for each observing day.

In each method's set of 1000 image samples covering a range of D-term calibration solutions, we study the azimuthal distribution of the polarization brightness ($p$) and EVPA ($\chi$) by performing intensity-weighted averages of these quantities over different angular sections along the ring. The width of the angular sections used in the averaging is set to $\Delta\varphi = 10°$ and the averages are computed from a position angle $\varphi = 0°$ to $\varphi = 360°$, in steps of $1°$.

Comparing angular averages of these quantities with a small moving window $\Delta\varphi$ avoids spurious features from the different pixel scales used in the different image reconstruction methods. The pixel coordinates of the image center are estimated (for each method) from the peak of the cross-correlation between the $\mathcal{I}$ images and the representative images of M87 used in the self-calibration. To avoid the effects of phase wrapping in the averaging (which biases the results for values of $\chi$ around $\pm 90°$), the quantity $\langle \chi \rangle$ is computed coherently within each angular section, i.e., the averages are defined as

$$\langle \mathcal{P} \rangle \equiv \frac{\langle \mathcal{I}\sqrt{\mathcal{Q}^2 + \mathcal{U}^2} \rangle}{\langle \mathcal{I} \rangle}, \tag{10}$$

$$\langle \chi \rangle \equiv \frac{1}{2} \arctan\left( \frac{\langle \mathcal{Q} \times \mathcal{I} \rangle}{\langle \mathcal{U} \times \mathcal{I} \rangle} \right). \tag{11}$$

In Figure 8, we show histograms of these quantities for two days, 2017 April 5 and 11, as a function of the orientation of the angular section used in the averaging (i.e., the position

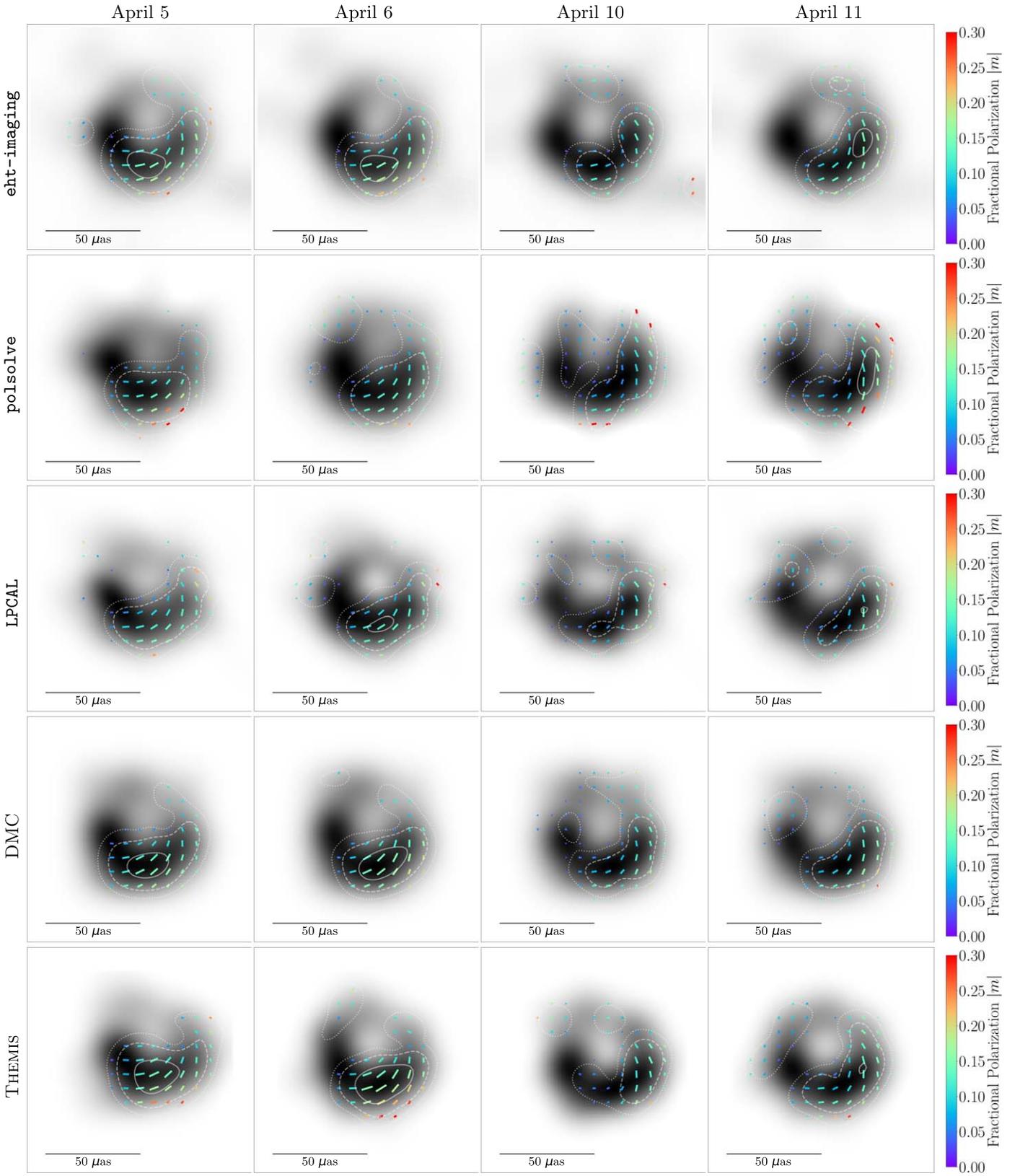

**Figure 6.** Fiducial polarimetric M87 images produced by five independent methods. The results from all imaging and posterior exploration pipelines are shown on the four M87 observation days for the low band data (the low- and high-band results are consistent, see Appendix I). Total intensity is shown in grayscale, polarization ticks indicate the EVPA, the tick length indicates linear polarization intensity magnitude (where a tick length of 10 $\mu$as corresponds to $\sim$30 $\mu$Jy $\mu$as$^{-2}$ of polarized flux density), and color indicates fractional linear polarization. The tick length is scaled according to the polarized brightness without renormalization to the maximum for each image. The contours mark the linear polarized intensity. The solid, dashed, and dotted contour levels correspond to linearly polarized intensity of 20, 10, and 5 $\mu$Jy $\mu$as$^{-2}$, respectively. Cuts were made to omit all regions in the images where Stokes $\mathcal{I} < 10\%$ of the peak brightness and $\mathcal{P} < 20\%$ of the peak polarized brightness. The images are all displayed with a field of view of 120 $\mu$as, and all images were brought to the same nominal resolution by convolution with the circular Gaussian kernel that maximized the cross-correlation of the blurred Stokes $\mathcal{I}$ image with the consensus Stokes $\mathcal{I}$ image of Paper IV.

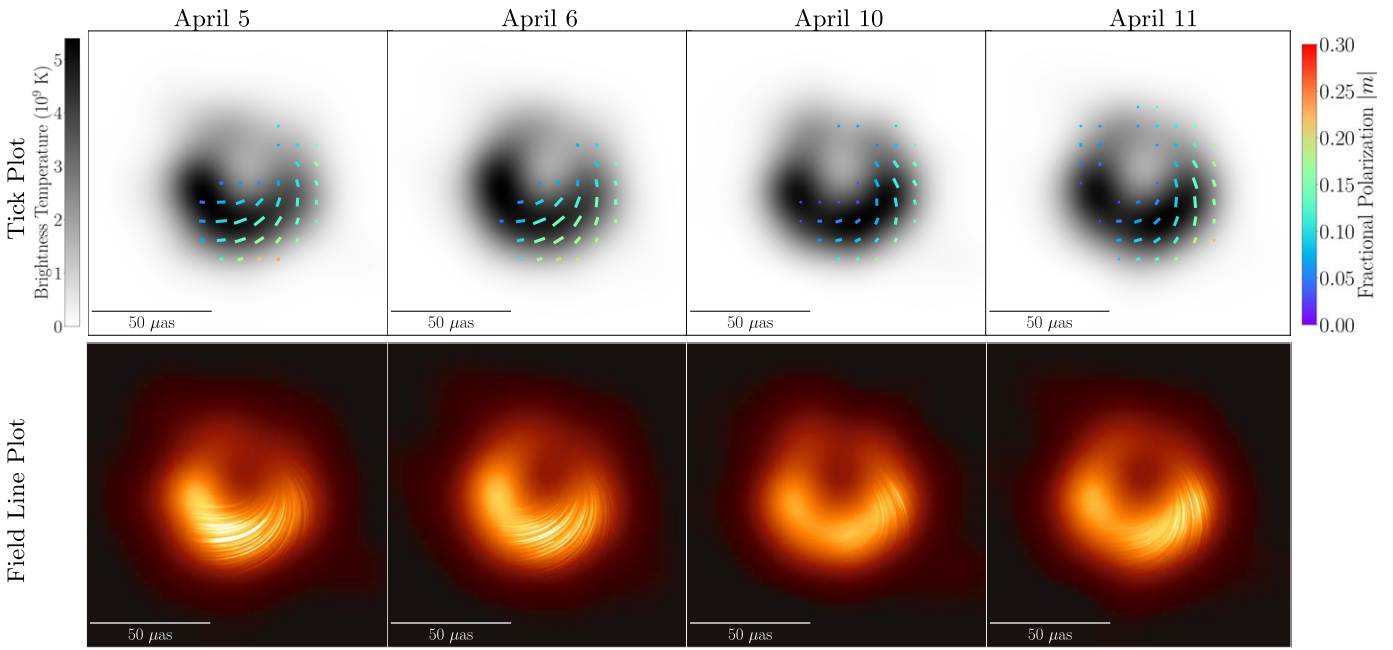

April 5      April 6      April 10      April 11

**Figure 7.** Fiducial M87 average images produced by averaging results from our five reconstruction methods (see Figure 6). Method-average images for all four M87 observation days are shown, from left to right. These images show the low-band results; for a comparison between these images and the high-band results, see Figure 28 in Appendix I. We employ here two visualization schemes (top and bottom rows) to display our four method-average images. The images are all displayed with a field of view of 120 $\mu$as. Top row: total intensity, polarization fraction, and EVPA are plotted in the same manner as in Figure 6. Bottom row: polarization "field lines" plotted atop an underlying total intensity image. Treating the linear polarization as a vector field, the sweeping lines in the images represent streamlines of this field and thus trace the EVPA patterns in the image. To emphasize the regions with stronger polarization detections, we have scaled the length and opacity of these streamlines as the square of the polarized intensity. This visualization is inspired in part by Line Integral Convolution (Cabral & Leedom 1993) representations of vector fields, and it aims to highlight the newly added polarization information on top of the standard visualization for our previously published Stokes $\mathcal{I}$ results (Papers I, IV).

angle around the ring). We consider these two days because they have the best $(u, v)$ coverage and span the full observation window; these results will thus include any effects of intrinsic source evolution in the recovered parameters. From Figure 8, it is evident that the difference in $\langle p \rangle$ between methods is larger than the widths of the $\langle p \rangle$ histograms in each method. This means that effects related to the residual instrumental polarization, giving rise to the dispersion seen in the histograms, are smaller than artifacts related to the deconvolution algorithms. In other words, the $\langle p \rangle$ images are limited by the image fidelity due to the sparse $(u, v)$ coverage rather than by the D-terms.

Even though there are differences among methods in the $p$ azimuthal distribution, some features are common to all our image reconstructions. The peak in the polarization brightness is located near the southwest on 2017 April 5 (at a position angle of $199° \pm 11°$, averaged among all methods) and close to the west on 2017 April 11 (position angle of $244° \pm 10°$). That is, the polarization peak appears to rotate counter-clockwise between the two observing days (see the dotted lines in Figure 8). On both days, the region of high polarization brightness is relatively wide, covering a large fraction of the southern portion of the image (position angles from around $100°-300°$).

In the azimuthal distribution of $\langle \chi \rangle$, all methods produce very similar values in the part of the image with the highest polarized brightness (the southwest region, between position angles of $180°$ and $270°$). The EVPA varies almost linearly, from around $\langle \chi \rangle = -80°$ (in the south) up to around $\langle \chi \rangle = 30°$ (in the east). The EVPAs on 2017 April 11 are slightly higher

(i.e., rotated counter-clockwise) compared to those on 2017 April 5. This difference is clearly seen for eht-imaging, polsolve, and THEMIS, though the difference is smaller for DMC and LPCAL. We notice, though, that the differences in the EVPAs between days could also be affected by small shifts in the estimates of the image center on each day. Outside of the region with high polarization, the EVPA distributions for all methods start to depart from each other. There is a hint of a constant EVPA $\langle \chi \rangle \sim 0°$ in the northern region (i.e., position angles around $0°-50°$) in polsolve and LPCAL on both days, but the other methods show larger uncertainties in this region.

The discrepancies in EVPA among all methods only appear in the regions with low brightness (i.e., around the northern part of the ring). Therefore, polarization quantities defined from intensity-weighted image averages, discussed in the next sections, will be dominated by the regions with higher brightness, for which all methods produce similar results. Image-averaged quantities are somewhat more robust to differences in the calibration and image reconstruction algorithms, though they are not immune to systematic errors.

### 5.3. Image-averaged Quantities

In comparing polarimetric images of M87, we are most interested in identifying acceptable ranges of three image-averaged parameters that are used to distinguish between different accretion models in Paper VIII: the net linear polarization fraction of the image $|m|_{\rm net}$, the average polarization fraction in the resolved image at 20 $\mu$as resolution $\langle |m| \rangle$, and the $m = 2$ coefficient of the azimuthal mode decomposition

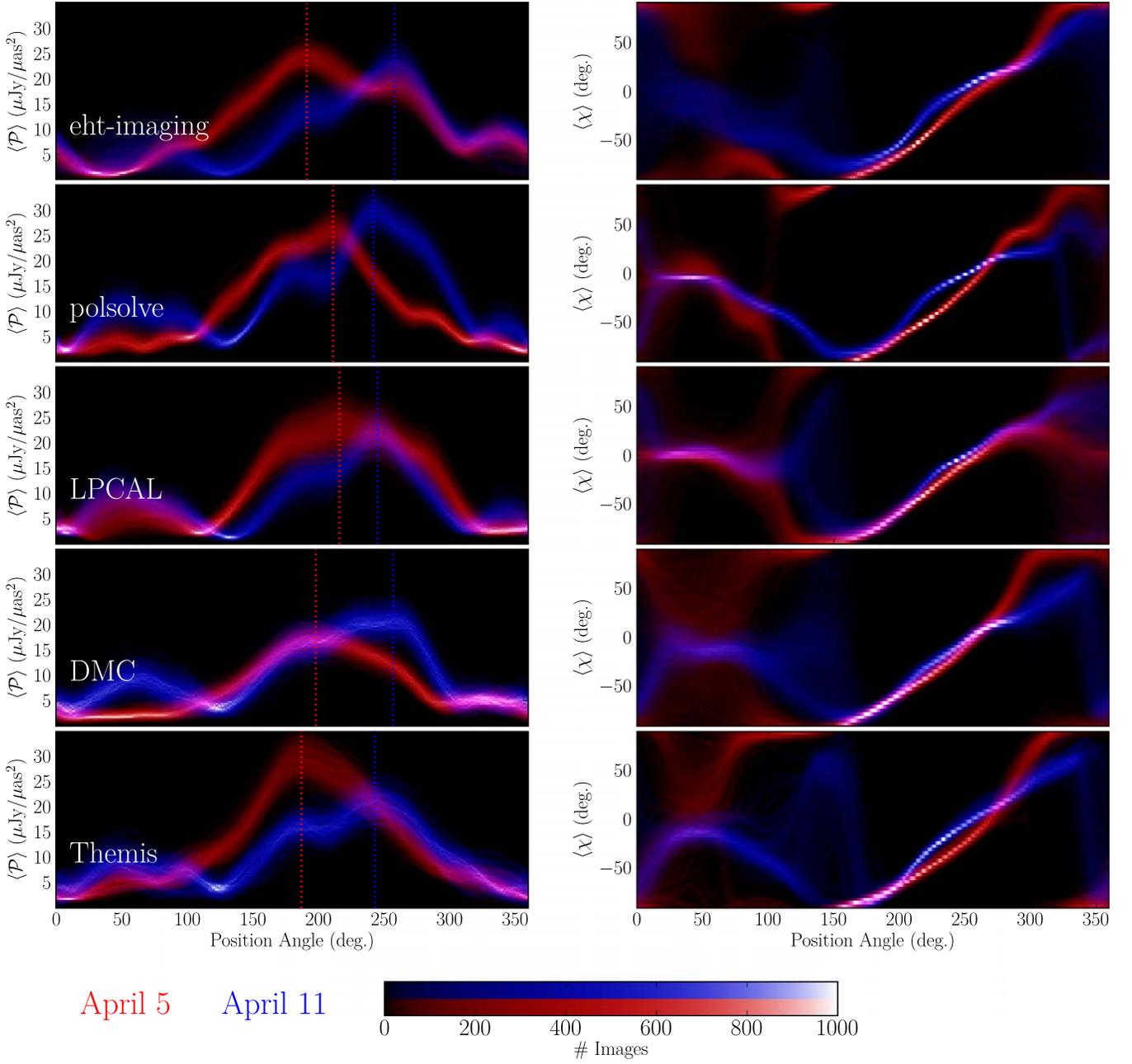

April 5    April 11

**Figure 8.** Histograms of the azimuthal distributions of polarized intensity (left panel) and EVPA (right panel) obtained from the low-band data in the survey of different D-term solutions with all five imaging and posterior exploration methods. These quantities are estimated as the intensity-weighted averages within an angular section of a width of 10°. The position angle is measured counter-clockwise, starting from the north. The position angles with the highest average polarization brightness are marked with dotted lines for each method and day.

of the polarized brightness $\beta_2$. These parameters are defined below.

First, the net linear polarization fraction of the image is

$$|m|_{\rm net} = \frac{\sqrt{\left(\sum_i Q_i\right)^2 + \left(\sum_i U_i\right)^2}}{\sum_i I_i}, \qquad (12)$$

where the sum is over the pixels indexed by $i$. ALMA measured $|m|_{\rm net} = 2.7\%$ on 2017 April 11 (Goddi et al. 2021), but this measurement includes emission at large scales outside of the 120 $\mu$as field of view of the EHT images. We also consider the intensity-weighted average polarization fraction

across the resolved EHT image:

$$\langle |m| \rangle = \frac{\sum_i \sqrt{Q_i^2 + U_i^2}}{\sum_i I_i}. \qquad (13)$$

The value of $\langle |m| \rangle$ is determined by the intensity of the polarized emission at each point in the image, and thus it is sensitive to the resolution of the image and the choice of restoring beam. Specifically, images restored with beams of larger FWHM will tend to be more locally depolarized and thus have lower $\langle |m| \rangle$ than images restored with beams of smaller FHWM. In contrast, the integrated polarization fraction $|m|_{\rm net}$ is insensitive to convolution.

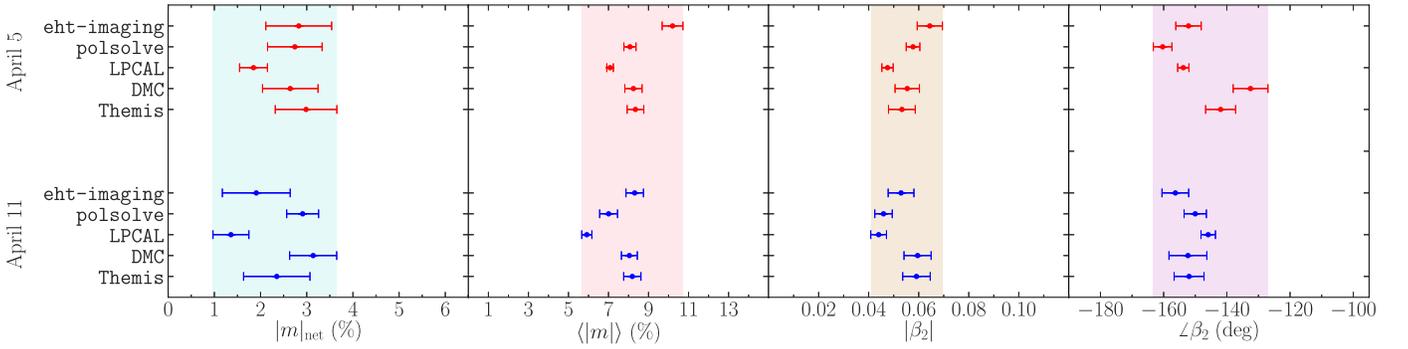

**Figure 9.** Summary of the key quantities used in Paper VIII measured by each method from the low-band data on both 2017 April 5 and 11. From left to right, the quantities are the integrated net polarization $|m|_{net}$ (Equation (12)), the average polarization fraction $\langle|m|\rangle$ (Equation (13)), and the amplitude $|\beta_2|$ and phase $\angle\beta_2$ of the $m = 2$ azimuthal mode of the complex polarization brightness distribution (Equation (14)). The shaded bands show the consensus ranges (Table 2) incorporating both uncertainties in these parameters from the D-term calibration and systematic discrepancies between image reconstruction methods. A comparison with analogous high-band results is provided in Figure 29.

We quantify the polarization *structure* with a decomposition into azimuthal modes. In particular, Paper VIII considers the complex amplitude $\beta_2$ of the $m = 2$ mode defined in Palumbo et al. (2020), who found this mode to be the most important in distinguishing different modes of accretion from 230 GHz images produced by different GRMHD simulations. The $\beta_2$ azimuthal mode decomposition coefficient is defined as

$$\beta_2 = \frac{1}{I_{ring}} \int_{\rho_{min}}^{\rho_{max}} \int_0^{2\pi} P(\rho, \varphi)\, e^{-2i\varphi}\, \rho\, d\varphi\, d\rho, \qquad (14)$$

where $(\rho, \varphi)$ are polar coordinates in the image plane, and $I_{ring}$ is the Stokes $\mathcal{I}$ flux density in the ring between the minimum radius $\rho_{min}$ and the maximum radius $\rho_{max}$. Because our reconstruction methods recover no significant extended brightness off the main ring, we take $\rho_{min} = 0$ and extend $\rho_{max}$ to encompass the full-image field of view, so $I_{ring}$ is equal to the total Stokes $\mathcal{I}$ flux density in the image.

Note that both the amplitude $|\beta_2|$ and phase $\angle\beta_2$ depend on the choice of image center, image resolution, and restoring beam size. In the comparisons that follow, we convolve images from each method with circular Gaussian beams with the FWHMs specified in Section 5.1 chosen to bring all images to the same resolution scale. Furthermore, we center each reconstruction by finding the pixel offset that maximizes the cross-correlation between the blurred Stokes $\mathcal{I}$ image and the 2017 April 11 consensus Stokes $\mathcal{I}$ image from Paper IV. In general, we find these offsets to be small, and our results do not change significantly if we do not apply any centering procedure in calculating $\beta_2$ from our reconstructed images.

From the sets of 1000 images generated by each method to explore variations of the image structure with the D-term solution, we compute distributions of each key metric—$|m|_{net}$, $\langle|m|\rangle$, $|\beta_2|$, and $\angle\beta_2$—that is used in Paper VIII for theoretical interpretation. These distributions are summarized in Figure 9, which displays the mean points and $1\sigma$ error bars from different D-term realizations for all four methods on both 2017 April 5 and 11. We present a more complete look at these distributions with histograms for each quantity from each method/day pair in Appendix H, Figures 25 and 26. Figure 9 shows results for low-band images only; we compare these results to results derived from the high-band data in Figure 29 in Appendix I. Because we derived and vetted our imaging procedures for the low-band data, we use only the low-band results in determining our final parameter measurements and use the high-band results in Appendix I as a consistency check.

On each observation day, the distributions of $|m|_{net}$ appear consistent between most pairs of reconstruction methods, with some notable exceptions. Many of the distributions of $|m|_{net}$ peak around the ALMA measured value of 2.7%, but the LPCAL distributions on both days and the eht-imaging distributions on 2017 April 11 are peaked closer to 1%. The distributions of $\langle|m|\rangle$ are peaked between 6% and 11% for all five methods across both days. On both days, the $\langle|m|\rangle$ distributions for eht-imaging, DMC, and THEMIS are peaked at values 2%–3% higher than the corresponding LPCAL or polsolve distributions. This systematic shift may indicate residual issues with bringing the reconstruction methods to the same resolution scale; in particular, the same circular Gaussian kernel was used to blur Stokes $\mathcal{I}$, $\mathcal{Q}$ and $\mathcal{U}$ in each method, while the intrinsic resolution of the reconstruction in $\mathcal{Q}$ and $\mathcal{U}$ may be lower than in total intensity. In each method, there appears to be a decrease in $\langle|m|\rangle$ of $\approx$1%–2% between 2017 April 5 and 11. Note that, because it is constrained to be positive and defined as the ratio of two uncertain flux densities, the mean of a distribution of fractional polarization $m$ can have a positive bias. In Figure 25, we see that the distributions of $|m|_{net}$ and $\langle|m|\rangle$ both can have long tails; this is most evident on 2017 April 10, when the image reconstructions are the most uncertain due to poor $(u, v)$ coverage. On 2017 April 5 and 11, we do not see prominent tails in the distributions of $|m|_{net}$ and $\langle|m|\rangle$. Furthermore, we expect any bias in the mean of these quantities in the measurement from a single method to be overwhelmed by the systematic uncertainty between different reconstruction methods.

The mean of the $|\beta_2|$ distribution is peaked between 0.04 and 0.07 for all methods on both days; however, the $|\beta_2|$ distributions from eht-imaging, DMC, and THEMIS have larger mean values on both days than the corresponding distributions for polsolve and LPCAL. Again, because $|\beta_2|$ is sensitive to the restoring beam size, this may be due to residual errors in bringing the polarized images to the same resolution scale. Similarly to the distributions of $\langle|m|\rangle$, there are indications of a shift downward in $|\beta_2|$ by an absolute value of $\approx$0.01 in all four methods between 2017 April 5 and 11. The distributions of the phase $\angle\beta_2$ are consistent between most pairs of methods with no obvious systematic difference

between the sub-component methods (`LPCAL`, `polsolve`) and those that use a continuous image representation (`eht-imaging`, DMC, and THEMIS). Furthermore, there is no apparent systematic difference in the $\angle\beta_2$ results between 2017 April 5 and 11.

To score different accretion models from GRMHD simulations against constraints from the EHT data, Paper VIII uses a range for each quantity that incorporates both the uncertainties in the parameters from the D-term calibration process (the error bars for each method in Figure 9) and the systematic uncertainty across methods (the scatter in the points across methods). The final ranges used in Paper VIII for each parameter were set by taking the minimum/maximum of the 10 mean values minus/plus the $1\sigma$ error bars across both days and all methods. These parameter ranges are denoted by colored bands in Figure 9 and are presented in Table 2.

## 6. Discussion

We discuss several important effects in the polarimetric emission from M87 that are relevant for our analysis of the 230 GHz linear polarization structure in this work. In particular, we discuss implications of our results for source variability and Faraday rotation in M87.

Figure 8 demonstrates variability in both the total intensity and polarimetric images of M87 between 2017 April 5 and 11. It is unlikely that the polarimetric variability in the reconstructed images is due to the different $(u, v)$ coverages on different days. The changes in the polarimetric images are consistent with signatures of the source intrinsic variability seen in the VLBI data themselves. In addition to the reconstructed images and calibrated data, we see variability in calibration-insensitive VLBI data products; these are introduced and discussed in Appendix B.

The total flux density from M87's inner arcseconds measured on EHT intra-site baselines (ALMA–APEX) and by ALMA alone is $F \sim 1.2$ Jy; this is a factor of 2 larger than the total flux density measured in the ring visible on EHT scales. Given that the net fractional polarization measured in the EHT images ($|m|_{net} \sim 1\%–3.7\%$) is consistent with that measured on arcseconds scales $|m| \sim 2.7\%$ (Goddi et al. 2021), the net fractional polarization of any other emission component(s) in the ALMA field of view should be comparable to that of the ring resolved by the EHT.

The polarimetric image stability analysis (Section 5.2) provides a measurement of the integrated EVPA and associated D-term calibration uncertainty for each reconstruction method and observing day. Figure 10 shows that the total EVPA integrated over the EHT images ranges from $\chi \sim -70°$ to $\chi \sim -55°$ on 2017 April 5 and from $\chi \sim -25°$ to $\chi \sim -10°$ on 2017 April 11, depending on the image reconstruction method. On all days, the EHT-measured EVPA in the core is significantly offset from the EVPA measured by ALMA on large scales. This offset implies that the extended component within the central arcseconds is polarized. We note that in both EHT and ALMA-only observations, the EVPA swings in a counter-clockwise direction from 2017 April 5 to 11.

Figure 10 also compares the image-integrated EVPA measured from the EHT high- and low-band images across all four days (see Appendix I for a full discussion of the high-band results). For 2017 April 6, 10, and 11, the high- and low-band net EVPAs are consistent for each method within the $1\sigma$ error bars derived from the D-term calibration sample. On 2017

April 5, we see a systematic net EVPA offset between the low- and high-band images: $\Delta\chi_{|HI-LO|} \approx 20°–30°$ in all five methods.

If we were to interpret this systematic EVPA offset on 2017 April 5 as Faraday rotation from an external screen, it would correspond to $|RM| \sim 1–2 \times 10^7$ rad m$^{-2}$. As this is the largest effect that we observe between the bands, we can adopt this number as a conservative upper limit on the resolved RM. While on the other three days (2017 April 6, 10, and 11) there is no signature of an offset in the image-integrated EVPA, we do see intriguing offsets in the EVPAs in some portions of the resolved images. Such non-uniform rotations may be indicative of Faraday rotation occurring internally in the compact source, but because of the low significance of these detections we make no further effort to interpret them here.

The full implications of these results for Faraday rotation in the source depend on the magnitude, location, and nature of the Faraday screen. Goddi et al. (2021) report contemporaneous ALMA measurements of the RM in M87 on arcseconds scales; these range from $1.5 \times 10^5$ rad m$^{-2}$ to $-0.4 \times 10^5$ rad m$^{-2}$. If we interpret the ALMA-only measurements as the result of a variable external Faraday screen, they would imply EVPA rotations from infinite frequency to 230 GHz of less than 15°, with day-to-day swings in the EVPA of up to 20° from variability in the external RM alone. However, Goddi et al. (2021) also consider a two-component model comprising compact and extended (arcseconds-scale) emission regions with separate, static Faraday screens. This two-component model is capable of reproducing both the magnitude and inter-day variability of the observed ALMA RMs, with the variability entirely in the underlying source, not the Faraday screen. In contrast to the direct ALMA measurement, this model suggests the RM relevant for the EHT images is of order $-5 \times 10^5$ rad m$^{-2}$. Intrinsic polarimetric evolution is also supported by the changes in Stokes $\mathcal{I}$ alone and by the changes in the distribution of polarized intensity in our polarimetric images; we do not see a simple uniform rotation of EVPAs between 2017 April 5 and 11 around the emission ring in Figure 7.

In reality, the net EVPA and resolved EVPA structure in both EHT bands are affected by the complicated interplay of intrinsic source structure and evolution, Faraday rotation internal to the emission region, and Faraday rotation from an external screen (if present). Paper VIII discusses all of these scenarios in more detail. Starting from 2018, the EHT observes simultaneously in 212.1–216.1 GHz and 226.1–230.1 GHz frequency bands (Paper II). This development should allow us to better quantify the resolved RM and to address the intrinsic polarimetric variability of M87 with better precision in the future.

**Table 2**
Final Parameter Ranges for the Quantities Used in Scoring GRMHD Models in Paper VIII

| Parameter | Min | Max |
|---|---|---|
| $|m|_{net}$ | 1.0% | 3.7% |
| $\langle|m|\rangle$ | 5.7% | 10.7% |
| $|\beta_2|$ | 0.04 | 0.07 |
| $\angle\beta_2$ | $-163°$ | $-127°$ |

**Note.** The ranges are taken from the bands plotted in Figure 9 incorporating the $\pm 1\sigma$ error from each method's D-term calibration survey.

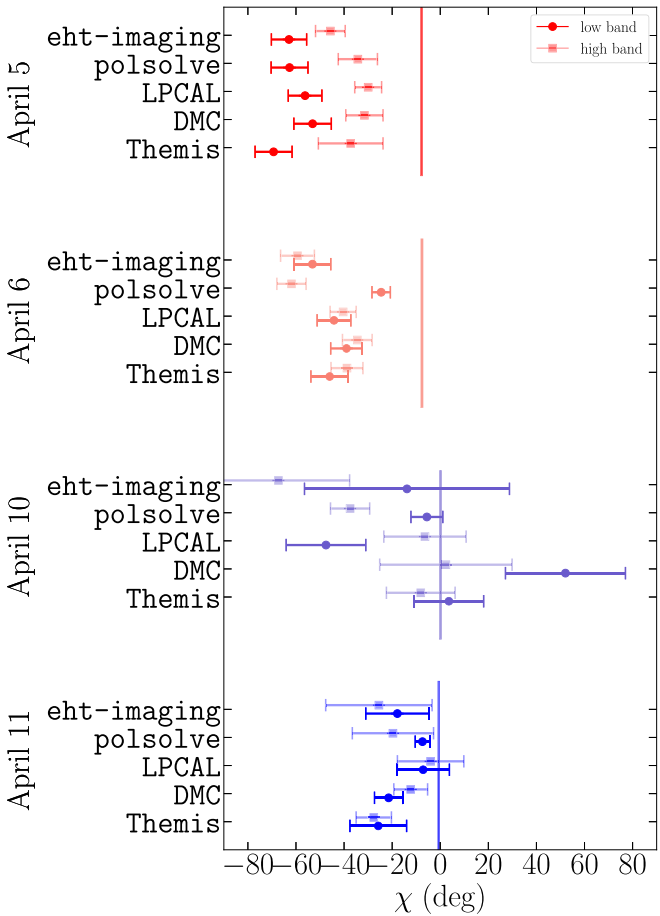

**Figure 10.** M87 net EVPA integrated within 120 μas on all four days for both high and low bands. The low-band results are indicated by circular markers and the high-band results are plotted in a lighter color with square markers. Vertical lines mark the ALMA-only EVPA measurements from Goddi et al. (2021), measured on arcseconds scales. (The error bars of ALMA-only measurements are very small so the shaded bands appear as vertical lines.) The image-integrated EVPAs are consistent between high and low bands on all days except 2017 April 5, where all methods show an offset of ~20°. Note that the image reconstructions and EVPA measurements on 2017 April 10 are poorly constrained due to the poor (u, v) coverage.

Finally, in this Letter, we discuss only the linear polarization images. Given its small magnitude, Stokes 𝒱 is significantly more sensitive to calibration choices and residual errors than the linear polarization components. For that reason a full analysis of the circular polarization structure in M87 will be presented separately.

## 7. Summary

We present polarimetric calibration and polarimetric imaging of the EHT 2017 data on the 230 GHz core of M87 on scales comparable to the supermassive black hole event horizon. Our analysis follows up on the M87 total intensity data calibration, image reconstructions, and model fits presented in Paper III, Paper IV, and Paper VI.

We employ multiple distinct methods for polarimetric calibration and polarimetric imaging. All methods were first tested on a suite of synthetic data. When applied to M87, they consistently show that the polarized emission is predominantly from the southwest quadrant. In all reconstructions, the polarization vectors are organized into a similar coherent pattern roughly oriented along the ring. In all reconstructions, both the image-integrated net linear polarization fraction and the average resolved polarization fraction on the ring are consistent to within a few percent. We observe signatures of evolution in the ring's polarization from 2017 April 5 to 11, the full length of the EHT 2017 observing campaign. In this work, we demonstrate that the main polarimetric characteristics of the M87 ring are robust to D-term calibration uncertainties and to the choice of image-reconstruction algorithm, though the detailed source structure (particularly in low brightness regions) is still limited by the EHT's very sparse (u, v) coverage and thus depends sensitively on choices made in the image reconstruction and calibration process.

The high-angular-resolution observation with the EHT, on unprecedented scales of ~20 μas ≈2.5 $R_S$, allows us for the first time to reconstruct the geometry of magnetic fields in the immediate vicinity of the event horizon of the M87 supermassive black hole. The physical interpretation of our polarimetric images and the full discussion of horizon-scale magnetic field geometries consistent with the EHT images are presented in Paper VIII.

The authors of the present Letter are extremely grateful to the anonymous referee for the thorough review and highly appreciate all comments and suggestions that significantly improved the quality of the manuscript.

The authors of the present Letter thank the following organizations and programs: the Academy of Finland (projects 274477, 284495, 312496, 315721); the Alexander von Humboldt Stiftung; Agencia Nacional de Investigación y Desarrollo (ANID), Chile via NCN19_058 (TITANs), and Fondecyt 3190878; an Alfred P. Sloan Research Fellowship; Allegro, the European ALMA Regional Centre node in the Netherlands, the NL astronomy research network NOVA and the astronomy institutes of the University of Amsterdam, Leiden University and Radboud University; the black hole Initiative at Harvard University, through a grant (60477) from the John Templeton Foundation; the China Scholarship Council; Consejo Nacional de Ciencia y Tecnología (CONACYT, Mexico, projects U0004-246083, U0004-259839, F0003-272050, M0037-279006, F0003-281692, 104497, 275201, 263356); the Delaney Family via the Delaney Family John A. Wheeler Chair at Perimeter Institute; Dirección General de Asuntos del Personal Académico-Universidad Nacional Autónoma de México (DGAPA-UNAM, projects IN112417 and IN112820); the EACOA Fellowship of the East Asia Core Observatories Association; the European Research Council Synergy Grant "BlackHoleCam: Imaging the Event Horizon of Black Holes" (grant 610058); the Generalitat Valenciana postdoctoral grant APOSTD/2018/177 and GenT Program (project CIDEGENT/2018/021); MICINN Research Project PID2019-108995GB-C22; the Gordon and Betty Moore Foundation (grants GBMF- 3561, GBMF-5278); the Istituto Nazionale di Fisica Nucleare (INFN) sezione di Napoli, iniziative specifiche TEONGRAV; the International Max Planck Research School for Astronomy and Astrophysics at the Universities of Bonn and Cologne; Joint Princeton/Flatiron and Joint Columbia/Flatiron Postdoctoral Fellowships, research at the Flatiron Institute is supported by the Simons Foundation; the Japanese Government (Monbukagakusho: MEXT) Scholarship; the Japan Society for the Promotion of Science (JSPS) Grant-in-Aid for JSPS Research Fellowship (JP17J08829); the Key Research Program of Frontier Sciences, Chinese Academy of Sciences (CAS, grants QYZDJ-SSW-SLH057, QYZDJSSW- SYS008, ZDBS-LY-SLH011); the Leverhulme Trust Early Career Research Fellowship; the Max-Planck-Gesellschaft (MPG); the Max Planck Partner Group of the MPG and the CAS; the MEXT/JSPS KAKENHI (grants 18KK0090, JP18K13594, JP18K03656, JP18H03721, 18K03709, 18H01245, 25120007); the Malaysian Fundamental Research Grant Scheme (FRGS) FRGS/1/2019/STG02/UM/02/6; the MIT International Science and Technology Initiatives (MISTI) Funds; the Ministry of Science and Technology (MOST) of Taiwan (105-2112-M-001-025-MY3, 106-2112-M-001-011, 106-2119- M-001-027, 107-2119-M-001-017, 107-2119-M-001-020, 107-2119-M-110-005, 108-2112-M-001-048, and 109-2124-M-001-005); the National Aeronautics and Space Administration (NASA, Fermi Guest Investigator grant 80NSSC20K1567 and 80NSSC20K1567, NASA Astrophysics Theory Program grant 80NSSC20K0527, NASA NuSTAR award 80NSSC20K0645, NASA grant NNX17AL82G, and Hubble Fellowship grant HST-HF2-51431.001-A awarded by the Space Telescope Science Institute, which is operated by the Association of Universities for Research in Astronomy, Inc., for NASA, under contract NAS5-26555); the National Institute of Natural Sciences (NINS) of Japan; the National Key Research and Development Program of China (grant 2016YFA0400704, 2016YFA0400702); the National Science Foundation (NSF, grants AST-0096454, AST-0352953, AST-0521233, AST-0705062, AST-0905844, AST-0922984, AST-1126433, AST-1140030, DGE-1144085, AST-1207704, AST-1207730, AST-1207752, MRI-1228509, OPP-1248097, AST-1310896, AST-1337663, AST-1440254, AST-1555365, AST-1615796, AST-1715061, AST-1716327, AST-1716536, OISE-1743747, AST-1816420, AST-1903847, AST-1935980, AST-2034306); the Natural Science Foundation of China (grants 11573051, 11633006, 11650110427, 10625314, 11721303, 11725312, 11933007, 11991052, 11991053); a fellowship of China Postdoctoral Science Foundation (2020M671266); the Natural Sciences and Engineering Research Council of Canada (NSERC, including a Discovery Grant and the NSERC Alexander Graham Bell Canada Graduate Scholarships-Doctoral Program); the National Research Foundation of Korea (the Global PhD Fellowship Grant: grants 2014H1A2A1018695, NRF-2015H1A2A1033752, 2015- R1D1A1A01056807, the Korea Research Fellowship Program: NRF-2015H1D3A1066561, Basic Research Support grant 2019R1F1A1059721); the Netherlands Organization for Scientific Research (NWO) VICI award (grant 639.043.513) and Spinoza Prize SPI 78-409; the New Scientific Frontiers with Precision Radio Interferometry Fellowship awarded by the South African Radio Astronomy Observatory (SARAO), which is a facility of the National Research Foundation (NRF), an agency of the Department of Science and Innovation (DSI) of South Africa; the South African Research Chairs Initiative of the Department of Science and Innovation and National Research Foundation; the Onsala Space Observatory (OSO) national infrastructure, for the provisioning of its facilities/observational support (OSO receives funding through the Swedish Research Council under grant 2017-00648) the Perimeter Institute for Theoretical Physics (research at Perimeter Institute is supported by the Government of Canada through the Department of Innovation, Science and Economic Development and by the Province of Ontario through the Ministry of Research, Innovation and Science); the Spanish Ministerio de Ciencia e Innovación (grants PGC2018-098915-B-C21, AYA2016-80889-P; PID2019-108995GB-C21, PGC2018-098915-B-C21); the State Agency for

Research of the Spanish MCIU through the "Center of Excellence Severo Ochoa" award for the Instituto de Astrofísica de Andalucía (SEV-2017-0709); the State Agency for Research of the Spanish MCIU through the "Center of Excellence Severo Ochoa" award for the Instituto de Astrofísica de Andalucía (SEV-2017-0709); the Toray Science Foundation; the Consejería de Economía, Conocimiento, Empresas y Universidad of the Junta de Andalucía (grant P18-FR-1769), the Consejo Superior de Investigaciones Científicas (grant 2019AEP112); the US Department of Energy (USDOE) through the Los Alamos National Laboratory (operated by Triad National Security, LLC, for the National Nuclear Security Administration of the USDOE (Contract 89233218CNA000001); the European Union's Horizon 2020 research and innovation program under grant agreement No 730562 RadioNet; ALMA North America Development Fund; the Academia Sinica; Chandra TM6-17006X; Chandra award DD7-18089X. This work used the Extreme Science and Engineering Discovery Environment (XSEDE), supported by NSF grant ACI-1548562, and CyVerse, supported by NSF grants DBI-0735191, DBI-1265383, and DBI-1743442. XSEDE Stampede2 resource at TACC was allocated through TG-AST170024 and TG-AST080026N. XSEDE JetStream resource at PTI and TACC was allocated through AST170028. The simulations were performed in part on the SuperMUC cluster at the LRZ in Garching, on the LOEWE cluster in CSC in Frankfurt, and on the HazelHen cluster at the HLRS in Stuttgart. This research was enabled in part by support provided by Compute Ontario (http://computeontario.ca), Calcul Quebec (http://www.calculquebec.ca) and Compute Canada (http://www.computecanada.ca). We thank the staff at the participating observatories, correlation centers, and institutions for their enthusiastic support. This paper makes use of the following ALMA data: ADS/JAO.ALMA#2016.1.01154.V. ALMA is a partnership of the European Southern Observatory (ESO; Europe, representing its member states), NSF, and National Institutes of Natural Sciences of Japan, together with National Research Council (Canada), Ministry of Science and Technology (MOST; Taiwan), Academia Sinica Institute of Astronomy and Astrophysics (ASIAA; Taiwan), and Korea Astronomy and Space Science Institute (KASI; Republic of Korea), in cooperation with the Republic of Chile. The Joint ALMA Observatory is operated by ESO, Associated Universities, Inc. (AUI)/NRAO, and the National Astronomical Observatory of Japan (NAOJ). The NRAO is a facility of the NSF operated under cooperative agreement by AUI. This paper has made use of the following APEX data: Project ID T-091.F-0006-2013. APEX is a collaboration between the Max-Planck-Institut für Radioastronomie (Germany), ESO, and the Onsala Space Observatory (Sweden). The SMA is a joint project between the SAO and ASIAA and is funded by the Smithsonian Institution and the Academia Sinica. The JCMT is operated by the East Asian Observatory on behalf of the NAOJ, ASIAA, and KASI, as well as the Ministry of Finance of China, Chinese Academy of Sciences, and the National Key R&D Program (No. 2017YFA0402700) of China. Additional funding support for the JCMT is provided by the Science and Technologies Facility Council (UK) and participating universities in the UK and Canada. The LMT is a project operated by the Instituto Nacional de Astrofísica, Óptica, y Electrónica (Mexico) and the University of Massachusetts at Amherst (USA), with financial support from the Consejo Nacional de Ciencia y Tecnología and the National Science Foundation. The IRAM 30-m telescope on Pico Veleta, Spain is operated by IRAM and supported by CNRS (Centre

National de la Recherche Scientifique, France), MPG (Max-Planck- Gesellschaft, Germany) and IGN (Instituto Geográfico Nacional, Spain). The SMT is operated by the Arizona Radio Observatory, a part of the Steward Observatory of the University of Arizona, with financial support of operations from the State of Arizona and financial support for instrumentation development from the NSF. The SPT is supported by the National Science Foundation through grant PLR- 1248097. Partial support is also provided by the NSF Physics Frontier Center grant PHY-1125897 to the Kavli Institute of Cosmological Physics at the University of Chicago, the Kavli Foundation and the Gordon and Betty Moore Foundation grant GBMF 947. The SPT hydrogen maser was provided on loan from the GLT, courtesy of ASIAA. The EHTC has received generous donations of FPGA chips from Xilinx Inc., under the Xilinx University Program. The EHTC has benefited from technology shared under open-source license by the Collaboration for Astronomy Signal Processing and Electronics Research (CASPER). The EHT project is grateful to T4Science and Microsemi for their assistance with Hydrogen Masers. This research has made use of NASA's Astrophysics Data System. We gratefully acknowledge the support provided by the extended staff of the ALMA, both from the inception of the ALMA Phasing Project through the observational campaigns of 2017 and 2018. We would like to thank A. Deller and W. Brisken for EHT-specific support with the use of DiFX. We acknowledge the significance that Maunakea, where the SMA and JCMT EHT stations are located, has for the indigenous Hawaiian people.

*Facilities:* EHT, ALMA, APEX, IRAM:30 m, JCMT, LMT, SMA, ARO:SMT, SPT.

*Software:* AIPS (Greisen 2003), ParselTongue (Kettenis et al. 2006), GNU Parallel (Tange 2011), eht-imaging (Chael et al. 2016), Difmap (Shepherd 2011), Numpy (van der Walt et al. 2011), Scipy (Jones et al. 2001), Pandas (McKinney 2010), Astropy (The Astropy Collaboration et al. 2013, 2018), Jupyter (Kluyver et al. 2016), Matplotlib (Hunter 2007), THEMIS (Broderick et al. 2020a), DMC (Pesce 2021), polsolve (Martí-Vidal et al. 2021), GPCAL (Park et al. 2021).

# Appendix A
# Polarimetric Data Issues

In this section we describe station-specific issues and present the results of a set of validation tests and refinements in the calibration that have been performed on the EHT data, prior to the calibration of the instrumental polarization and the final reconstruction of the full-Stokes EHT images.

## A.1. Instrumental Polarization of ALMA in VLBI Mode

Phased ALMA records the VLBI signals in a basis of linear polarization, which need a special treatment after the correlation (Martí-Vidal et al. 2016; Matthews et al. 2018). The post-correlation conversion of the ALMA data from a linear basis into a circular basis has implications for the kind of instrumental polarization left after fringe fitting. As discussed in Goddi et al. (2019), any offset in the estimate of the phase difference between the $X$ and $Y$ signals of the ALMA antenna used as the phasing reference (an offset likely related to the presence of a non-zero Stokes $\mathcal{V}$ in the polarization calibrator) maps into a post-conversion polarization leakage that can be modeled as a symmetric, pure-imaginary D-term matrix (i.e., $D_R = D_L = i\Delta$). The amplitude of the ALMA D-terms, $\Delta$, can be approximated (to a first order) as the value of the phase offset between $X$ and $Y$ in radians (Goddi et al. 2019). Hence, we expect the $D_R$ and $D_L$ estimates for ALMA to be found along the imaginary axis and to be of similar amplitude.

Furthermore, the ALMA feeds in Band 6 (the frequency band used in the EHT observations) are rotated by 45° with respect to their projection on the focal plane. This introduces a phase offset between the RCP and LCP post-converted signals that has to be corrected after the fringe fitting. This offset can be applied as a global phase added (subtracted) to the $RL^{*}$ ($LR^{*}$) correlation products in all baselines (because ALMA has been used as the reference antenna in the construction of the global fringe-fitting solutions). We have applied this 45° rotation to all the visibilities before performing the analysis described in this Letter. Hence, the absolute position angles of the electric vectors (EVPA) derived from our EHT observations are properly rotated into the sky frame. This property of the ALMA–VLBI observations (see Appendix D) gives us absolute EVPA values instantaneously.

## A.2. Instrumental Polarization of the LMT

The LMT shows an unexpectedly high leakage signal with a large delay of ∼1.5 ns, which affects the cross-polarization phase spectra of the baselines related to the LMT. As a consequence, all the baselines related to the LMT show spurious instrumental fringes in the $RL^{*}$ and $LR^{*}$ correlations, with amplitudes similar to (and even higher than, for the case of sources with low intrinsic polarization) that of the main fringe. These instrumental fringes are smallest in the parallel-hand correlations ($RR^{*}$ and $LL^{*}$), but relatively high in the cross-polarization hands and are related to strong polarization leakage likely due to reflections in the optical setup of the LMT receiver used in 2017 (Paper III). For the EHT observations on year 2018 and beyond, the special-purpose interim receiver used at the LMT was replaced by a dual-polarization sideband-separating 1.3 mm receiver, with better stability and full 64 Gbps sampling as for the rest of the EHT (Paper II), so future polarimetry analyses of the EHT may be free of this instrumental effect from the LMT.

If we take the frequency average over all intermediate-frequency (IF) sub-bands (the results presented in Paper I–Paper VI are based on this averaging), the effect of this leaked fringe is smeared out, as the average is equivalent to taking the value of the visibility at the peak of the main fringe. This main peak is only affected by the sidelobe of the delayed leaked fringe, with a relative amplitude that we estimate to be 10%–20% of the cross-polarization main fringe. Therefore, the effect of the leaked fringe is small in comparison to the contribution from the ordinary instrumental polarization, which can especially dominate the cross-polarization signal for observations of sources with low polarization like M87, and can be ignored.

## A.3. Instrumental Polarization of the SMA

The dual-polarization observations performed by the SMA use two independent receivers at each antenna to register the RCP and LCP signals. However, the visibility matrices of the baselines related to the SMA are built from the combination of the RCP and LCP streams as if they were registered with one single receiver. Therefore, some of the assumptions made in the RIME (see Equation (6)) for the polarimetry calibration (e.g., stable relative phases and amplitude between polarizations) may not apply for the SMA-related visibilities. However, the fringe fitting of the parallel-hand correlations related to the SMA, as well as the absolute amplitude calibration (both described in Paper III) did account for the drifts in cross-polarization phase and amplitude between the SMA receivers, which makes it possible to model the instrumental polarization using ordinary leakage matrices.

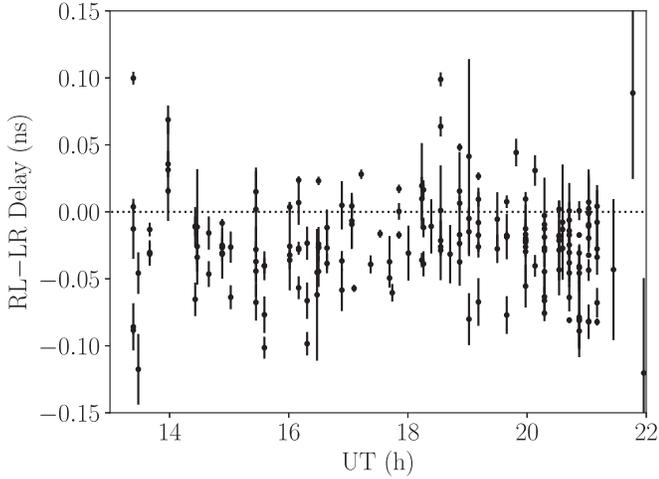

**Figure 11.** Differences between the $RL^{*}$ and $LR^{*}$ delays for all baselines and scans with S/N higher than 10 (only scans when ALMA is observing are shown). The level of zero delay is shown as a dashed line.

One extra correction that has to be applied to the D-terms of the SMA is a phase rotation between the RCP and LCP leakages, to account for the 45° rotation of the antenna feed with respect to the mount axes. The D-terms shown in Section 4.2 and in Appendix D are corrected for this rotation.

### A.4. Instrumental Polarization of the JCMT

The JCMT was equipped with a single-polarization receiver for these observations, so that only one of the two polarizations can be used at each epoch. Therefore, only one of the two cross-polarization correlations can be computed in all the baselines related to the JCMT; depending on which product is computed, we can only solve for one of the two D-terms of the JCMT (i.e., $D_L$ if RCP is recorded; $D_R$ otherwise).

### A.5. Cross-polarization Delays

As explained in Martí-Vidal et al. (2016), a byproduct of the use of `polconvert` in VLBI is the calibration of the absolute cross-polarization delays and phases in the stations with polconverted data, which allow for the reconstruction of the absolute EVPAs of the observed sources. The only condition to have this absolute R/L delay and phase calibration is to use the polconverted station (i.e., ALMA, in the case of the EHT) as the reference antenna in the fringe fitting.

In Figure 11, we show the difference of multi-band delays between $RL^{*}$ and $LR^{*}$ after the GFF calibration described in Paper III. All source scans and baselines with an S/N higher than 10 are shown for times when ALMA was participating in the observations. According to Martí-Vidal et al. (2016), the delay difference between $RL^{*}$ and $LR^{*}$ should be around zero when ALMA is the reference antenna. We see, though, hints of a small global residual delay difference after the GFF calibration (the points are not symmetrically distributed around zero). The weighted average of all the delay differences shown in Figure 11 is $\Delta\tau = -28 \pm 1$ ps. This is a very small delay in absolute value (the amplitude losses due to this delay in each correlation product is lower than 1%), but still detectable at the remarkably high S/N level of the EHT observations.

## Appendix B
## Closure Traces

### B.1. Definition

Closure traces (Broderick & Pesce 2020) are calibration-insensitive quantities constructed on station quadrangles from the coherency matrices $\rho_{jk}$ defined in Equation (2):

$$\mathcal{T}_{ijkl} = \frac{1}{2}\mathrm{tr}(\rho_{ij}\rho_{kj}^{-1}\rho_{kl}\rho_{il}^{-1}). \qquad (B1)$$

These data products are a superset of the more familiar closure quantities (closure phases and closure amplitudes), with the additional property that they are independent of instrumental polarization. The closure traces are also independent of any other station-based effects that can be described in a Jones matrix formalism, including the definition of the polarization basis (e.g., the representation of the polarized quantities in terms of linear or circular feeds). The closure traces thus provide a powerful tool with which to make calibration-independent statements regarding polarimetric data and intrinsic source structure.

By analogy with trivial closure phases (see Paper III), trivial closure traces may be constructed on "boomerang" quadrangles, i.e., quadrangles in which a station is effectively repeated in such a way as to make the quadrangle area vanish (Broderick & Pesce 2020). Given two co-located stations $i$ and $i'$, the closure trace $\mathcal{T}_{iji'k}$ reduces to unity.

Each quadrangle $ijkl$ has an associated "conjugate" quadrangle $ilkj$, constructed by reordering the baselines within the coherency matrix product.[139] Conjugate closure trace products can be expressed as

$$
\begin{aligned}
\mathcal{C}_{ijkl} \equiv \mathcal{T}_{ijkl}\mathcal{T}_{ilkj} = 1 + & \ (\breve{q}_{ij} - \breve{q}_{kj} + \breve{q}_{kl} - \breve{q}_{il})^2 \\
+ & \ (\breve{u}_{ij} - \breve{u}_{kj} + \breve{u}_{kl} - \breve{u}_{il})^2 \\
+ & \ (\breve{v}_{ij} - \breve{v}_{kj} + \breve{v}_{kl} - \breve{v}_{il})^2 \\
+ & \ \mathcal{O}(\breve{q}^3, \breve{u}^3, \breve{v}^3),
\end{aligned}
\qquad (B2)
$$

where $\breve{q}_{ij} \equiv \tilde{\mathcal{Q}}_{ij}/\tilde{I}_{ij}$, $\breve{u}_{ij} \equiv \tilde{\mathcal{U}}_{ij}/\tilde{I}_{ij}$, and $\breve{v}_{ij} \equiv \tilde{\mathcal{V}}_{ij}/\tilde{I}_{ij}$. The $\mathcal{C}_{ijkl}$ are identically unity in the absence of intrinsic source polarization and for point sources. Deviations from unity require non-constant interferometric polarization fractions on baselines in the quadrangle $ijkl$, and therefore closure trace products are a robust indicator of polarized source structures (Broderick & Pesce 2020).

### B.2. Implications for Polarimetric Data Quality

For EHT observations, boomerang quadrangles are formed using the redundant baselines presented by ALMA and APEX.[140] In the top panel of Figure 12, the phases of all of the trivial closure traces are shown for M87 (blue) and the calibrators (J1924−2914, NRAO 530, 3C 279; gray), constructed from scan-averaged visibility data. The values of these phases are clustered about zero, consistent with the expectation that the trivial closure traces are unity.

---

[139] This conjugate quadrangle is identical to the degenerate quadrangle formed by inverting the numerator and denominator in a standard closure amplitude.
[140] The redundant baselines to SMA and JCMT cannot be used as a result of the single-polarization observations at the latter.

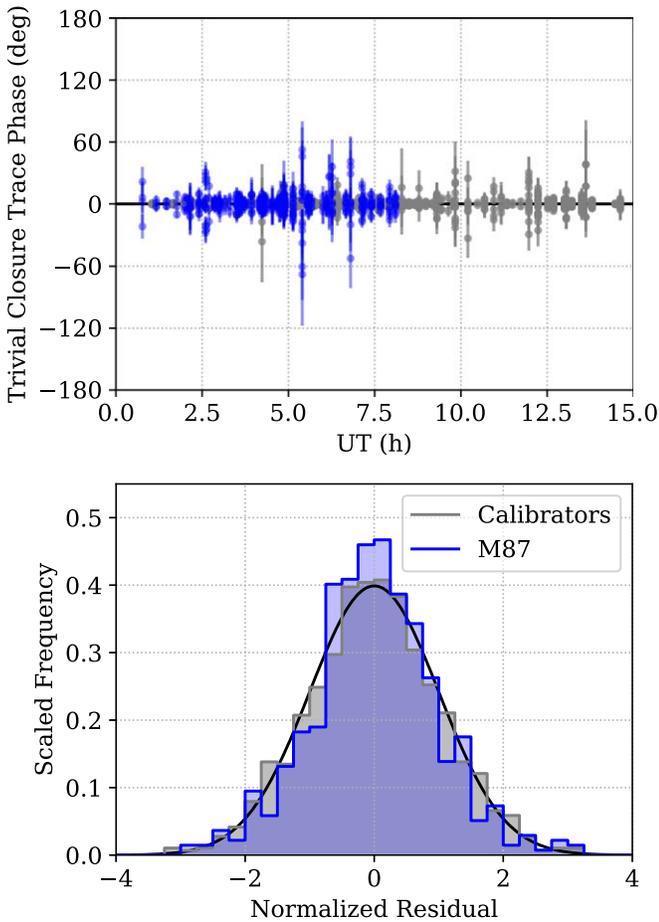

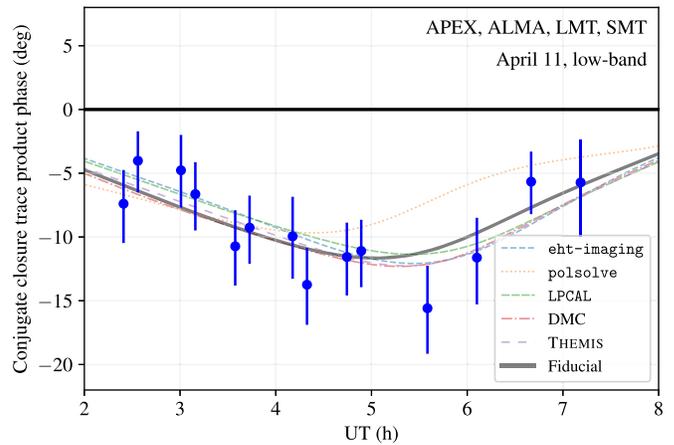

**Figure 13.** Phase of the conjugate closure trace product constructed from the 2017 April 11 low-band observations on the APEX–ALMA–LMT–SMT and APEX–SMT–LMT–ALMA quadrangles. This quantity is sensitive solely to polarization structure; deviations from zero indicate the presence of non-trivial polarization structure (i.e., a polarization fraction that is not constant across the source; Broderick & Pesce 2020). The colored lines show the same conjugate closure trace product phase for each of the image reconstructions (see Figure 6), and the dark gray solid line shows the same for the fiducial image (see Figure 7).

**Figure 12.** Top panel: phases of "boomerang" closure traces for M87 (blue) and calibrators (J1924−2914, NRAO 530, 3C 279; gray), i.e., those with a repeated station (here ALMA/APEX) and thus expected to trivially vanish. Bottom panel: normalized residuals of the trivial closure traces on M87 and the calibrator sources in comparison to a unit-variance normal distribution. In both panels, high- and low-band values are shown for scan-averaged data. We see that these boomerang closure trace phases exhibit the expected clustering around zero.

The distribution of the normalized residuals provide a direct assessment of the systematic error budget of the polarimetric data independent of the gain and leakage calibration. These residuals are shown in the bottom panel of Figure 12 for both M87 and calibrators. We find that the data match the anticipated unit-variance Gaussian, which is consistent with an absence of unidentified systematic uncertainties in the polarimetric data.

### B.3. Calibration-insensitive Detection of Polarization

Figure 13 shows the phase of the conjugate closure trace product, $\mathcal{C}_{ijkl}$, for a quadrangle pair ALMA–APEX–LMT–SMT and ALMA–SMT–LMT–APEX. The presence of non-zero $\mathcal{C}_{ijkl}$ is a calibration-insensitive indicator of significant polarized structures in the Stokes map. Because the uncertainties of the closure traces on conjugate quadrangles are correlated, the resulting uncertainty in the conjugate closure trace product is typically smaller than would be estimated from assuming independent uncertainties in the individual closure traces. The errors shown have been estimated using Monte Carlo sampling of the constituent visibilities.

### B.4. Calibration-insensitive Detection of Evolving Source Structure

Closure trace phases are shown on a handful of non-trivial quadrangles in Figure 14 for M87. These phases are clearly non-zero and exhibit variations throughout the observing night, which is consistent with non-trivial source structure. The behavior of the closure trace evolution is similar across neighboring observation days (e.g., 2017 April 5/6, April 10/11) and consistent between quadrangles constructed using ALMA (filled markers) and APEX (open markers). The behavior of the closure trace evolution is dissimilar between the 2017 April 5/6 and April 10/11 observations, providing direct evidence for an evolving source structure in M87 independent of all station-based corrupting effects, including polarization leakage. Because ALMA, LMT, and SMT are nearly co-linear as seen from M87 for much of the observations, and because the closure traces are presumably track primarily the Stokes $\mathcal{I}$ emission, the closure trace phases in Figure 14 are very similar to the Stokes $\mathcal{I}$ closure phases shown in Figure 14 of Paper III. Note that the points plotted in Figure 14 have been averaged across both the high and low bands.

### B.5. Calibration-insensitive Probe of Evolving Polarimetric Source Structure

Conjugate closure trace product phases are shown in Figure 15 for each observation day for the ALMA–PV–LMT–SMT and ALMA–SMT–LMT–PV quadrangle pair. There is the appearance of temporal evolution from April 5/6 to April 10/11, with an attendant implication for an evolution in the polarization map of M87 between those periods. However, the paucity of quadrangles exhibiting significant evolution renders this conclusion suggestive at best. Note that the points plotted in Figure 15 have been averaged across both the high and low bands.

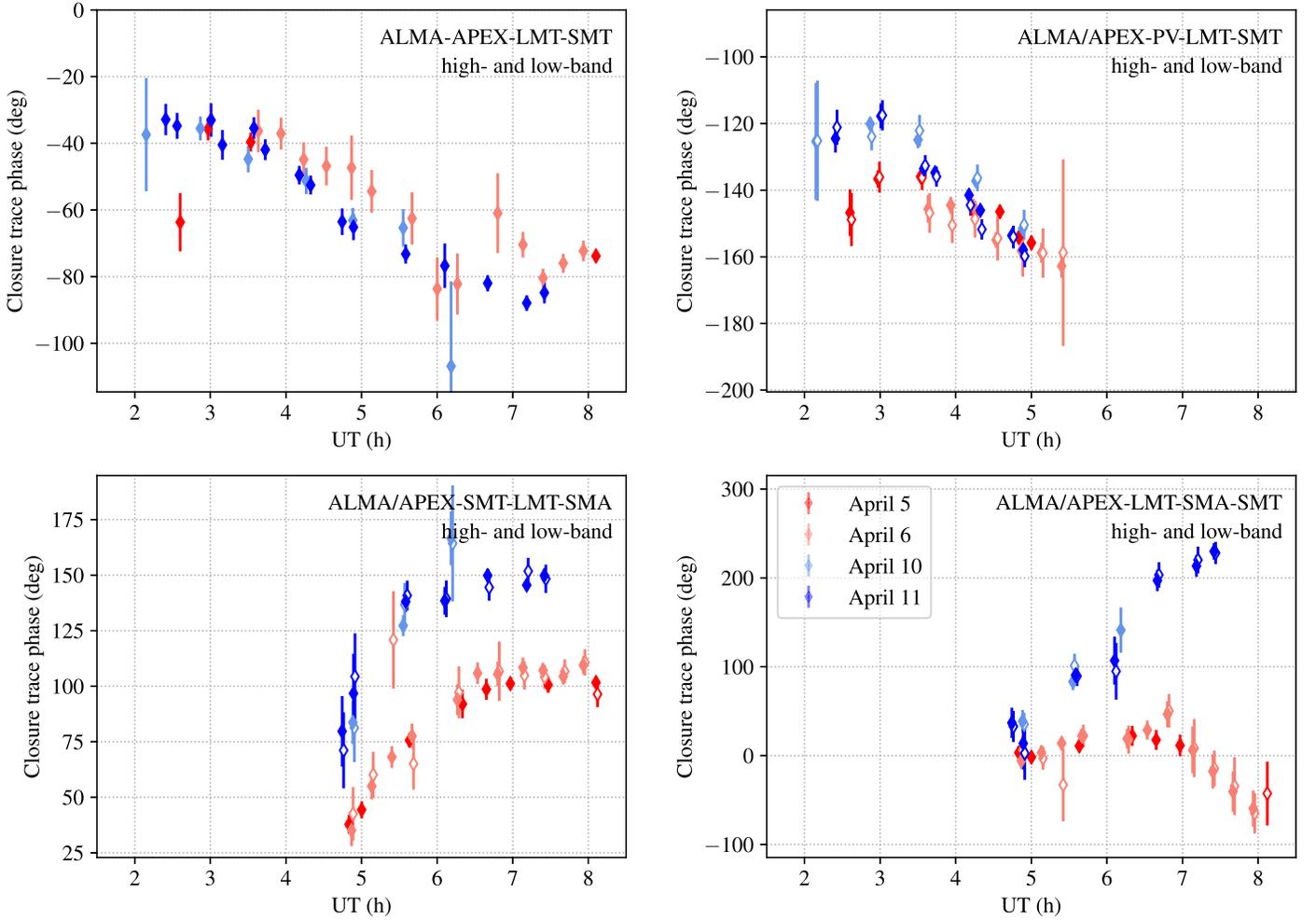

**Figure 14.** Phases of M87 closure traces on four illustrative quadrangles for the different observation days. Closure traces constructed with ALMA and APEX are shown by filled and open points, respectively. The plotted closure traces represent an average across high- and low-band data. An S/N > 1 selection in the final closure trace phase has been applied.

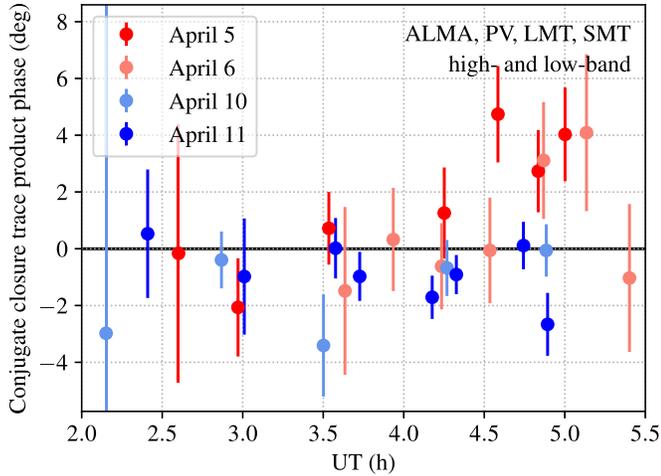

**Figure 15.** Phase of conjugate closure trace products constructed on the ALMA–PV–LMT–SMT and ALMA–SMT–LMT–PV quadrangles, averaged across high and low bands for each of the days on which M87 was observed.

# Appendix C
# Detailed Description of Algorithms for the Calibration of Instrumental Polarization

In this Appendix we provide a brief description of each of the five image reconstruction and D-term calibration pipelines that we employ in this work. More complete descriptions of each method can be found in the individual method papers referenced for each individual pipeline.

## C.1. Polarimetric Imaging via Sub-component Fitting: polsolve, LPCAL, and GPCAL

In the sub-component fitting method for polarimetric modeling, the instrumental polarization (i.e., the complex D-terms) and the source polarized brightness distribution are estimated simultaneously from the interferometric observables and a fixed estimate of the source brightness distribution $\mathcal{I}(\boldsymbol{x})$. The sub-component fitting calibration algorithms estimate the D-terms from Equations (4) and (7) by modeling the polarized source structure as a disjoint set of $N$ "polarization sub-components," $\mathcal{I}_i(\boldsymbol{x})$, such that:

$$\mathcal{I}(\boldsymbol{x}) = \sum_{i}^{N} \mathcal{I}_i(\boldsymbol{x}). \qquad \text{(C1)}$$

The fractional polarization of each sub-component is assumed to be constant, implying that:

$$
\begin{aligned}
\mathcal{Q}(\boldsymbol{x}) &= \sum_i^N q_i \, \mathcal{I}_i(\boldsymbol{x}) \\
\mathcal{U}(\boldsymbol{x}) &= \sum_i^N u_i \, \mathcal{I}_i(\boldsymbol{x}),
\end{aligned} \tag{C2}
$$

where $q_i$ and $u_i$ are real-valued constants. In the sub-component fitting method, we therefore assume that the polarized brightness is exactly proportional to $\mathcal{I}_i$ for each source sub-component. This condition is known as the "similarity approximation" and may produce inaccurate estimates of the instrumental polarization for cases of strongly polarized and resolved calibrators (e.g., Cotton 1993) and/or if the sub-division of $\mathcal{I}$ into sub-components is not performed properly. Discussions about the self-similarity assumption can be found in Appendix K.

The polsolve, LPCAL, and GPCAL algorithms determine which values of $q_i$, $u_i$, $D_R$, and $D_L$ minimize the difference between the calibrated visibility matrix and the Fourier-transformed model brightness matrix (Equations (6) and (4)). The total number of parameters used in this fit is equal to two times the number of source sub-components (i.e., $2N$, which correspond to $q_i$ and $u_i$ in Equation (C2)) plus four times the number of antennas (i.e., $4N_a$, accounting for the real and imaginary parts of the $D_R$ and $D_L$ of each antenna). The error function to be minimized is the sum of the $\chi^2$ values computed for the cross-polarization matrix elements of the RIME, i.e.,

$$
\begin{aligned}
\chi^2 &= \sum_{m=1}^{N_v} w_m \, |RL^{*c}_{kl,m} - (\tilde{\mathcal{Q}} + i\tilde{\mathcal{U}})_m|^2 \\
&+ \sum_{m=1}^{N_v} w_m \, |LR^{*c}_{kl,m} - (\tilde{\mathcal{Q}} - i\tilde{\mathcal{U}})_m|^2,
\end{aligned} \tag{C3}
$$

where $w_m$ is the weight of the $m$th visibility, the index $c$ stands for calibrated visibilities (corrected both for station gains and for instrumental polarization using the current estimate of the D-terms) and $N_v$ is the number of visibilities.

The calibrated visibilities, $RL^{*c}_{kl,m}$ and $LR^{*c}_{kl,m}$, depend on $D_R^k$, $D_L^k$, $D_R^l$ and $D_L^l$ (Equation (4)), whereas $\tilde{\mathcal{Q}}$ and $\tilde{\mathcal{U}}$ depend on $q_i$ and $u_i$. The $\chi^2$ minimization solves for the intrinsic source Stokes parameters and the instrumental polarization simultaneously. We note that the effects of instrumental polarization are constant in the frame of the antenna feed for Cassegrain-mounted feeds, whereas the intrinsic source polarization is defined in the sky frame; as a consequence, the changing feed angle of each antenna across the observations (i.e., the Earth rotation during the extent of the observations) allows the model fitting to decouple the antenna D-terms from the Stokes parameters of the source sub-components. For Nasmyth-mounted feeds, there is an additional rotation between cross-polarization introduced by the feed/optics and the telescope itself. In this case, we assume a minimal contribution from the antennas themselves, a reasonable assumption for these on-axis telescopes. Equation (C3) implies that there are several implicit assumptions in the polarimetric modeling of polsolve and LPCAL. First, $RL^{*c}$ and $LR^{*c}$ are computed by setting $\mathcal{V} = 0$ (i.e., any circular polarization in the calibrators is neglected, compared to Stokes $\mathcal{I}$). Furthermore, the real and the imaginary parts of the residual visibilities (i.e., either $RL^{*c}$ or $LR^{*c}$ minus the Fourier transforms of the corresponding model brightness distributions) are assumed to be statistically independent.

If the linear polarization structures of calibrators are not similar to their total intensity structures, a breakdown of the similarity approximation can occur. This can be a source of uncertainties in D-term estimation. In Appendix K, we discuss this effect on our results for different polarization calibrators reported in this Letter.

### C.2. Polarimetric Imaging via Regularized Maximum Likelihood: eht-imaging

The package eht-imaging (Chael et al. 2016, 2018) implements polarimetric image reconstruction via RML. eht-imaging solves for an image $X$ by minimizing an objective function via gradient descent. The objective function $J(X)$ is a weighted sum of data-consistency log-likelihood terms and *regularizer* terms that favor or penalize certain image features. That is, to find an image (in either total intensity, polarization, or both) we minimize

$$
J(X) = \sum_{\text{data terms } i} \alpha_i \chi_i^2(X) + \sum_{\text{regularizers } j} \beta_j S_j(X). \tag{C4}
$$

Picking optimal values of the "hyperparameter" weights $\alpha_i$ and $\beta_j$ in Equation (C4) is an essential task in RML imaging. Here we describe the data terms and regularizers that we use for polarimetric imaging, and in Appendix G we describe our method for determining the hyperparameters using parameter surveys.

For polarized image reconstructions, we follow the method laid out in Chael et al. (2016), with the addition of iterative self-calibration of any uncorrected station D-terms. We start with data that has had the overall time-dependent station amplitude and phase gains calibrated using the SMILI fiducial image from Paper IV, and the ALMA, APEX, SMA, and JCMT D-terms have been corrected using the zero-baseline solutions described in Section 4.2. The data are scan-averaged. We then reconstruct the Stokes $\mathcal{I}$ using the same fiducial imaging script for eht-imaging developed in Paper IV. We fix the image field of view at 120 $\mu$as and solve for a grid of $64 \times 64$ pixels. In the Stokes $\mathcal{I}$ imaging, the total flux density is constrained to be 0.6 Jy. We next (re)self-calibrate the station amplitude and phase gains (assuming $G_R = G_L$) to our final Stokes I image. Having extensively explored the imaging parameter space for Stokes $\mathcal{I}$ imaging in Paper IV, we do not vary these parameters in our polarimetric imaging surveys. After self-calibrating to our final total intensity image, we drop zero baselines for the polarimetric imaging stage.

In defining an objective function of the form in Equation (C4) for the *polarized* image reconstruction, we consider two log-likelihood $\chi^2$ terms; one computed using the RL$^*$ polarimetric visibility $\mathcal{P} = \tilde{\mathcal{Q}} + i\tilde{\mathcal{U}}$, and one using the visibility domain polarimetric ratio $\breve{m} = \mathcal{P}/\tilde{\mathcal{I}}$. $\chi_{\breve{m}}^2$ is immune to any residual station gain error left over from Stokes $\mathcal{I}$ imaging, while $\chi_{\tilde{p}}^2$ is not. We use two regularizers on the polarized intensity. First, the Holdaway-Wardle (Holdaway & Wardle 1990) regularizer $S_{\mathrm{HW}}$ (Equation (13) of Chael et al. 2016) acts like an entropy term that prefers image pixels take a value less than $m_{\max} = 0.75$. This regularizer encourages image pixels to stay below the theoretical maximum polarization fraction for synchrotron radiation, but it is not a hard limit. Second, the total variation (TV) regularizer $S_{\mathrm{TV}}$ (Rudin et al. 1992) acts to minimize pixel-to-pixel image gradients in both the real and imaginary parts of the complex polarization brightness distribution (Equation (15) of Chael et al. 2016).

Taken together, the objective function we minimize in polarimetric imaging is

$$
J_{\mathrm{pol}}(\mathcal{Q}, \mathcal{U}) = \alpha_p \chi_{\tilde{p}}^2 = \alpha_m \chi_{\breve{m}}^2 - \beta_{\mathrm{HW}} S_{\mathrm{HW}} - \beta_{\mathrm{TV}} S_{\mathrm{TV}}. \tag{C5}
$$

The relative weighting between the data constraints and the regularizer terms is set by the four hyperparameters $\alpha_P$, $\alpha_m$, $\beta_{HW}$, and $\beta_{TV}$.

We solve for the polarized intensity distribution that minimizes Equation (C5) parameterized by the fractional polarization $m$ and EVPA $\xi$ in each pixel. The Stokes $\mathcal{I}$ image is fixed in the polarimetric imaging step and defines the region where polarized emission is allowed. To ensure our solution respects $\mathcal{Q}^2 + \mathcal{U}^2 < \mathcal{I}^2$ everywhere, we transform the fractional polarization $m$ in each pixel from the range $m \in [0, 1]$ to $\kappa \in (-\infty, \infty)$ and solve for $\kappa$ (See Appendix D of Chael et al. 2016). In the `eht-imaging` script for EHT M87 observations, we solve for the pixel values of $\kappa$ and $\xi$ that minimize the objective function by gradient descent, and we then transform $(\mathcal{I}, \kappa, \xi) \rightarrow (\mathcal{I}, \mathcal{Q}, \mathcal{U})$. We often restart the gradient descent process several times, using the output of the previous round of imaging blurred by a 20 $\mu$as Gaussian kernel as the new initial point.

In between rounds of polarimetric imaging with `eht-imaging`, we iteratively solve for the remaining D-terms by minimizing the $\chi^2$ between the real (gain-calibrated) data and sampled data from the current image reconstruction corrupted with Jones matrices (Equation (4)). We do not use any linearized approximations of the effects of the Jones matrices when solving for the D-terms, but throughout we assume the model image has no circular polarization ($\mathcal{V} = 0$). The `eht-imaging` pipeline thus alternates between rounds of polarimetric imaging and D-term calibration; often it takes many successive rounds of imaging and D-term calibration ($n_{iter} \approx 50$–100) for the process to converge on a stable D-term solution.

### C.3. Polarimetric Imaging as Posterior Exploration: DMC and THEMIS

In this section we describe two MCMC schemes developed for polarimetric imaging. Both MCMC codes model the polarized emission structure on a Cartesian grid of intensity points, with the Stokes vector parameterized using a spherical (Poincaré) representation,

$$\begin{pmatrix} \mathcal{I}_i \\ \mathcal{Q}_i \\ \mathcal{U}_i \\ \mathcal{V}_i \end{pmatrix} = \mathcal{I}_i \begin{pmatrix} 1 \\ \ell_i \cos(\xi_i)\sin(\varsigma_i) \\ \ell_i \sin(\xi_i)\sin(\varsigma_i) \\ \ell_i \cos(\varsigma_i) \end{pmatrix}, \qquad (C6)$$

where the index $i$ runs over individual grid points, where we solve for the total intensity $\mathcal{I}_i$, the fractional polarization $\ell_i$, and the two angles $\xi_i$, $\varsigma_i$. Stokes visibilities are generated from the gridded emission structure via a direct Fourier transform (i.e., treating each grid point as a point source), and the visibilities are then multiplied with a smoothing kernel to impose image continuity. The parallel- and cross-hand visibilities on each baseline are then computed from the Stokes visibilities using Equation (7), and the gains and leakage terms are applied to the model visibilities using a Jones matrix formalism (see Equation (6)). The model and data visibilities are ultimately compared via complex Gaussian likelihood functions for each of the parallel- and cross-hand data products independently, with the total likelihood taken to be the product of the individual likelihoods for all parallel- and cross-hand data products.

### C.3.1. DMC

We introduce a new DMC that utilizes the Hamiltonian Monte Carlo (HMC) sampler implemented in the PyMC3 probabilistic programming Python package (Salvatier et al. 2016) to perform posterior exploration. We briefly describe the relevant aspects of the DMC analysis in this section; a more thorough description of the software is provided in Pesce (2021). Prior to fitting, we coherently average the visibility data on a per-scan basis and flag the intra-site baselines.

Within the DMC framework, the $\mathcal{I}$ image axes are aligned with the equatorial coordinate axes. The pixel intensities are constrained to sum to a total flux density via the imposition of a flat Dirichlet prior, and the total flux density is restricted to be positive via a uniform prior on the range (0,2) Jy. The radial Stokes parameter ($\ell_i$ in Equation (C6)) is sampled from a unit uniform prior, and the angular Stokes parameters ($\xi_i$ and $\varsigma_i$ in Equation (C6)) are uniformly sampled on the sphere. We multiply the model visibilities by a circular Gaussian kernel to impose image smoothness.

In DMC, both the right- and left-hand complex station gains are modeled independently on every scan, save for a single reference station (chosen to be ALMA) that is constrained to have zero right- and left-hand gain phase at all times. We impose log-normal priors on the gain amplitudes and wrapped uniform priors[141] on the gain phases. The right- and left-hand leakage amplitudes are sampled from a unit uniform prior, and the leakage phases are sampled from a wrapped uniform prior.

The DMC likelihood variances are set to the quadrature sum of the data thermal noise variances and a systematic component that is modeled as the square of a fraction of the Stokes $\mathcal{I}$ visibility amplitude; this fractional uncertainty parameter is sampled from a unit uniform prior.

### C.3.2. THEMIS

The existing method for imaging via posterior exploration described in Broderick et al. (2020b) has been extended to polarization reconstructions. This makes use of a deterministic even–odd swap tempering scheme (Syed et al. 2019) using the HMC sampling kernel from the Stan package (Carpenter et al. 2017). Here we briefly summarize the implementation and assumptions underlying the THEMIS polarization map reconstructions; more detail on these points will be presented elsewhere (A. E. Broderick et al. 2021, in preparation).

As with DMC, all THEMIS analyses are performed on coherently scan-averaged visibility data. Unlike the DMC analysis, intra-site baselines are included to facilitate gain and leakage calibration. This is enabled by the inclusion of a large, uniformly polarized Gaussian to model the milliarcsecond-scale structure (see, e.g., Broderick et al. 2020b; Paper IV)

THEMIS models the polarized image as a small number of control points located on a rectilinear grid, from which the fields $\mathcal{I}$, $\ell$, $\xi$, and $\cos(\varsigma)$ are constructed via an approximate cubic spline in a fashion similar to Broderick et al. (2020b). The field of view and orientation of the rectilinear grid are fit parameters and permitted to vary. In this way the effective resolution is reconstructed from the data itself. Logarithmic priors are adopted on $\mathcal{I}$ and $\ell$, flat priors are adopted on $\xi$ and $\cos(\varsigma)$ with the natural limits.

---

[141] A "wrapped" or "circular" uniform distribution is defined on the unit circle and has constant probability density for all angles. That is, an angular variable drawn from a wrapped uniform distribution is being sampled uniformly on the unit circle.

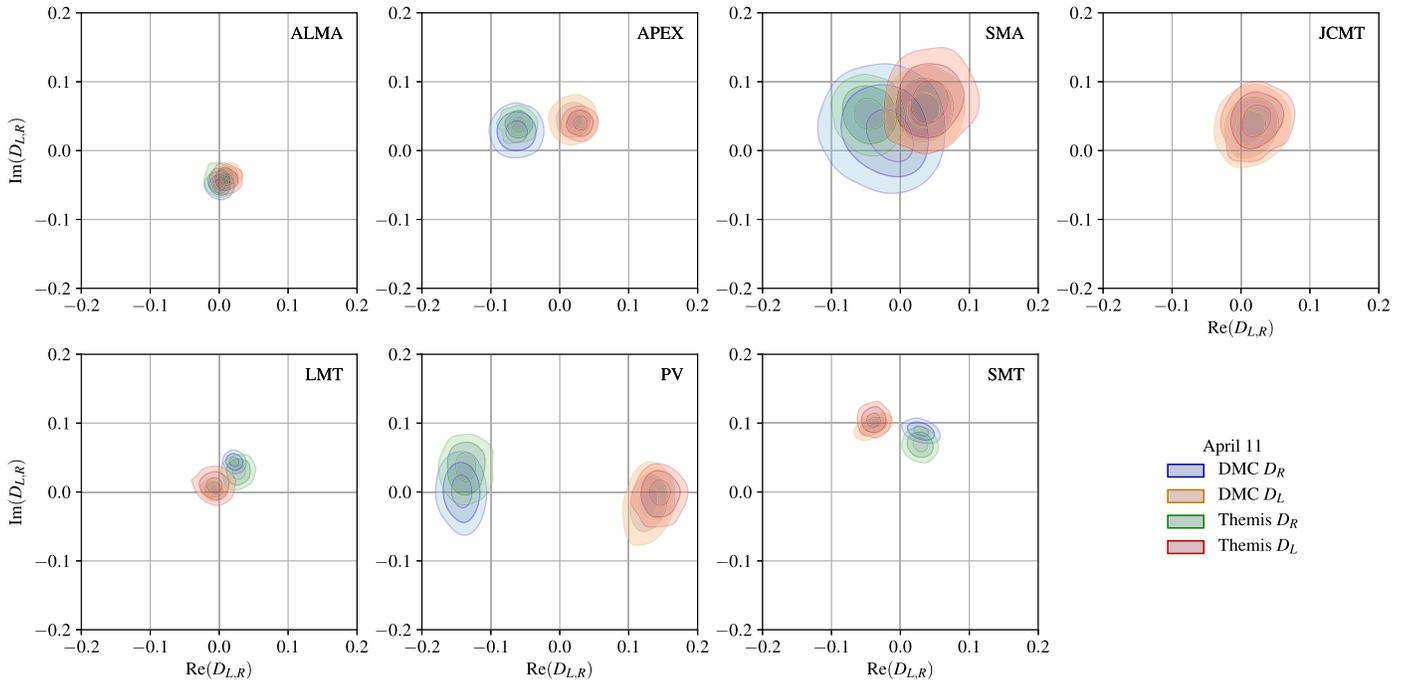

**Figure 16.** Leakage posteriors for individual stations from DMC and THEMIS reconstructions of M87 on 2017 April 11. Because JCMT only records a single polarization, only left-hand D-terms are shown. The plotted contours enclose 50%, 90%, and 99% of the posterior probability, and show a large degree of overlap for all stations despite considerable differences in the underlying model specifications.

Complex station gains are reconstructed via the Laplace approximation (see Section 6.8 of Broderick et al. 2020a). The right- and left-hand complex station gains are constrained to be equal, and permitted to vary independently on every scan. Lognormal priors are imposed on the station gain amplitudes. The real and imaginary components of the right- and left-hand leakages are treated as additional model parameters, with each component sampled uniformly on $[-1,1]$.

Unless otherwise indicated, THEMIS analyses shown here used a $5 \times 5$ raster grid, which is consistent with that typically necessary to capture features on the scale of the EHT beam within the field of view imposed by the shortest intersite baselines. A 3% systematic noise component was added in quadrature to the thermal uncertainties to capture non-closing errors in the scan-averaged visibilities. These are similar to the magnitude of fractional systematic error inferred from DMC analyses.

*C.3.3. THEMIS−DMC Leakage Posterior Comparison*

In Figure 16 we show a comparison between the leakage posteriors for all stations, as determined by DMC and THEMIS fits to the 2017 April 11 observations of M87. Despite the various different assumptions and model specifications, we find excellent agreement in both the means and shapes of the posterior distributions recovered from both methods. The modest discrepancies between the posteriors shown in Figure 16 are associated with the different treatment of systematic uncertainty and right/left gain ratios between the two methods; when these model choices are homogenized, the DMC and THEMIS fits to both synthetic data sets and to the M87 data return indistinguishable posteriors. Notably, both model treatments of the Stokes map are comparably capable of capturing the source structure.

## Appendix D
## Intra-site D-term Validation

We present the final D-terms for ALMA in Table 3 and for APEX, JCMT, and SMA in Table 4. These D-terms were fit to EHT intra-site baseline data from multiple calibrators simultaneously using the method described in Section 4.2, implemented in `polsolve`.

JCMT can only record one of two polarization channels at a given time; see Appendix A.4. Therefore, the coherency matrix given in Equation (2) is incomplete for the JCMT–SMA baseline; the missing cross-polarization components on all baselines to the JCMT imply that the relation between visibilities and polarized brightness distribution is an underdetermined problem. Fortunately, when fitted in combination with the ALMA–APEX baseline, the Stokes parameters of the unresolved source are determined by the latter. This information is used simultaneously to fit for the JCMT and SMA D-terms. In this fit, only the D-terms affecting the observed cross-polarization product can be estimated, which means that, for each JCMT polarization configuration, only one of the two D-terms of each station (SMA and JCMT) can be determined.

Most D-terms, being instrumental properties, are expected to remain constant across observations of different target sources and observations carried out across multiple days. In the case of ALMA, however, D-terms are generated from an offset in the relative phase calibration between the $X$ and $Y$ linear polarizations of the reference ALMA antenna in the VLBI phasing procedure (Martí-Vidal et al. 2016; Matthews et al. 2018; Goddi et al. 2019). The estimated relative $X$-$Y$ phase at ALMA may change between epochs due to, e.g., a change in the reference antenna,[142] a resetting of the ALMA delay calibration, or the use of different

---

[142] In case the reference antenna is changed within one epoch, the APP scripts can re-reference the polarizer phases to any other antenna, though with some loss of precision (Goddi et al. 2019).

**Table 3**
Daily Average D-terms for ALMA Derived via the Multi-source Intra-site Method

| Date | Band | $D_R$ (%) | $D_L$ (%) |
|------|------|-----------|-----------|
| 2017 Apr 5 | low | $0.30 - 2.80i$ ($\pm 0.70$) | $-1.42 - 3.74i$ ($\pm 0.70$) |
| | high | $-0.17 - 4.10i$ ($\pm 0.60$) | $-1.09 - 4.02i$ ($\pm 0.60$) |
| 2017 Apr 6 | low | $0.60 - 5.45i$ ($\pm 0.40$) | $-0.53 - 6.08i$ ($\pm 0.40$) |
| | high | $-0.09 - 1.52i$ ($\pm 0.30$) | $-0.75 - 1.66i$ ($\pm 0.30$) |
| 2017 Apr 7 | low | $1.12 - 7.10i$ ($\pm 0.70$) | $-0.46 - 5.77i$ ($\pm 0.70$) |
| | high | $1.25 - 4.93i$ ($\pm 0.70$) | $-0.37 - 4.00i$ ($\pm 0.70$) |
| 2017 Apr 10 | low | $0.78 - 2.61i$ ($\pm 0.30$) | $-0.40 - 2.82i$ ($\pm 0.30$) |
| | high | $-0.02 - 3.04i$ ($\pm 0.30$) | $-0.56 - 3.92i$ ($\pm 0.30$) |
| 2017 Apr 11 | low | $-0.15 - 6.33i$ ($\pm 0.50$) | $-0.80 - 6.09i$ ($\pm 0.50$) |
| | high | $-0.29 - 5.19i$ ($\pm 0.40$) | $-0.76 - 5.07i$ ($\pm 0.40$) |

**Note.** The D-term posterior distributions are assumed to be circular Gaussians in the complex plane.

**Table 4**
Campaign-average D-terms for APEX, JCMT, and SMA Derived via the Multi-source Intra-site Method

| Station | $D_R$ (%) | $D_L$ (%) |
|---------|-----------|-----------|
| APEX | $-8.67 + 2.96i$ ($\pm 0.70$) | $4.66 + 4.58i$ ($\pm 1.20$) |
| JCMT | $-0.09 - 2.29i$ ($\pm 1.80$) | $-0.46 + 3.34i$ ($\pm 0.60$) |
| SMA | $-1.73 + 4.81i$ ($\pm 1.00$) | $2.79 + 4.00i$ ($\pm 2.20$) |

calibrators in the polarization calibration process. Therefore, day-to-day variations in ALMA D-terms are expected. D-terms are expected to have a frequency dependence for all stations, hence we obtain separate estimates for each 2 GHz band.

The D-terms fitted for ALMA are dominated by an imaginary component and indicate day-to-day variation along the imaginary axis, as expected from the physical understanding of the leakage origin (Martí-Vidal et al. 2016; Matthews et al. 2018; Goddi et al. 2019). The dispersion in D-term estimates between days and bands is remarkably low for APEX. Thus the fitting is consistent among days as expected: the APEX hardware appears stable across the whole EHT campaign. Similar to APEX, SMA and JCMT should not have varying D-terms across epochs. Therefore, we derive campaign-average D-terms for these three stations from the day-by-day estimates, combining bands. For ALMA, per-day/band D-term estimates are used.

We validate our D-term calibration via intra-site baseline properties using three methods: comparing intra-site baseline source properties to interferometric-ALMA measurements; comparing SMA intra-site leakage estimates to interferometric-SMA estimates; and comparing `polsolve` leakage estimation to point-source polarimetric modeling in the `eht-imaging` library and DMC. We additionally motivate leakage calibration using band-averaged products from intra-band leakage studies for ALMA–APEX.

Simultaneously to our VLBI observations, ALMA also observes as an interferometric array (referred to as ALMA-only) in a linear-polarization basis. This array data is used for ALMA–VLBI calibration in the Quality Assurance process at ALMA (QA2; Goddi et al. 2019), and provides source-integrated information for calibration refinement and validation, such as total flux densities or polarization properties. Given that our intra-site baselines do not resolve the observed sources, the source-integrated properties from ALMA–APEX, SMA–JCMT, and the core component of ALMA-only should match. We show our validation of the derived source polarimetric properties from the intra-site D-term fitting against QA2 ALMA-only estimates in the top (for $\mathcal{Q}$) and bottom (for $\mathcal{U}$) panels of Figure 17. There is a strong correlation between the Stokes parameters of all sources derived from the ALMA-only observations (Goddi et al. 2019) and the estimates from the ALMA–APEX intra-site VLBI baseline. This correlation can be seen as a further validation test of the quality of the EHT polarimetric calibration.

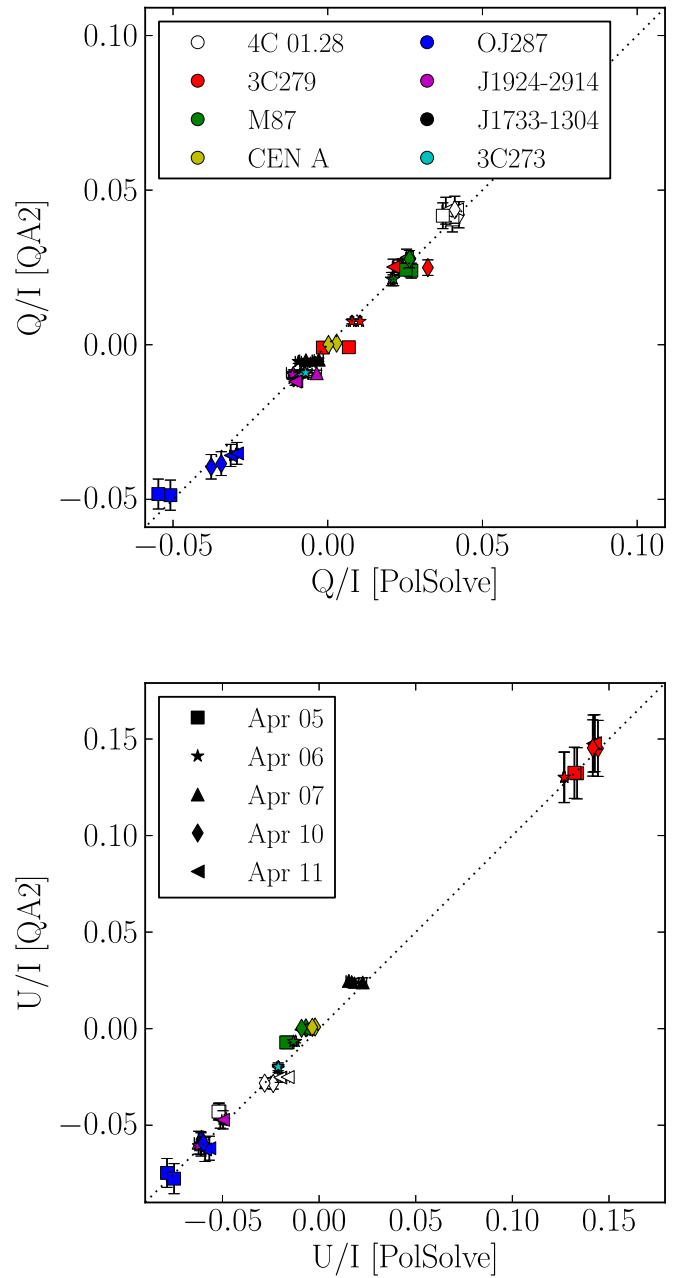

**Figure 17.** Comparison of source-integrated Stokes $\mathcal{Q}$ and $\mathcal{U}$ estimates from intra-site EHT baselines using the multi-source fitting mode of `polsolve` to those from ALMA-only observations (Goddi et al. 2019).

The polarimetric leakage of the SMA is well characterized, with D-terms of only a few percent expected for observations near the 233.0 GHz tuned frequency of the quarter-wave plates (Marrone 2006; Marrone & Rao 2008). In addition to historical measurements of leakage, near-in-time polarimetric observations of sources with the SMA also allowed us to compute quasi-simultaneous leakage estimates that can be compared with our intra-site method estimates. Observations of 3C 454.3, M87, and 3C 279 within a month of our EHT campaign provided an upper limit of 10% for D-terms, which is consistent with our intra-site method estimates, and stable across days at the 1% level.

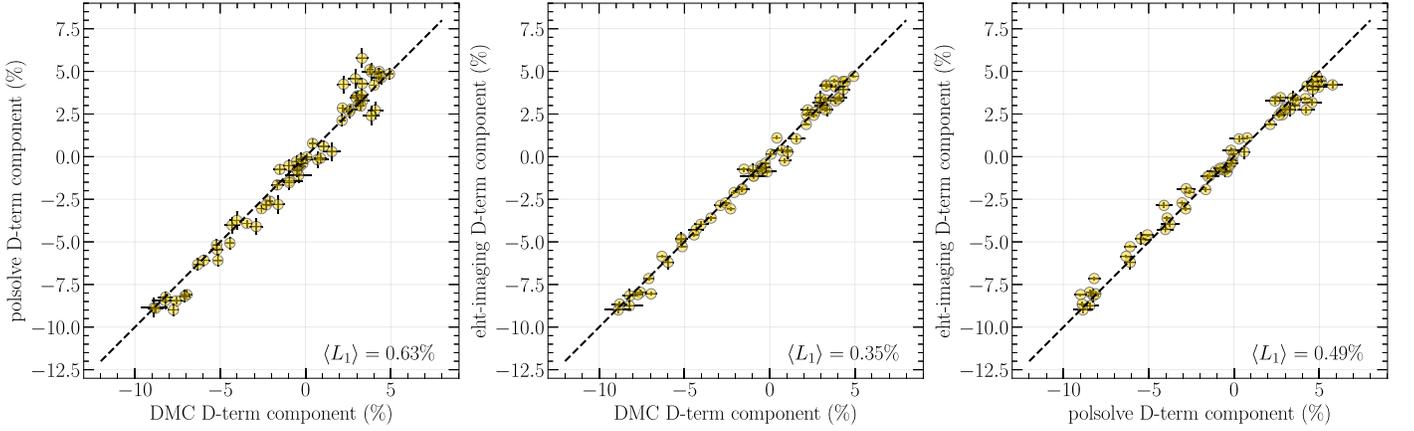

**Figure 18.** Pairwise comparisons of leakage estimates for ALMA and APEX, obtained from point-source modeling of the intra-site method with `polsolve`, `eht-imaging`, and DMC. Both `polsolve` and `eht-imaging` leakages are derived from multi-source fits, while the DMC leakages are derived from fitting to 3C 279 only. Each panel aggregates leakage estimates from both stations (ALMA and APEX), both bands, and all four observing days. Values quoted in the lower right-hand corner of each panel are the uncertainty-weighted mean absolute deviation for the corresponding pair of fits. The dashed line on each plot marks where $y = x$.

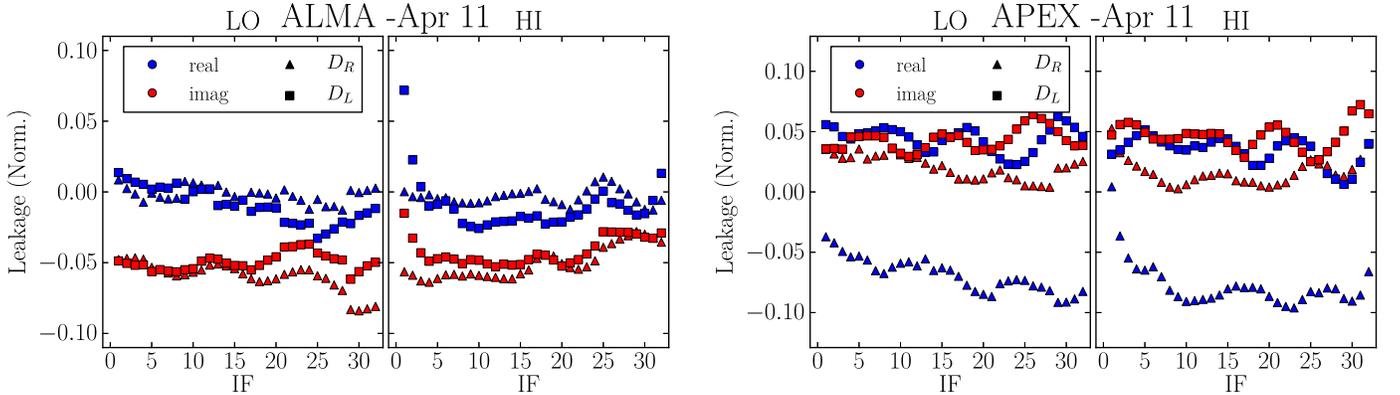

**Figure 19.** ALMA (left panels) and APEX (right panels) D-term spectra recovered on 2017 April 11 using `polsolve`. Each band has a width of 2 GHz and is divided into 32 intermediate-frequency sub-bands (IFs) of equal width.

We also validated the `polsolve` leakage estimates using point-source modeling within the `eht-imaging` and DMC libraries. Both modeling schemes assume a constant polarization fraction and EVPA for the point source. The DMC model fits to cross-hand and parallel-hand visibilities by incorporating right- and left-hand station gains as model parameters, while the `eht-imaging` model fits to gain-independent "polarimetric closure" data products consisting of the ratio of cross-hand visibilities to parallel-hand visibilities on a single baseline (see, e.g., Blackburn et al. 2020),

$$\mathcal{K}_{jk} = \frac{R_j L_k^* \times L_j R_k^*}{R_j R_k^* \times L_j L_k^*}. \tag{D1}$$

Both models use Gaussian likelihood functions for their respective data products.

In Figure 18, we compare the multi-source `polsolve` leakage estimates to multi-source `eht-imaging` fits and single-source (using 3C 279) DMC fits; both `eht-imaging` and DMC have fit only to the ALMA–APEX baseline, while the `polsolve` estimates additionally fit to the SMA–JCMT baseline. We find that the leakage terms recovered by all three methods are consistent with one another, with an uncertainty-weighted mean absolute deviation across all days and bands of <1% in absolute leakage between any two methods.

Furthermore, our leakage estimation methods used band-averaged data products. Given the high S/N of the detections between ALMA and APEX, there are strong detections in all four correlation products (i.e., $RR^*$, $RL^*$, $LR^*$, and $LL^*$) at each intermediate-frequency band.[143] Therefore, we can use the high S/N on ALMA–APEX to estimate the D-terms at each intermediate-frequency band in a `polsolve` fit and study the possible frequency dependence of the instrumental polarization of ALMA and APEX. The results are shown in Figure 19. This test showed very stable D-term estimates across the entire band, motivating our use of band-averaging for our final results in Table 3.

## Appendix E
## Fiducial Leakage D-terms from M87 Imaging

We provide fiducial M87 D-term estimates for each imaging or posterior exploration method in Table 5. These D-term results for LMT, SMT, and PV are depicted in the main text in the right panel of Figure 2. In Figure 20 we show an example of one-to-one software comparisons of the campaign-average D-terms for LMT, PV, and SMT.

---

[143] In the VLBI correlation, each 2 GHz band is divided into 32 contiguous sub-bands of equal width, which are called "intermediate-frequency bands," or IFs.

**Table 5**
Fiducial Set of Low-band D-terms for Each Station as Derived from M87 Data Via Polarimetric Imaging with Fiducial Polarimetric Survey Parameters

| Method | 2017 April 5 | | 2017 April 6 | | 2017 April 10 | | 2017 April 11 | |
|---|---|---|---|---|---|---|---|---|
| | $D_R$ (%) | $D_L$ (%) | $D_R$ (%) | $D_L$ (%) | $D_R$ (%) | $D_L$ (%) | $D_R$ (%) | $D_L$ (%) |
| LMT | | | | | | | | |
| eht-imaging | $0.48 + 3.16i$ | $-0.47 + 2.17i$ | $1.29 + 5.37i$ | $0.56 + 2.23i$ | $-2.04 + 2.99i$ | $-5.02 + 0.64i$ | $1.43 + 3.12i$ | $-0.44 + 0.52i$ |
| polsolve | $1.5 + 2.92i$ | $-2.26 + 0.12i$ | $1.63 + 5.1i$ | $-0.38 + 1.48i$ | $1.24 + 0.69i$ | $-3.65 - 2.69i$ | $2.48 + 0.47i$ | $-1.12 + 0.25i$ |
| LPCAL | $0.6 + 2.11i$ | $-3.0 + 0.0i$ | $-0.2 + 4.9i$ | $-1.35 + 0.73i$ | $-1.79 + 2.96i$ | $-3.88 - 2.11i$ | $0.74 + 2.42i$ | $-1.37 - 0.51i$ |
| DMC | $1.8 + 3.3i$ | $-1.5 + 1.7i$ | $2.5 + 6.4i$ | $-0.8 + 2.4i$ | $-2.0 + 4.3i$ | $-5.2 + 2.3i$ | $2.2 + 4.4i$ | $-0.8 + 0.3i$ |
| | $(0.6 + 0.6i)$ | $(0.6 + 0.6i)$ | $(0.6 + 0.6i)$ | $(0.6 + 0.6i)$ | $(1.1 + 1.0i)$ | $(1.4 + 1.4i)$ | $(0.5 + 0.5i)$ | $(0.5 + 0.4i)$ |
| THEMIS | $2.3 + 1.7i$ | $-1.0 + 2.9i$ | $3.1 + 5.5i$ | $-0.2 + 3.0i$ | $-0.2 + 3.5i$ | $-4.8 + 0.8i$ | $2.8 + 2.9i$ | $-0.7 + 0.9i$ |
| | $(0.7 + 0.8i)$ | $(0.7 + 0.7i)$ | $(0.6 + 0.6i)$ | $(0.7 + 0.8i)$ | $(1.4 + 1.3i)$ | $(1.5 + 1.6i)$ | $(0.7 + 0.7i)$ | $(0.8 + 0.8i)$ |
| SMT | | | | | | | | |
| eht-imaging | $3.94 + 7.51i$ | $-4.84 + 9.46i$ | $3.36 + 7.78i$ | $-4.22 + 8.19i$ | $4.88 + 9.26i$ | $-3.01 + 9.2i$ | $3.84 + 7.46i$ | $-4.38 + 10.46i$ |
| polsolve | $3.99 + 7.45i$ | $-5.42 + 9.17i$ | $3.66 + 7.62i$ | $-4.52 + 7.49i$ | $5.61 + 8i$ | $-5.3 + 9.68i$ | $4.07 + 7.48i$ | $-5.88 + 10.95i$ |
| LPCAL | $3.24 + 8.23i$ | $-5.68 + 9.04i$ | $3.14 + 8.29i$ | $-3.94 + 7.22i$ | $4.51 + 9.54i$ | $-5.6 + 6.64i$ | $4.0 + 7.56i$ | $-5.56 + 9.3i$ |
| DMC | $2.0 + 9.4i$ | $-3.1 + 10.7i$ | $1.7 + 10.3i$ | $-2.6 + 8.6i$ | $3.1 + 9.3i$ | $-2.9 + 8.9i$ | $3.0 + 8.8i$ | $-4.0 + 9.5i$ |
| | $(0.9 + 0.6i)$ | $(0.9 + 0.7i)$ | $(0.9 + 0.6i)$ | $(0.9 + 0.6i)$ | $(1.3 + 1.2i)$ | $(1.2 + 0.9i)$ | $(0.8 + 0.5i)$ | $(0.8 + 0.5i)$ |
| THEMIS | $2.7 + 8.3i$ | $-4.4 + 9.6i$ | $2.8 + 9.1i$ | $-3.1 + 8.7i$ | $1.0 + 9.3i$ | $-3.6 + 9.7i$ | $2.9 + 6.9i$ | $-3.9 + 10.5i$ |
| | $(0.8 + 0.8i)$ | $(0.8 + 0.8i)$ | $(0.8 + 0.7i)$ | $(0.8 + 0.7i)$ | $(1.3 + 1.3i)$ | $(1.3 + 1.4i)$ | $(0.7 + 0.7i)$ | $(0.7 + 0.7i)$ |
| PV | | | | | | | | |
| eht-imaging | $-14.57 + 2.34i$ | $15.77 + 1.21i$ | $-13.01 + 3.51i$ | $13.16 + 1.75i$ | $-10.82 - 2.25i$ | $14.68 + 1.71i$ | $-12.70 - 1.21i$ | $14.86 + 0.57i$ |
| polsolve | $-10.64 + 1.2i$ | $13.23 + 2.83i$ | $-13.1 + 3.99i$ | $11.53 + 2.35i$ | $-6.38 - 1.08i$ | $16.64 + 5.17i$ | $-11.32 - 0.63i$ | $13.96 + 0.84i$ |
| LPCAL | $-9.98 + 0.67i$ | $16.4 + 2.22i$ | $-11.66 + 1.5i$ | $14.61 + 2.14i$ | $-12.42 - 3.85i$ | $16.56 + 3.66i$ | $-11.54 + 0.18i$ | $16.16 + 1.56i$ |
| DMC | $-14.0 + 1.7i$ | $18.2 - 0.4i$ | $-11.9 + 4.0i$ | $12.8 - 1.1i$ | $-11.1 - 1.0i$ | $12.7 + 3.3i$ | $-14.2 + 0.1i$ | $12.9 - 1.6i$ |
| | $(1.5 + 1.9i)$ | $(1.6 + 2.2i)$ | $(1.3 + 1.6i)$ | $(1.3 + 1.6i)$ | $(2.6 + 2.5i)$ | $(3.1 + 3.1i)$ | $(1.0 + 1.7i)$ | $(1.0 + 1.6i)$ |
| THEMIS | $-13.9 + 5.2i$ | $17.7 - 2.4i$ | $-10.6 + 6.3i$ | $13.8 + 1.0i$ | $-13.4 + 0.2i$ | $17.1 + 1.7i$ | $-13.6 + 3.6i$ | $14.5 - 0.3i$ |
| | $(1.5 + 3.3i)$ | $(1.5 + 1.7i)$ | $(2.2 + 1.5i)$ | $(1.3 + 1.4i)$ | $(2.4 + 1.9i)$ | $(2.5 + 2.4i)$ | $(1.1 + 1.5i)$ | $(1.1 + 1.3i)$ |
| Residual Leakage ALMA | | | | | | | | |
| LPCAL | $0.91 + 0.81i$ | $0.17 + 1.82i$ | $-0.48 + 1.18i$ | $-0.62 - 0.58i$ | $-2.16 + 0.94i$ | $-1.11 - 0.74i$ | $-0.65 + 0.83i$ | $0.16 + 0.43i$ |
| DMC | $0.4 - 0.5i$ | $-0.4 + 3.5i$ | $0.1 + 0.7i$ | $0.2 + 4.1i$ | $-0.7 + 2.5i$ | $2.5 + 1.3i$ | $0.3 + 1.3i$ | $2.4 + 2.5i$ |
| | $(0.7 + 0.8i)$ | $(0.6 + 0.5i)$ | $(0.7 + 0.7i)$ | $(0.5 + 0.6i)$ | $(0.7 + 0.7i)$ | $(1.1 + 1.0i)$ | $(0.6 + 0.6i)$ | $(0.5 + 0.5i)$ |
| THEMIS | $0.9 + 1.4i$ | $-0.4 + 3.1i$ | $0.7 + 1.5i$ | $0.4 + 3.6i$ | $0.3 + 0.6i$ | $-0.6 + 2.5i$ | $0.1 + 1.9i$ | $1.9 + 1.9i$ |
| | $(0.7 + 0.8i)$ | $(0.6 + 0.6i)$ | $(0.6 + 0.7i)$ | $(0.6 + 0.7i)$ | $(1.3 + 1.1i)$ | $(1.2 + 0.9i)$ | $(0.6 + 0.7i)$ | $(0.7 + 0.6i)$ |
| Residual Leakage APEX | | | | | | | | |
| LPCAL | $1.47 + 0.22i$ | $-0.62 - 1.69i$ | $-0.13 - 0.40i$ | $-1.27 - 0.56i$ | $0.03 - 1.00i$ | $-0.97 + 1.31i$ | $1.12 - 0.26i$ | $-0.45 - 0.01i$ |
| DMC | $-4.3 + 1.0i$ | $-0.8 - 0.5i$ | $2.8 + 0.5i$ | $-1.1 - 1.3i$ | $7.3 - 2.6i$ | $1.9 + 2.1i$ | $2.4 + -0.2i$ | $-2.7 - 0.1i$ |
| | $(2.1 + 2.1i)$ | $(2.0 + 2.1i)$ | $(1.3 + 1.3i)$ | $(1.2 + 1.2i)$ | $(1.7 + 1.4i)$ | $(2.0 + 2.1i)$ | $(1.1 + 1.1i)$ | $(1.0 + 1.0i)$ |
| THEMIS | $0.6 - 0.4i$ | $-1.7 - 0.4i$ | $3.6 + 0.2i$ | $-1.6 - 1.8i$ | $3.7 - 1.7i$ | $0.5 - 0.6i$ | $2.7 + 0.8i$ | $-1.7 - 0.6i$ |
| | $(1.2 + 1.3i)$ | $(1.3 + 1.2i)$ | $(0.9 + 0.8i)$ | $(0.8 + 0.8i)$ | $(1.3 + 1.6i)$ | $(1.4 + 1.6i)$ | $(0.8 + 0.8i)$ | $(0.7 + 0.7i)$ |
| Residual Leakage SMA | | | | | | | | |
| LPCAL | $-1.59 + 8.08i$ | $7.46 + 9.51i$ | $3.02 + 8.02i$ | $12.24 + 5.44i$ | $9.36 + 11.73i$ | $13.42 + 14.64i$ | $0.21 + 4.53i$ | $14.01 + 8.40i$ |
| DMC | $0.8 - 3.4i$ | $-4.9 - 1.0i$ | $-1.5 - 2.1i$ | $0.2 + 0.3i$ | $5.5 + 5.4i$ | $-1.8 + 2.5i$ | $-0.2 - 2.2i$ | $0.7 + 1.9i$ |
| | $(1.1 + 1.2i)$ | $(2.4 + 2.5i)$ | $(1.3 + 1.4i)$ | $(2.7 + 2.8i)$ | $(4.5 + 5.5i)$ | $(3.1 + 3.3i)$ | $(2.5 + 2.6i)$ | $(1.6 + 1.8i)$ |
| THEMIS | $2.4 - 2.9i$ | $-1.4 - 0.8i$ | $-0.4 + 0.7i$ | $0.3 + 1.4i$ | $2.3 + 1.0i$ | $-4.9 + 7.1i$ | $-2.8 + 0.4i$ | $1.4 + 3.2i$ |
| | $(1.3 + 1.2i)$ | $(1.2 + 1.3i)$ | $(1.5 + 1.6i)$ | $(1.3 + 1.7i)$ | $(2.8 + 2.9i)$ | $(3.2 + 3.3i)$ | $(1.6 + 1.7i)$ | $(1.9 + 2.1i)$ |

**Table 5**
(Continued)

| Method | 2017 April 5 | | 2017 April 6 | | 2017 April 10 | | 2017 April 11 | |
|---|---|---|---|---|---|---|---|---|
| | $D_R$ (%) | $D_L$ (%) | $D_R$ (%) | $D_L$ (%) | $D_R$ (%) | $D_L$ (%) | $D_R$ (%) | $D_L$ (%) |
| | | | | Residual Leakage JCMT | | | | |
| DMC | $1.5 - 0.1i$ | ⋯ | $1.1 - 1.8i$ | ⋯ | ⋯ | $-1.3 + 1.1i$ | ⋯ | $2.0 + 0.1i$ |
| | $(1.2 + 1.2i)$ | ⋯ | $(1.3 + 1.4i)$ | ⋯ | ⋯ | $(2.9 + 3.2i)$ | ⋯ | $(1.5 + 1.7i)$ |
| THEMIS | $2.3 - 0.4i$ | ⋯ | $0.8 + 0.3i$ | ⋯ | ⋯ | $-0.6 + 6.9i$ | ⋯ | $2.7 + 1.1i$ |
| | $(1.2 + 1.3i)$ | ⋯ | $(1.4 + 1.5i)$ | ⋯ | ⋯ | $(2.7 + 2.6i)$ | ⋯ | $(1.5 + 1.6i)$ |

**Note.** The imaging pipelines pre-calibrate the ALMA, APEX, SMA, and JCMT D-terms using intra-site baseline fitting (see Section 4.2), and so only the D-terms for stations forming long baselines (i.e., LMT, SMT, and PV) are reported for these approaches. LPCAL is an exception due to its inability to fix D-terms for the pre-corrected stations: derived residual D-terms are shown here. The posterior exploration pipelines do not pre-calibrate the zero-baseline D-terms (see Appendix C.3), and we report here "residual" leakage values—i.e., the excess leakage, as determined by the posterior exploration pipelines, over that obtained from zero-baseline fitting (given in Table 3 for ALMA and Table 4 for APEX, JCMT, and SMA). The uncertainties for each of the posterior exploration leakage estimates are quoted in parenthesis. D-terms for LMT, SMT, and PV are depicted in Figure 2.

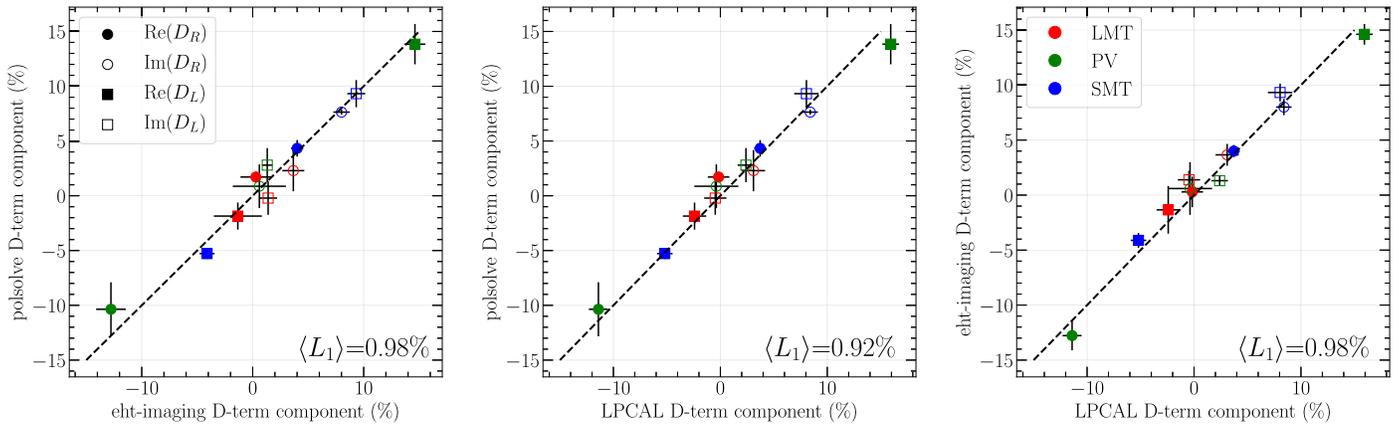

**Figure 20.** Example one-to-one software comparisons of the campaign-average D-term estimates for LMT, PV, and SMT. Left panel: comparison of `eht-imaging` estimates against `polsolve` estimates. Middle panel: comparison of `LPCAL` estimates against `polsolve` estimates. Right panel: comparison of `LPCAL` estimates against `eht-imaging` estimates. The $L_1$ norm is averaged over the left/right and real/imaginary components of the D-terms and over all three EHT stations shown. See Section 4.2 for a discussion of the D-term $L_1$ norm values between `eht-imaging/polsolve/LPCAL` and THEMIS/DMC.

## Appendix F
## Preliminary Imaging Results for M87

In this Appendix we present the preliminary polarimetric results on M87 obtained using the three imaging methods. These preliminary images were generated "by hand," with manual tuning of free parameters in the imaging and calibration process, before full parameter surveys were done to choose parameters more objectively and evaluate uncertainties. These results are not blind tests in analogy to the initial stage of total intensity imaging (see Paper IV). Nonetheless, in this early stage of manual imaging we found a high degree of similarity in the recovered structure and D-terms between methods; these results guided the design of our synthetic data tests and parameter survey strategy that we pursued to obtain our final polarimetric images of M87.

In Figure 21, we present our recovered total intensity and preliminary polarimetric images of M87 on 2017 April 11

produced by the three methods available when preliminary image reconstructions were conducted. In Figure 21, we also show the D-terms associated with these images. Each method reproduces consistent D-terms for all three remaining long-baseline EHT stations. The preliminary polarimetric images are roughly consistent across methods. In all images, the M87 ring-like structure is predominantly polarized mostly in the south-west part with a fractional polarization up to $|m| \sim 15\%$. The EVPAs are organized into a coherent pattern along the ring. However, small differences in fractional polarization and polarized flux density are evident between the three packages.

The preliminary results in Figure 21 revealed the main structure of the linearly polarized source and suggested consistency between different imaging methods. They strongly motivate the need for conducting full parameter surveys for each method to optimize the chosen imaging parameters and validate the results on synthetic data.

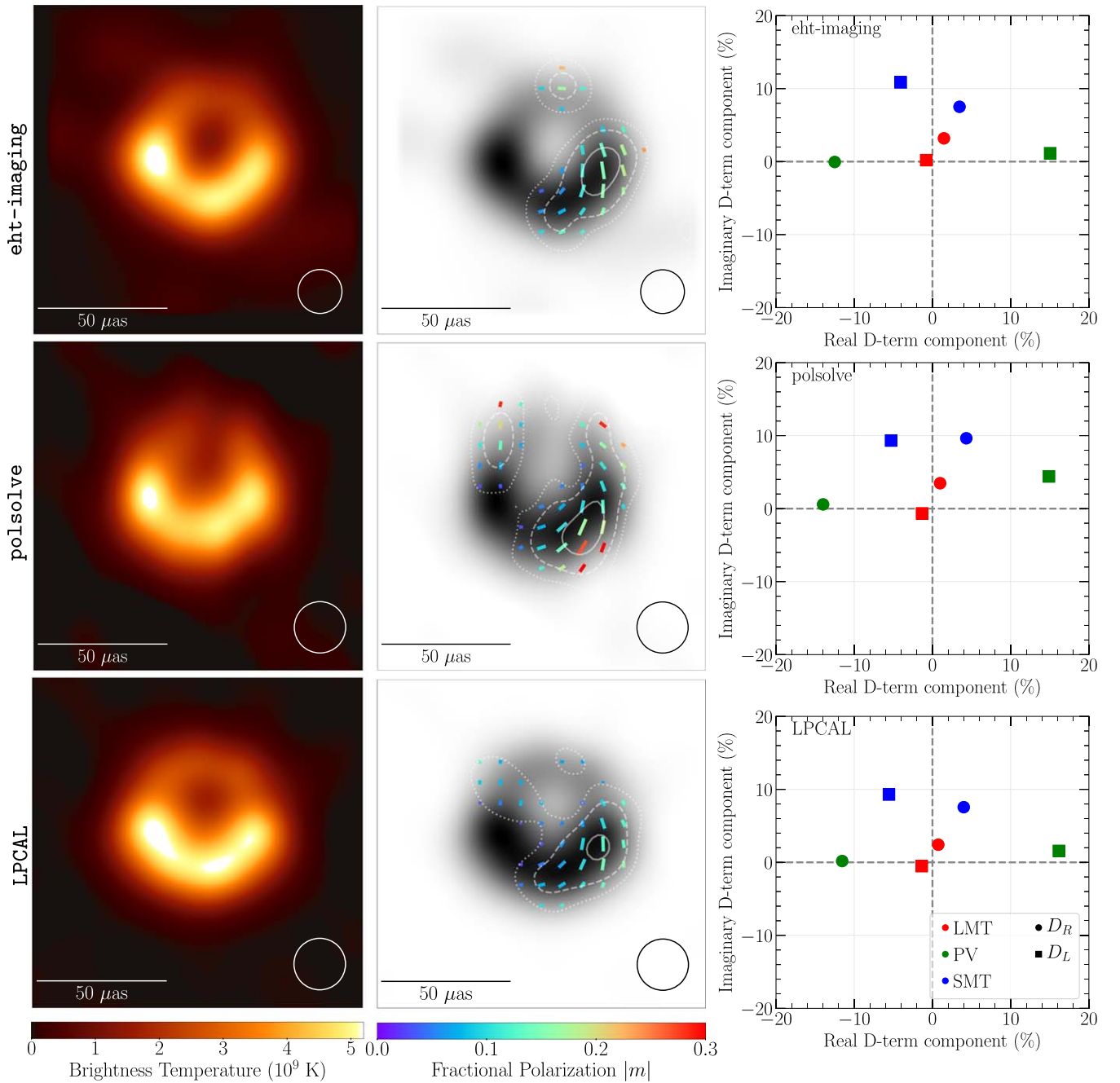

**Figure 21.** Left column: preliminary total intensity images reconstructed with `eht-imaging`, `polsolve`, and `LPCAL` on 2017 April 11 low-band data. `eht-imaging` images are blurred with a 17.1 μas circular Gaussian, to obtain an equivalent resolution to the `polsolve` and `LPCAL` CLEAN images restored with a 20 μas circular Gaussian. Middle column: corresponding polarimetric reconstructions obtained as a result of the full-array leakage calibration. Total intensity is shown in the background in grayscale. Polarization ticks indicate the EVPA, the tick length is proportional to the linear polarization intensity magnitude, and color indicates fractional linear polarization. The contours mark linear polarized intensity. The solid, dashed, and dotted contour levels correspond to linearly polarized intensity of 20, 10, and 5 μJy μas$^{-2}$. Cuts were made to omit all regions in the images where Stokes $\mathcal{I} < 10\%$ of the peak flux density and $\mathcal{P} < 20\%$ of the peak polarized flux density. In all reconstructions, the region with the highest linear polarization fraction and polarized intensity is predominantly in the southwest portion of the ring. Right column: preliminary D-terms for SMT, PV, and LMT derived via leakage calibration through `eht-imaging`, `polsolve`, and `LPCAL` polarimetric imaging.

## Appendix G
## Description of the Parameter Survey and Scoring for Each Method

In what follows, we describe each method's approach to surveying the space of free parameters available to it, scoring the results using the six synthetic data models introduced in Section 4.3, and from these scores, determining a fiducial set of parameters. We use each method's fiducial parameter settings to obtain the final calibrated D-terms and M87 images.

### G.1. `eht-imaging` Parameter Survey

The polarimetric imaging procedure alternates between imaging via minimization of the objective function (Equation (C5)) and D-term calibration, as described in

**Table 6**
The Parameters Surveyed by eht-imaging

| Parameter | Values | | | | |
|---|---|---|---|---|---|
| $\alpha_P$ | 0.01 | 0.1 | **1** | 10 | 100 |
| $\beta_{HW}$ | 0.01 | 0.1 | 1 | 10 | **100** |
| $\beta_{TV}$ | 0.01 | 0.1 | **1** | 10 | 100 |
| $n_{iter}$ | 1 | 10 | **50** | ... | ... |
| $f_{sys}$ | 0 | 0.002 | **0.005** | 0.01 | ... |

**Note.** The selected fiducial parameters are displayed in bold.

Section C.2. In the imaging stage, the critical parameters that influence the final reconstruction include the four hyperparameters $\alpha_P$, $\alpha_m$, $\beta_{HW}$, and $\beta_{TV}$ that set the relative weighting in the objective function between the different data constraints and regularizing terms. In surveying different parameters in the eht-imaging survey, we fix $\alpha_m = 1$ and vary the other three hyperparameter weights.

In addition to the hyperparameter weights, an additional free parameter in our objective function is the amount of additional systematic noise to add to the data as a budget for non-leakage sources of systematic error (see Papers III, IV). To account for these systematics, we add a term equal to $f_{sys} \times |\tilde{\mathcal{I}}|$ in quadrature to our baseline thermal noise estimates, where $f_{sys}$ is an overall multiplicative factor. Finally, we also include as a parameter in our surveys the number of iterations $n_{iter}$ of alternating between imaging and calibrating the station D-terms. The full list of parameters that we vary in the eht-imaging parameter survey is listed in Table 6, with the fiducial parameters used in reconstructing images of M87 denoted in bold.

To select a fiducial set of parameters that performs best on all six synthetic data tests, we assign each parameter set $\boldsymbol{p}$ two scores on each synthetic data set $a$; one scoring the fidelity of the polarized image reconstruction $s_{fid,a}(\boldsymbol{p})$, and one scoring the accuracy of the D-term estimation $s_{dterm,a}(\boldsymbol{p})$. First, we score the image fidelity by computing the normalized cross-correlation $\rho_{NX}$ between the reconstructed and ground-truth polarimetric intensity distribution. That is, we use Equation (15) of Paper IV on the images of $\sqrt{\mathcal{Q}^2 + \mathcal{U}^2}$. Then the fidelity score for the parameter set $\boldsymbol{p}$ on the synthetic data set $a$ is simply

$$s_{fid,a}(\boldsymbol{p}) = \rho_{NX}. \tag{G1}$$

We compute the D-term estimation accuracy metric by first calculating the average $\ell_2$ distance $d_D$ between the reconstructed D-terms and the ground truth for the data set. For M87, we average this distance over the three stations that we calibrate at this stage: SMT, LMT, and PV. Then we transform this distance to a score between 0 (bad) and 1 (good) by using a sigmoid function;

$$s_{dterm,a}(\boldsymbol{p}) = 1 - \mathrm{Erf}(d_D / \sqrt{2} \, d_{tol}), \tag{G2}$$

where $d_{tol}$ is a threshold for the average distance between the ground truth and recovered D-terms beyond which we begin to heavily penalize the reconstruction. We set $d_{tol} = 5\%$.

Finally, having computed the two scores $s_{fid,a}$ and $s_{dterm,a}$ for the parameter set $\boldsymbol{p}$ on the synthetic data set $a$, we compute a final score $s(\boldsymbol{p})$ for the parameter set by multiplying these

individual scores together on all six synthetic data sets $a$:

$$s(\boldsymbol{p}) = \prod_a s_{fid,a}(\boldsymbol{p}) s_{dterm,a}(\boldsymbol{p}). \tag{G3}$$

We then have a final score $s$ for each parameter set incorporating its performance in accurately reconstructing the polarized flux distribution and input D-terms on six synthetic data sets. We take the parameter set $\boldsymbol{p}$ with the highest score as our fiducial parameter set. The fiducial set is indicated in bold in Table 6.

### G.2. polsolve Parameter Survey

The polsolve algorithm is characterized by several degrees of freedom: the pixel's angular size (smaller values increase the astrometric accuracy of the CLEAN components and, hence, the quality of the deconvolution), the field of view (the effect of this parameter is minimum if a CLEAN masking is applied), the visibility weighting (mainly defined with the Briggs "robustness" parameter, $r$; Briggs 1995), and the method for division of the Stokes $\mathcal{I}$ model into subcomponents of constant fractional polarization (see Section C.1).

The first step in the polsolve procedure is to generate a first version of the $\mathcal{I}$ image (using the CASA task clean). Several iterations of phase and amplitude self-calibration (using tasks gaincal and applycal) may be applied to the data, in order to optimize the dynamic range of the $\mathcal{I}$ model. The self-calibration gains are forced to be equal for the $R$ and $L$ polarizations at all antennas.[144] Then, the $\mathcal{I}$ model is split into several sub-components and formatted for its use in polsolve, using the CASA task CCextract.[145] Finally, polsolve estimates the D-terms of LMT, SMT, and PV, together with the fractional polarizations and EVPAs of all the source sub-components. The estimated D-terms are applied to the data and a final run of clean is peformed, to generate the final version of the full-polarization images.

In the polsolve parameter survey, the $\mathcal{I}$ division is done in two ways. In one approach, a centered square mask of $50 \times 50$ $\mu$as is created and divided into a regular grid of $n \times n$ cells (in this case, cells that do not contain CLEAN components are not used in the fit). Alternatively, a centered circular mask of $40-80$ $\mu$as diameter is created and divided azimuthally into a regular set of $n$ pieces. The full parameter survey with polsolve consists of an exploration of the following.

1. Both types of model sub-divisions (i.e., either regular grid or azimuthal cuts), with $n$ running from 1 to 12.
2. Robustness parameter, from $r = -2$ to $r = 2$ (i.e., from uniform to natural weighting) in steps of 0.2.
3. Relative weight of the ALMA antenna (which affects the shape of the point-spread function (PSF) of the instrument, especially for values of $r$ far from $-2$), running from 0.1 to 1.0 in steps of 0.1.
4. In all images, a circular CLEAN mask of diameter varying from 40 to 80 $\mu$as (in steps of 5 $\mu$as) is used. The size of the CLEAN mask is the same as the mask to define the $\mathcal{I}$ model sub-division.

---

[144] A necessary assumption at this stage, as polarization-dependent gains do not commute with the Dterm corrections already applied to ALMA, APEX, and SMA.

[145] Part of the CASA-poltools software package: https://code.launchpad.net/casa-poltools.

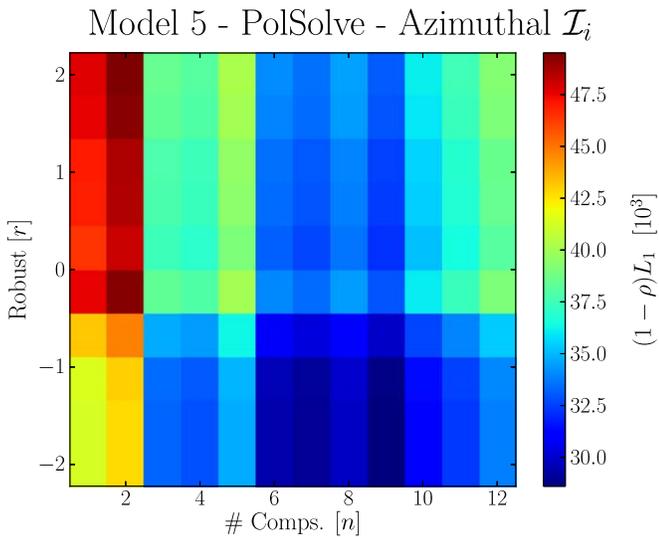

Model 5 - PolSolve - Azimuthal $\mathcal{I}_i$

**Figure 22.** Figure of merit $(1 - \rho)L_1$ (lower values mean better results) for the polsolve algorithm (running on synthetic data Model 5), as a function of its two main degrees of freedom (i.e., $\mathcal{I}$ sub-division and visibility weighting).

The pixel size is fixed to 1 $\mu$as and the image size covers $256 \times 256$ $\mu$as. We note that for models with extended emission (i.e., Models 1 and 2, see Figure 4) an additional CLEAN mask is added at the southern part of the image.

For each combination of parameters in the survey, we compute the $L_1$ norm of the differences between true and estimated D-terms, as well as the correlation coefficients, $\rho_s$ (for each Stokes parameter, $s$) between the CLEAN image reconstructions and true-source structures, properly convolved with the same beam. These quantities can be used to select the best combination of parameters for D-term calibration and image reconstruction (the fiducial imaging parameters for polsolve). Depending on the relative weight that is given to $L_1$ and the average image correlation, $\rho = (\rho_\mathcal{I} + \rho_\mathcal{Q} + \rho_\mathcal{U})/3$, we obtain slightly different fiducial parameters.

In Figure 22, we show an example plot from our polsolve parameter survey for Model 5 (see Figure 4; similar results are found for the rest of models in the survey). The chosen figure of merit is $(1 - \rho)L_1$, which we show as a function of the robustness parameter $r$ and the number of slices $n$ in the azimuthal sub-division. From Figure 22, we note that the dependence of the figure of merit with $n$ becomes weak for values of $n$ larger than 3–4 and robustness parameters between $-1$ and $-2$. This behavior also happens if the regular gridding is used to generate the $\mathcal{I}_i$ sub-components. A qualitative explanation of this effect may be that large values of $n$ translate into sub-components of sizes smaller than the synthesized resolution. Therefore, increasing the number of sub-components does not improve the fit, because the small separations between neighboring sub-components correspond to spatial frequencies that are not sampled by the interferometer.

Conversely, the fitted fractional polarizations of close by sub-components become highly correlated in the fit, and the $L_1$ norm of the D-terms saturate around a minimum value.

Based on the combined analysis of all the synthetic data sets, we determine the fiducial parameters for polsolve: a robustness parameter of $-1$ (though $-2$ produces similar results, especially for models 4 to 6, see Figure 4), a circular slicing with $n = 8$ sub-components (which produces results similar to 9–10 sub-components, and also similar to those from a regular gridding, $n \times n$, with $n = 3$–5), relative ALMA weights of 1.0 (which produces similar results as for values between 0.5–1.0), and a circular CLEAN mask of 50 $\mu$as diameter.

### G.3. LPCAL Parameter Survey

The standard procedure for D-term calibration using LPCAL is as follows. 1.) Select a calibrator that has either a low fractional polarization or a simple polarization structure, and a wide range of parallactic angle coverage. This is M87 in our case. 2.) Produce a total intensity CLEAN map of the calibrator with e.g., Difmap. 3.) Split the CLEAN model into several sub-models with the task CCEDT in AIPS. LPCAL assumes that each sub-model has a constant fractional polarization and EVPA. Therefore, the more sub-models that we use to divide the Stokes $\mathcal{I}$ image, the more degrees of freedom we have for modeling the source linear polarization. 4.) Run LPCAL using the sub-models.

We follow this standard procedure for the D-term calibration. In this work, we consider an additional parameter, the ALMA weight-scaling factor. Down-weighting the ALMA data can be useful when there are significant systematic uncertainties in the ALMA visibilities. In this case, the solutions for other stations can be distorted as the fitting would be dominated by the ALMA baselines due to its high sensitivity and the corresponding smaller formal error bars. In addition to the ALMA weight-scaling factor, we consider the number of CLEAN sub-models as the main parameter that may significantly affect D-term estimation with LPCAL.

We first performed a manual parameter survey using the synthetic data. We reconstructed D-terms with LPCAL by using different numbers of sub-models and ALMA weight-scaling factors, and compared with the ground-truth values. We conclude that using a relatively large number of sub-models ($\gtrsim 10$) gives better reconstructions, while the results do not change much when more than 10 sub-models are used. Also, we find that the results are not sensitive to the ALMA weight re-scaling. The strategy and parameters that we adopted could reproduce the ground-truth D-terms in the synthetic data within an accuracy of $\sim 1\%$ (Figure 5).

Next, we analyzed the low-band M87 data. We used a common approach as for the synthetic data tests, but we allowed several users to independently calibrate and image the real data with different parameter settings to test the robustness of the method. First, each user reconstructed the Stokes $\mathcal{I}$ image with CLEAN in `Difmap`. The CLEAN components were divided into several sub-models by the task CCEDT in AIPS. Each user then used `LPCAL` in AIPS to solve for D-terms for all antennas. This includes the possible residual D-terms of ALMA, APEX, and SMA (as `LPCAL` cannot set any station D-terms to zero). We let each user test different schemes for CLEAN and use different parameters that were not seen as sensitively impacting results in the synthetic data survey. However, all users split their CLEAN models into a number of sub-models ($\gtrsim$10), in accordance with our synthetic data parameter survey results. Different $(u, v)$ weighting parameters, CLEAN windows, and CLEAN cutoffs were used. Users could choose to downweight ALMA baselines through ParselTongue (Kettenis et al. 2006), a Python interface to AIPS, or average the data in time for `LPCAL`. We selected fiducial D-terms for the `LPCAL` pipeline by taking the median of the real and imaginary parts of the surveyed D-terms of each station. This approach allows us to take into account the uncertainties in `LPCAL` that may be associated with the Stokes $\mathcal{I}$ image reconstruction and the parameters used for the different tasks. To get the final M87 image for `LPCAL`, we applied these fiducial D-terms to the data and imaged in. For the high-band results (Appendix I), we obtained D-term solutions and images from a single, best-bet pipeline using representative parameters (15 CLEAN sub-models, ALMA weight = 1.0) from our survey on the low-band data.

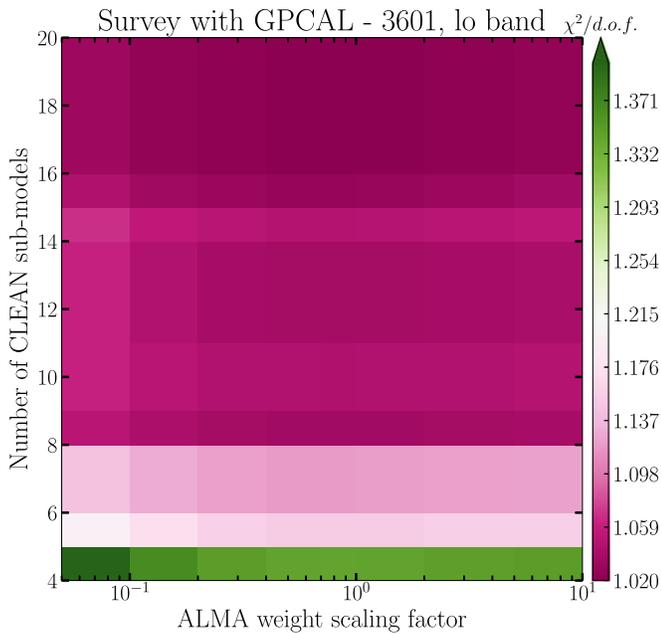

**Figure 23.** Distribution of the reduced chi-squares of the fits to M87 2017 April 11 low-band data with `GPCAL`, depending on the ALMA weight-scaling factor and the number of CLEAN sub-models. `GPCAL` parameters that produce nearly identical results to `LPCAL` were used. A larger number of sub-models provides a better fit, while the results are insensitive to the ALMA weight-scaling factors.

Additionally, we investigated the effects of the parameter selection on the real data with `GPCAL`, using the same parameters as we identify for `LPCAL`,[146] to examine the improvements of the fit statistics when changing parameters.

---

[146] `GPCAL` and `LPCAL` produce nearly identical results, but `GPCAL` reports the reduced chi-square ($\chi^2_{red}$) of the fits.

Figure 23 shows the distribution of $\chi^2_{red}$ for the two parameters from the survey on the M87 data for 2017 April 11. We explored the number of sub-models from five to 20 with an increment of one and the ALMA weight-scaling factor of (0.1, 0.2, 0.4, 0.7, 1.0, 2.0, 5.0, 10.0). Similar to our conclusions from the synthetic data analysis, we found that the fit statistics on the M87 data improve with a larger number of sub-models up to about 10 sub-models. The result is insensitive to changing the ALMA weight-scaling factors. This trend was seen for the M87 data for the other observation days as well. Therefore, we conclude that the parameters that we used for the real data analysis with `LPCAL` are reasonable.

### G.4. DMC Parameter Survey

For the DMC image reconstructions, we surveyed two hyperparameters: the pixel separation and the image field of view. Because the DMC method fits for a systematic uncertainty term in addition to the image and calibration parameters, all fits are formally "good" from the perspective of, e.g., a $\chi^2$ metric. We thus determine an acceptable fit for a particular data set to be the one that minimizes the number of model parameters (i.e., a combination of largest pixel separation and smallest image field of view) while recovering the expected level of systematic uncertainty (i.e., 0% for the synthetic data sets and 2% for the M87 data sets; see Paper III) within some threshold (taken to be the $3\sigma$ bounds determined by the posterior distribution).

### G.5. THEMIS Parameter Survey

Associated with `THEMIS` reconstructions are two hyperparameters corresponding to the two-dimensional number of control points. These have natural values set by the number of independent modes that may be reconstructed; for the EHT, $5 \times 5$. Because this resulted in formally acceptable fits, i.e., reduced-$\chi^2$ near unity, and based on similar considerations from Stokes $\mathcal{I}$ image reconstructions (see, e.g., Broderick et al. 2020b), no additional exploration was required for M87. For the Model 1 and 2 synthetic data reconstructions, the number of control points were incrementally increased until acceptable fits were obtained.

## Appendix H
## D-term Monte Carlo Survey Details

In Sections 5.2 and Section 5.3 we generate a sample of 1000 images with different D-term calibration solutions for each method on each observing day to assess our uncertainty in the polarimetric image structure. In this Appendix, we discuss the procedure that we use to generate these samples and provide detailed histograms of the results for the various image-integrated quantities.

### H.1. Method

To sample the effects of uncertainties in the D-terms on the reconstructed images in the posterior exploration methods (DMC and THEMIS), we can simply draw our 1000 image sample randomly directly from the full posterior. For the three non-MCMC imaging pipelines (`eht-imaging`, `LPCAL`, `polsolve`), our method is to use a simple Monte Carlo approach, similar to the analysis of Martí-Vidal et al. (2012). For each method, we draw 1000 random sets of D-terms from normal distributions with means and (co)variances determined by considering the fiducial results from Table 5 across the four observing days. We assume for this test that the uncertainties in the D-terms are uncorrelated from station to station and between LCP and RCP. This represents a worst-case scenario test, as correlations between the D-terms would reduce the volume of the D-term parameter space surveyed by each method. The full sample of 1000 D-term sets sampled for SMT, LMT, and PV on 2017 April 11 for the `eht-imaging`, `polsolve`, and `LPCAL` pipelines are shown in Figure 24. In addition to the distributions shown in Figure 24, we also include D-terms on ALMA, APEX, SMA, and JCMT drawn from circular complex Gaussians with 1% standard deviation; these represent residual uncertainties left over from the zero-baseline D-term calibration of these stations.

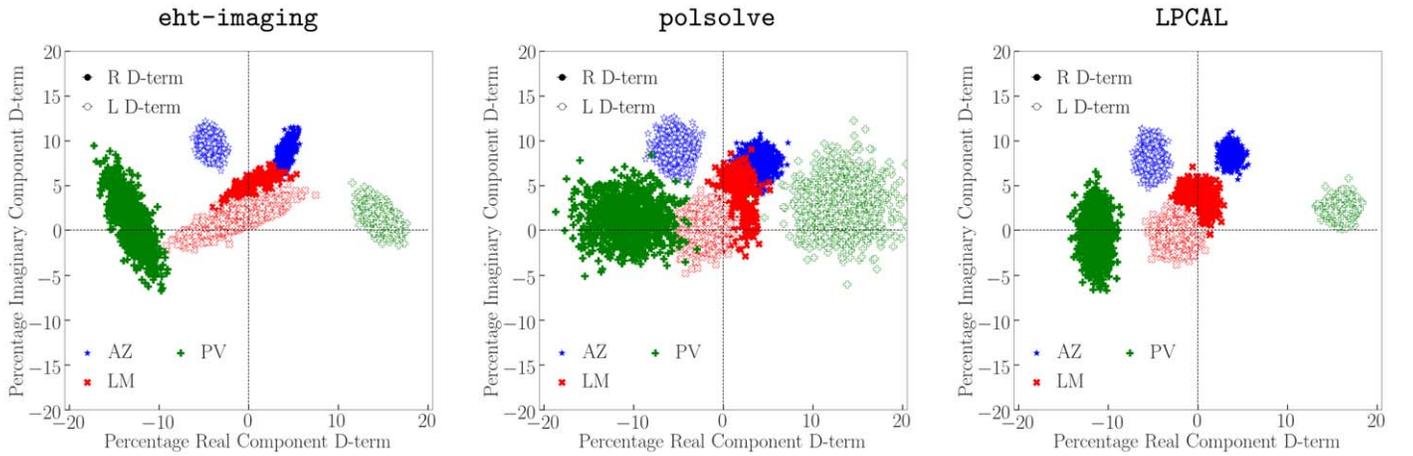

**Figure 24.** Samples of 1000 SMT, LMT, and PV D-terms applied to the data and used in each of the three image reconstruction algorithms in the Monte Carlo procedure described in Section 5.2. Each D-term was drawn from a normal distribution with no correlations between the D-terms. The means and (co)variances of these distributions for each method were determined from the set of four fiducial D-term solutions found across the four observing days for each method. `eht-imaging` included covariance between the real and imaginary parts of each D-term in its approach, while `LPCAL` and `polsolve` did not.

After drawing a given set of random D-terms, we apply this calibration solution to the data and reconstruct a polarized image using the same fiducial imaging procedure described for each method in Appendix G. Our imaging scripts in this stage do not involve any simultaneous leakage calibration, but only reconstruct the Stokes $\mathcal{Q}$ and $\mathcal{U}$ from the visibilities with the assumed calibration solution already applied.[147] That is, in this procedure we draw a set of random D-terms from distributions reflecting our uncertainty in the recovered D-terms from the earlier stage of calibration and imaging, and then we reconstruct an image assuming that this D-term calibration is perfect with no need for further leakage calibration.

### H.2. Distributions of Image-averaged Parameters

In Figures 25 and 26, we show histograms over each imaging method's sample of 1000 images with different D-term calibration solutions on all four observing days of the four image-integrated quantities used in Section 5.3. Figure 25 shows histograms of the image net polarization fraction $|m|_{\mathrm{net}}$ (Equation (12), plotted in red) and the intensity-weighted average polarization fraction at 20 $\mu$as resolution $\langle|m|\rangle$ (Equation (13), plotted in green). Figure 26 shows the amplitude $|\beta_2|$ (plotted in brown) and phase $\angle\beta_2$ (plotted in purple) of the $\beta_2$ coefficient of the azimuthal decomposition defined in Equation (14). Because the observations on 2017 April 5 and 11 have the highest-quality $(u, v)$ coverage and bracket the observed time evolution of the source, we choose to define acceptable ranges for these parameters (the shaded bars in Figures 25, 26, taken from Table 2) using only these two days. In particular, the poor quality of the 2017 April 10 $(u, v)$ coverage leads to broader distributions of the four key quantities with large systematic uncertainties between imaging methods (third columns of Figures 25, 26). The distributions on 2017 April 5 and 11 are summarized with mean and $1\sigma$ error bars in the main text Figure 9, and are discussed in Section 5.3.

---

[147] Note that for the `polsolve` and `LPCAL` pipelines, the actual `polsolve` and `LPCAL` software is used only in solving for the D-terms, and final images for these pipelines are produced in CASA and `Difmap`, respectively. Even though we do not solve for D-terms in this stage, we continue to refer to the imaging part of these pipelines as `polsolve` and `LPCAL` in this section for consistency.

Finally, Table 7 presents ranges of the image-integrated Stokes parameters $\mathcal{I}, \mathcal{Q}, \mathcal{U}$ derived from the surveys over different D-term calibration solutions from each day of observations. The ranges in Table 7 were calculated by taking the minimum mean $- 1\sigma$ and maximum mean $+ 1\sigma$ point from the five individual method surveys on each day.

**Table 7**
Ranges of Image-integrated Stokes $\mathcal{I}, \mathcal{Q}, \mathcal{U}$ Obtained from Each Method's D-term Calibration Survey on the Low-band Data

| Stokes (Jy) | Min | Max | Min | Max |
| --- | --- | --- | --- | --- |
| | 2017 April 5 | | 2017 April 6 | |
| $\mathcal{I}$ | 0.419 | 0.512 | 0.376 | 0.508 |
| $\mathcal{Q}$ | −0.0136 | −0.0002 | −0.0056 | 0.0075 |
| $\mathcal{U}$ | −0.0157 | −0.0055 | −0.0167 | −0.0063 |
| | 2017 April 10 | | 2017 April 11 | |
| $\mathcal{I}$ | 0.381 | 0.545 | 0.407 | 0.565 |
| $\mathcal{Q}$ | −0.0067 | 0.0158 | 0.0030 | 0.0180 |
| $\mathcal{U}$ | −0.0074 | 0.0115 | −0.0113 | 0.0007 |

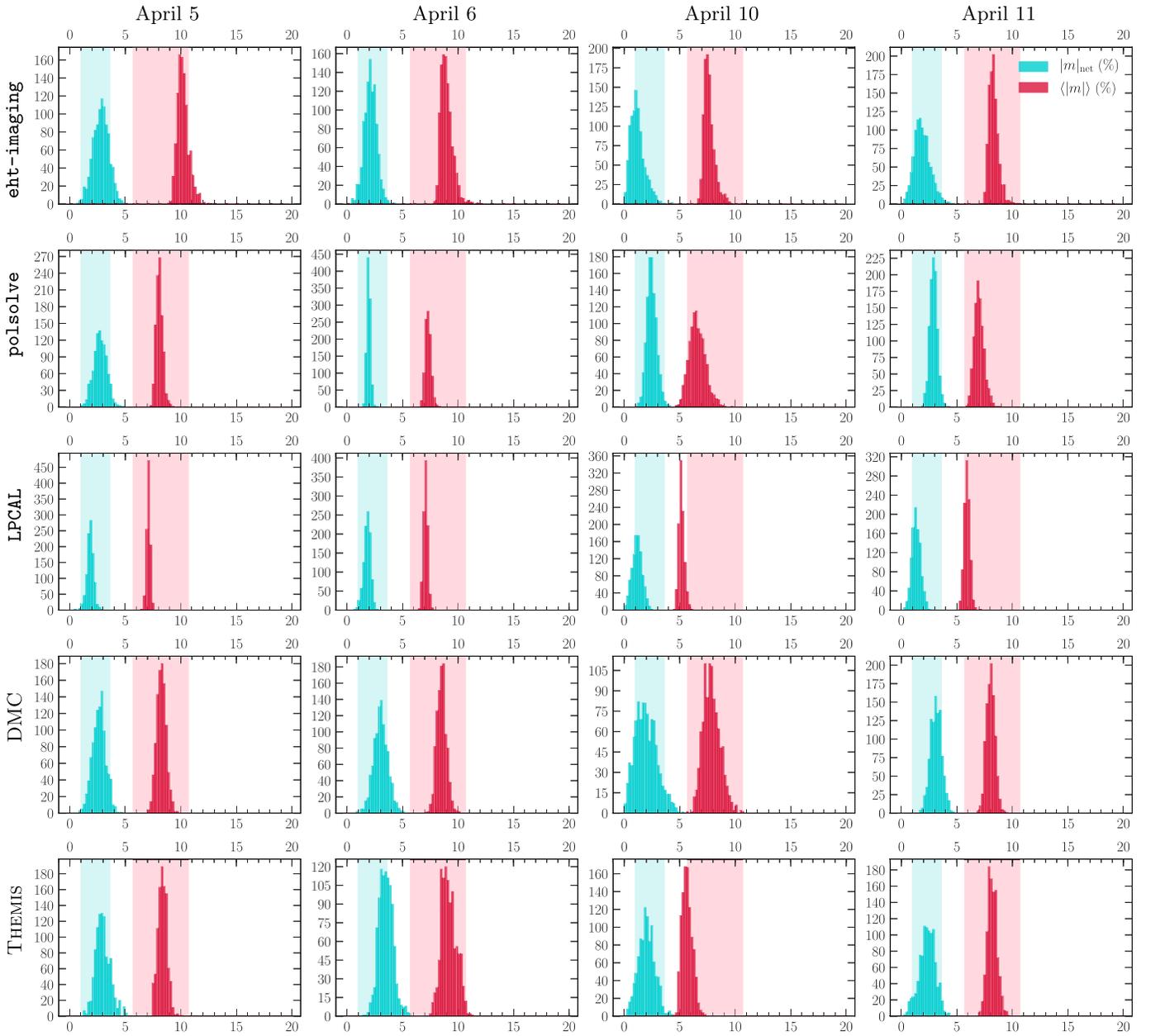

**Figure 25.** Histograms of the net polarization fraction $|m|_{net}$ (green: Equation (12)) and the image-averaged polarization fraction $\langle|m|\rangle$ (red: Equation (13)) from each method's survey over random D-term calibration solutions performed on the low-band data. From left to right, the four columns show histograms for 2017 April 5, 6, 10, and 11. In all panels the shaded bands represent the final parameter ranges reported in this work, incorporating the uncertainty both across D-term realizations and reconstruction methods. These ranges are presented in Table 2. Note that as a consequence of the poor $(u, v)$ coverage and parallactic angle sampling, the 2017 April 10 image reconstructions from all methods show more systematic uncertainty in the derived parameters than on the other days.

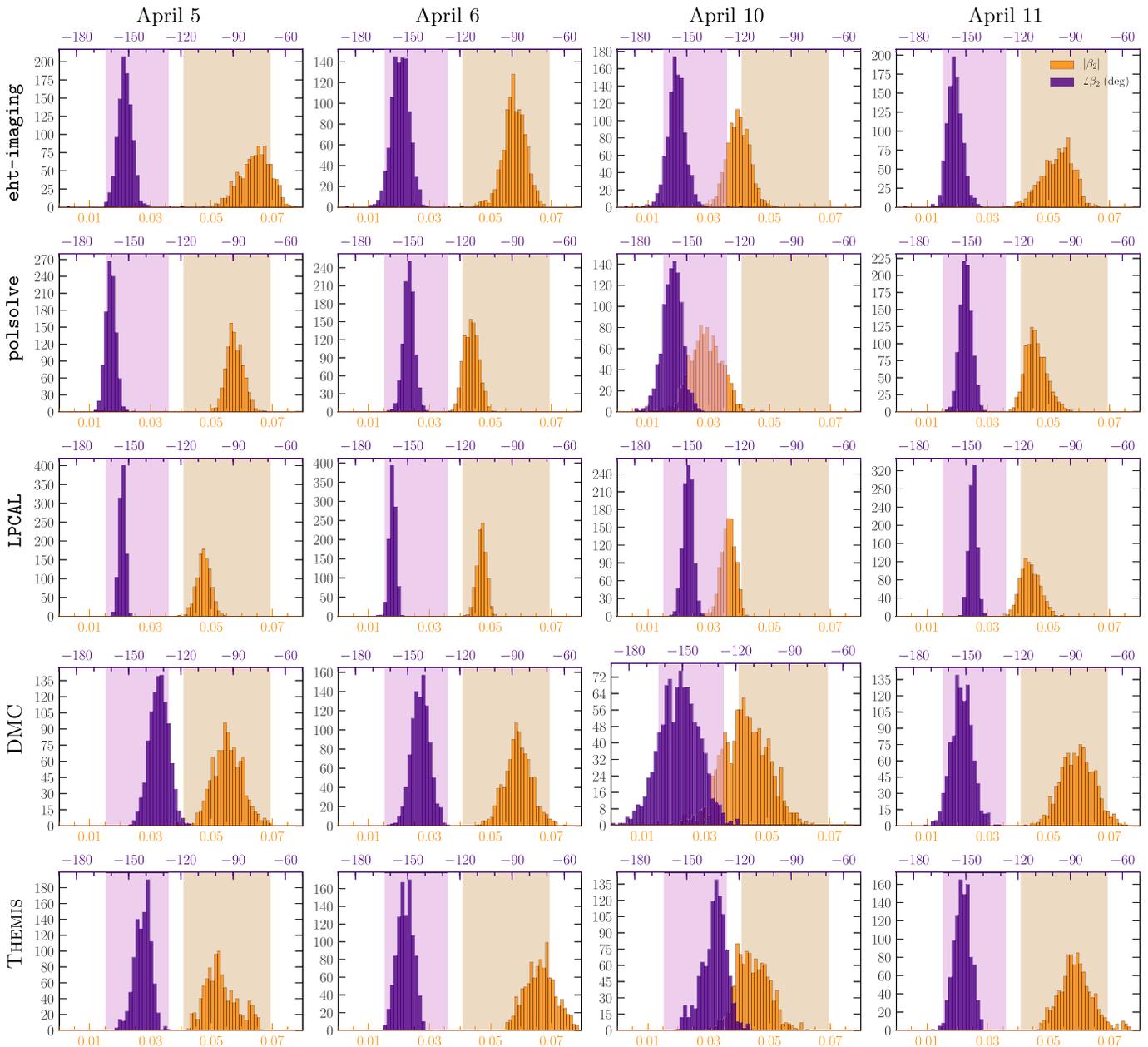

**Figure 26.** Histograms of the amplitude $|\beta_2|$ (brown: bottom axis) and phase (purple: top axis) $\angle\beta_2$ of the $m = 2$ azimuthal mode of the complex polarization brightness distribution (Equation (14)) from each method's survey over random D-term calibration solutions performed on the low-band data. From left to right, the four columns show histograms for 2017 April 5, 6, 10, and 11. In all panels the shaded bands represent the final parameter ranges presented in Table 2. Note that as a consequence of the poor $(u, v)$ coverage, the 2017 April 10 image reconstructions from all methods show more systematic uncertainty in the derived parameters than on the other days.

## Appendix I
## Consistency of Low- and High-band Results for M87

The results shown in the main text were obtained for the EHT low-band data (centered at 227.1 GHz). In this Appendix, we verify the consistency of these results with the EHT high-band data (centered at 229.1 GHz) by repeating several steps of the low-band analysis with the same methodology.

### I.1. Fiducial High-band Images and D-terms

We first compare fiducial D-terms and fiducial polarimetric images derived from the high-band data with the low-band results reported in the main text (Section 5.1). To produce the high-band images, each imaging method used the same imaging procedure as for the low-band images shown in Figures 6–7—we did not re-derive parameters specific to each imaging method or repeat the synthetic data surveys described in Section 4.3 (and Appendix G). Using the same parameters tests whether the methods, when not tuned to high-band data, are able to reproduce our most robust results.

Figure 27 shows the low- and high-band D-terms for LMT, PV, and SMT derived for all five methods. The high band D-terms sit within the systematic scatter among methods in the low-band results. Figure 28 compares the method-averaged high- and low-band consensus images on all four days. The overall level of linear polarization and azimuthal

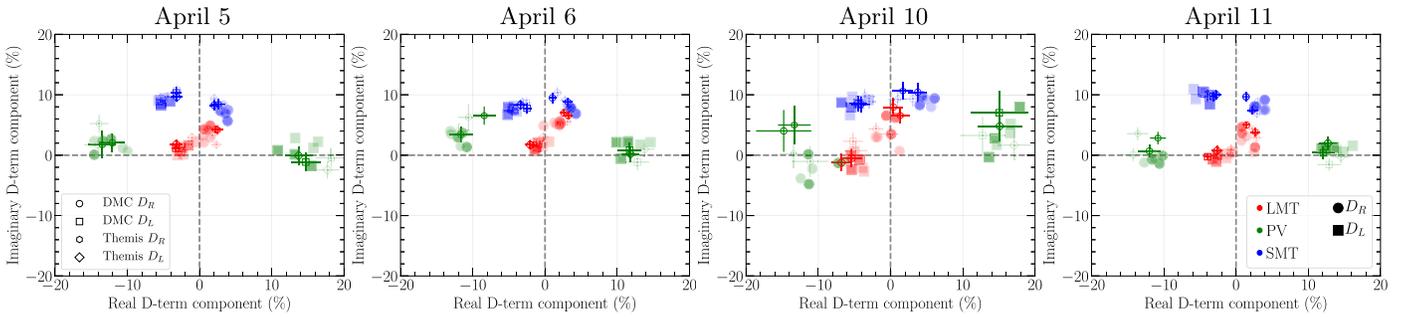

**Figure 27.** Fiducial D-terms for LMT, PV, and SMT derived via leakage calibration in the `eht-imaging`, `polsolve`, `LPCAL` polarimetric imaging pipelines and the DMC/THEMIS posterior exploration of M87 data. The D-terms derived from the low band (lighter points) and the high band (heavier points) are consistent with one another within the systematic scatter among methods seen in the low-band results. All D-terms are displayed in the same manner as in the right panels in Figure 2.

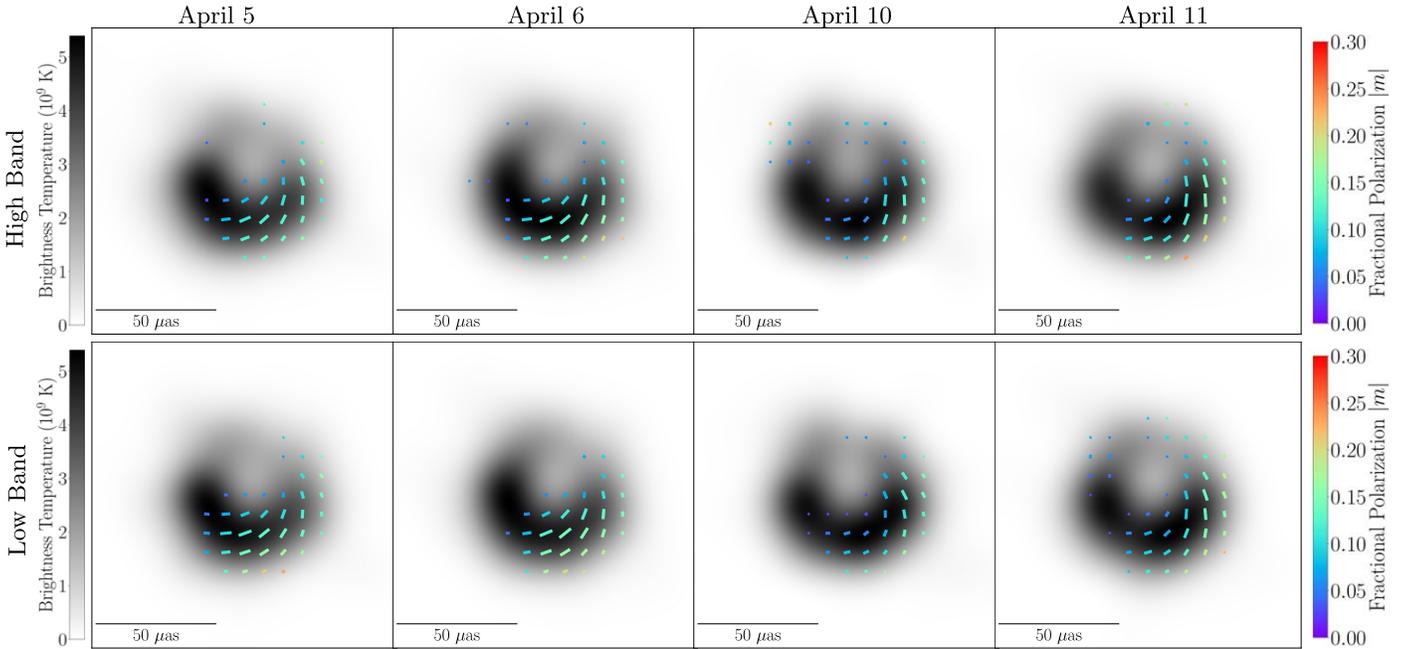

**Figure 28.** Fiducial M87 average images per day produced by averaging results from our five methods (see Figure 6). Method-average images for all four M87 observation days are shown, from left to right. The top and bottom rows show high- and low-band results, respectively. The images are all displayed with a field of view of 120 μas, and all images were brought to the same nominal resolution by convolution with the circular Gaussian kernel that maximized the cross-correlation of the blurred Stokes $\mathcal{I}$ image with the consensus Stokes $\mathcal{I}$ image of Paper IV. Total intensity, polarization fraction, and EVPA are plotted in the same manner as in Figure 7.

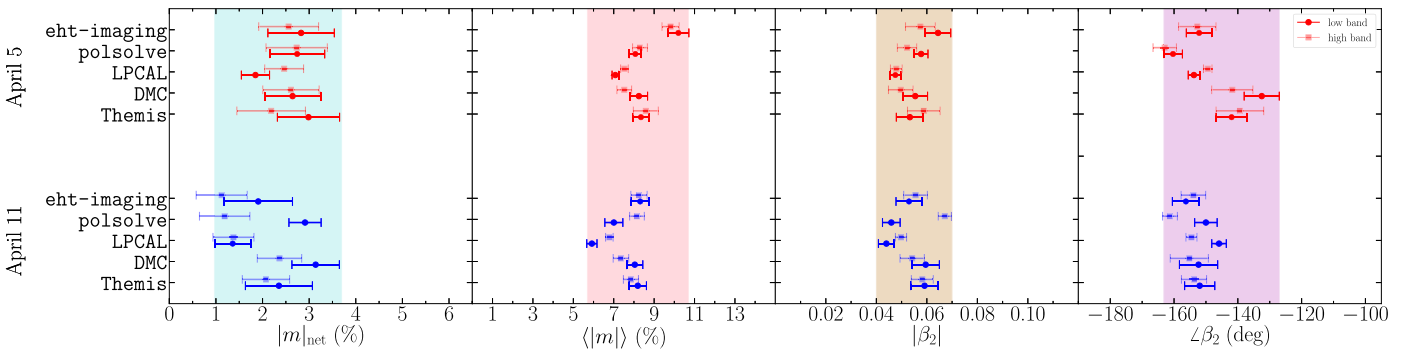

**Figure 29.** Comparison of high- and low-band results for the key quantities used in Paper VIII on 2017 April 5 and 11. The low-band results are indicated by circular markers, and the high-band results are plotted in a lighter color with square markers. The low-band results were presented in the main text in Figure 9. The vertical bands indicate the derived parameter ranges from the low-band results presented in Table 2.

polarization pattern is consistent between the bands on each day.

### I.2. Image-averaged Quantities

To evaluate the consistency of the image-averaged quantities, we extend the analysis presented in Section 5.3 to the high-band data. In particular, we generate a sample of 1000 images from the high-band data for each method, exploring a range of different D-term calibration solutions (this procedure is described in Appendix H).

Figure 29 compares results for the key image-integrated metrics ($|m|_{\rm net}$, $\langle |m| \rangle$, $\beta_2$; see definitions in Section 5.3) derived from such image samples. High-band results for a given method are generally consistent with their low-band counterparts within $1\sigma$. The only notable exception to the overall consistency in the low- and high-band results are the 2017 April 11 polsolve measurements; in particular, $|m|_{\rm net}$, $|\beta_2|$ and $\angle\beta_2$ show deviations of 2–5$\sigma$ between the low and high band. We note that the reported error bars in Figure 29 are derived only from sampling uncertainties in the applied D-terms; they do not include systematic error in the choice of imaging hyperparameters, which were derived only once for each imaging method, on the low-band data.

The mean of the high-band results for all methods fall within the ranges established using the low-band images (Table 2), but the high-band mean $-1$ $\sigma$ lower limits fall outside the established ranges for the eht-imaging and polsolve $|m|_{\rm net}$ measurements on 2017 April 11, and for the polsolve measurement of $|\beta_2|$ on 2017 April 5. Because the imaging procedures and results for the low band were extensively validated with synthetic data tests, we choose to use the low-band results only in defining the parameter ranges in Table 2. Note that $|m|_{\rm net}$ in particular can be quite sensitive to the choice of imaging hyperparameters used, and that these parameters were not re-derived for the high-band data.

In addition to the image-integrated quantities reported in Figure 29, we also computed image-integrated EVPAs for each method from the high-band D-term calibration survey. These are discussed along with the corresponding low-band results in the main text in Section 6.

## Appendix J
## LMT, SMT, and PV D-terms using Calibrator Data: Synthetic Data Tests, Expected Uncertainties, and Convergence with M87 Results

Together with M87, full-array polarimetric calibration and imaging was also attempted on three other sources: 3C 279, observed contemporaneously with M87; and J1924–2914 and NRAO 530, observed contemporaneously with our second EHT primary target, Sgr A*, in the second half of each observing day. 3C 279 was observed on the same four days as M87, with the latter two days having the best $(u, v)$ coverage with the addition of SPT. J1924–2914 was observed on all five days of the EHT campaign (the same four days as M87 with the addition of 2017 April 7), and NRAO 530 was observed on the first three days of the campaign (2017 April 5–7). Coverage and data quality vary from day to day, depending on the structures of the observations and, in the case of the Sgr A* calibrators, whether ALMA is observing. For optimal calibration and imaging, we make an initial cut based on $(u, v)$ and field rotation angle coverage, and the presence of ALMA in the array. We exclude J1924–2914 and NRAO 530 observations on 2017 April 5, which do not have ALMA, and the 2017 April 10 two-scan snapshot observations of J1924–2914, which severely lack $(u, v)$ coverage.

In Figure 30, bottom row, we show the field angle coverage on the three calibrators for their best-coverage day (2017 April 11 for 3C 279 and J1924–2914, and 2017 April 7 for NRAO 530). For comparison, the field angle coverage for M87 on 2017 April 11 is also shown. Compared to M87, the three calibrators are at sufficiently low decl. to also be observed by the SPT, but the elevation stays constant for sources viewed from the South Pole and only a constant field angle is sampled. In Figure 30, top row, we present the $|\tilde{m}|$ structure in the $(u, v)$ plane prior to D-term calibration for the best-coverage days of the calibrators; 2017 April 11 M87 is also shown for reference. High-polarization fractions are expected in M87 on baselines that probe our visibility minima in total intensity, but the source overall is weakly polarized. 3C 279, on the other hand, has multiple baselines exhibiting high polarization fraction. The recovery of D-terms for a highly polarized source like 3C 279 would require an extremely accurate source model in both total intensity and polarization. However, 3C 279's complex structure in both total and polarized intensity add to the difficulty of imaging and calibrating the source (Kim et al. 2020). Furthermore, interferometric-ALMA measurements taken contemporaneously to our EHT campaign found that 3C 279 may have non-negligible Stokes $\mathcal{V}$ (see Goddi et al. 2021), which breaks the Stokes $\mathcal{V} = 0$ assumptions made in most of our calibration and imaging pipelines. Based on these findings, 3C 279 is thus not the best choice for D-term comparisons with M87.

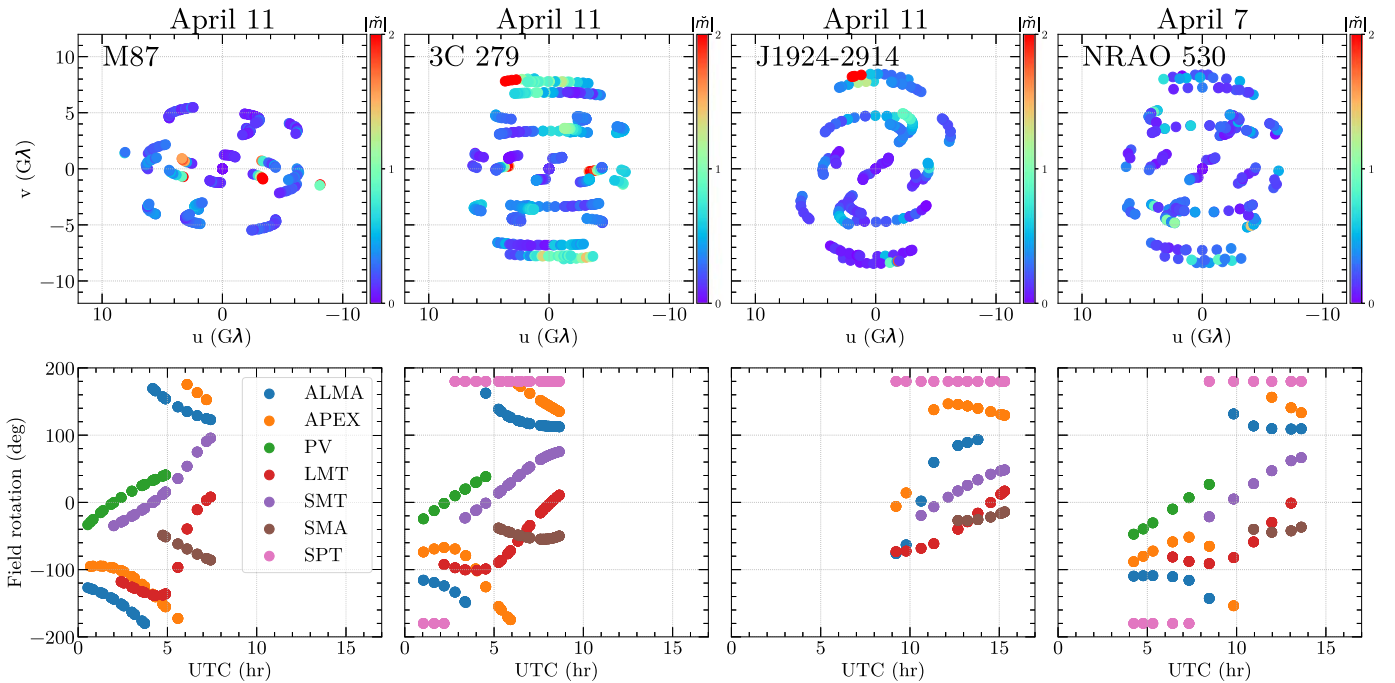

**Figure 30.** Top row: comparison of the polarization, $(u, v)$ coverage, and field rotation angle coverage of the main target and the calibrators. 2017 April 11 is shown for M87, 3C 279, J1924–2914 while 2017 April 7 is shown for NRAO 530. Color scales indicates fractional polarization amplitude $|\breve{m}|$ in the range from 0 to 2. Bottom row: sources field rotation angle $\phi$ for each station as a function of time. The figure is analogous to Figure 1 for M87 on 2017 April 5, 6, 10 and 11.

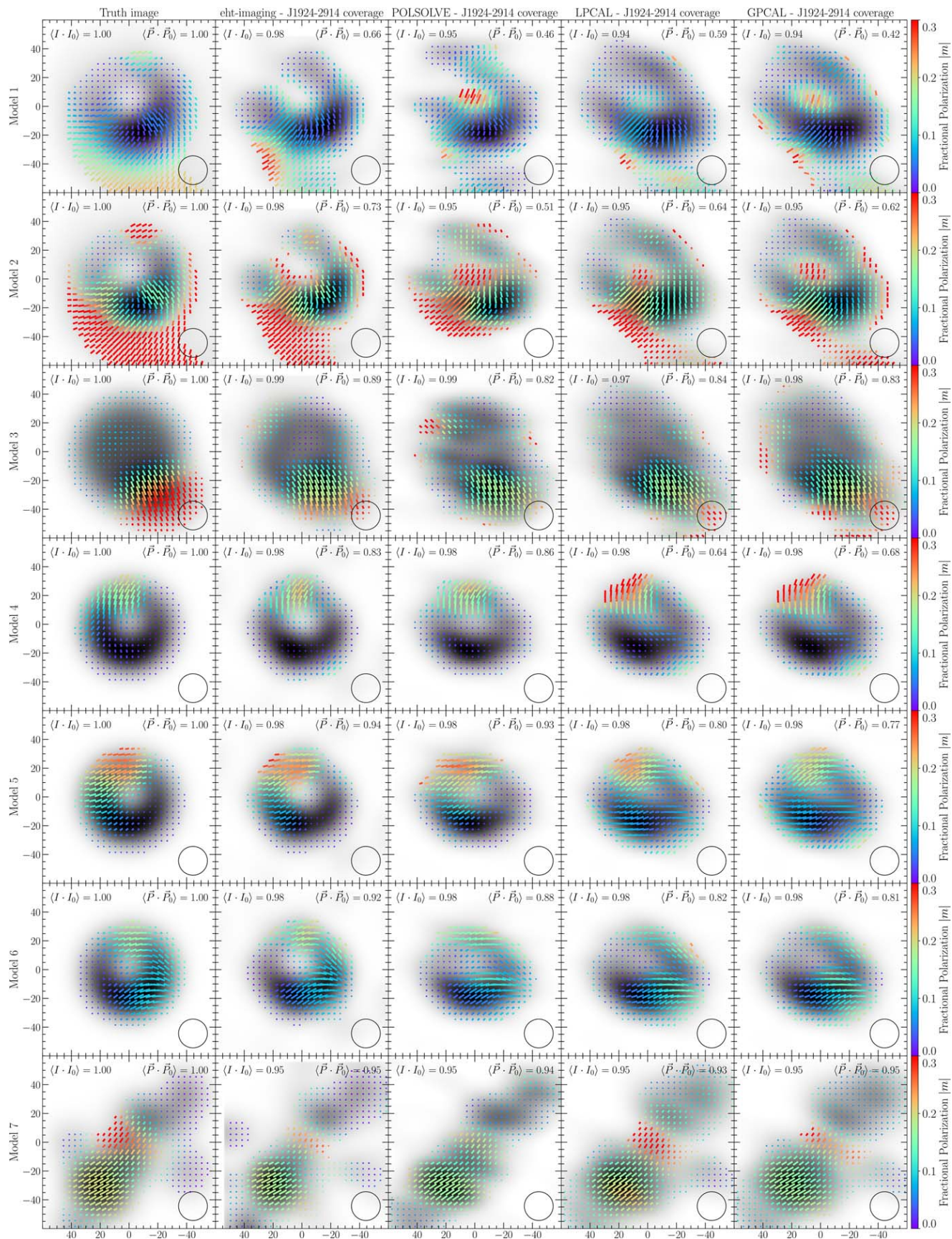

**Figure 31.** Fiducial images from synthetic data model reconstructions using J1924–2914 low-band (*u*, *v*) coverage on 2017 April 11. Polarization tick length reflects total polarization, while color reflects fractional polarization from 0 to 0.3. Normalized overlap is calculated against the respective ground-truth image, and for the case of total intensity is mean-subtracted.

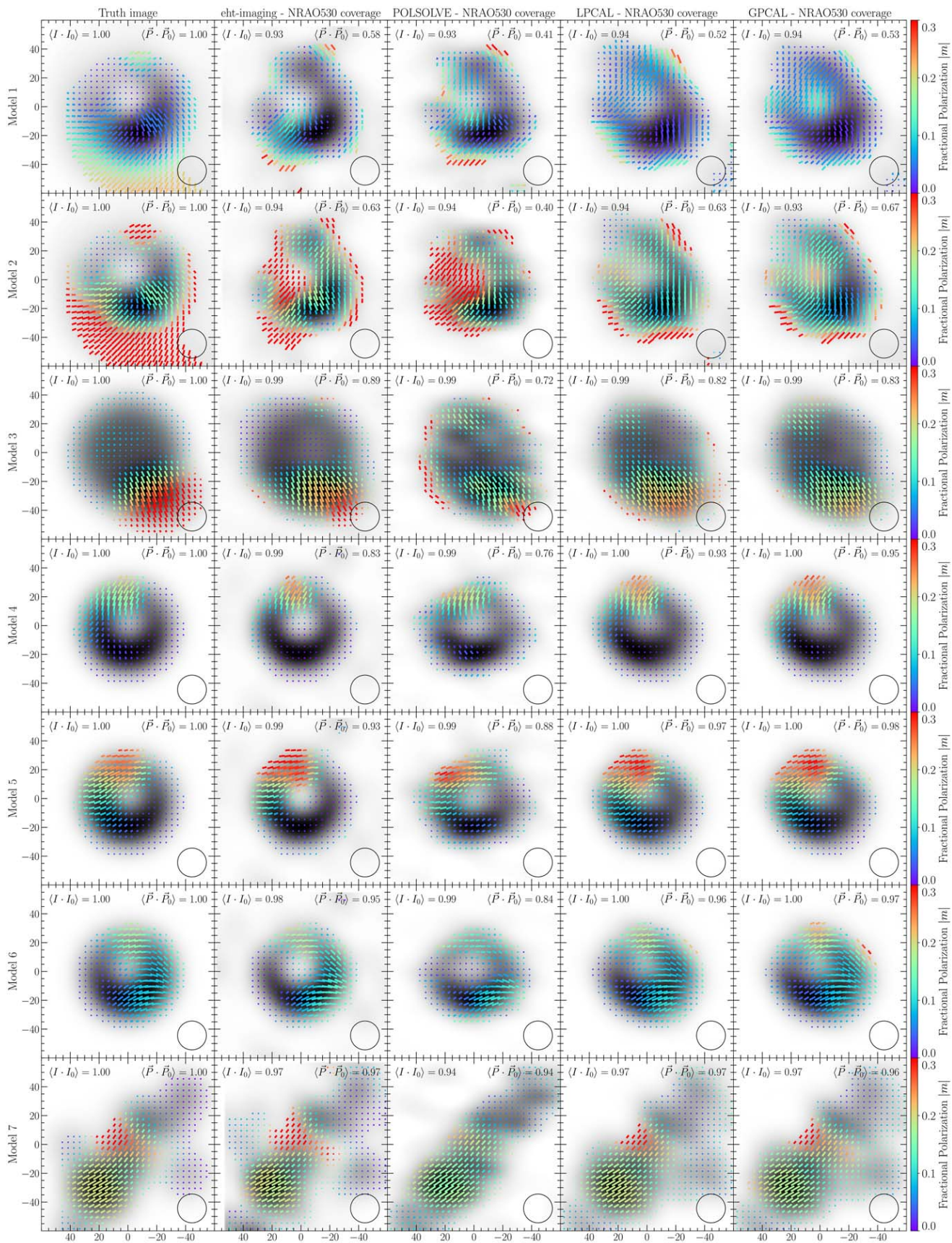

**Figure 32.** Fiducial images from synthetic data model reconstructions using NRAO 530 low-band ($u$, $v$) coverage on 2017 April 7. Polarization tick length reflects total polarization, while color reflects fractional polarization from 0 to 0.3. Normalized overlap is calculated against the respective ground-truth image, and for the case of total intensity is mean-subtracted.

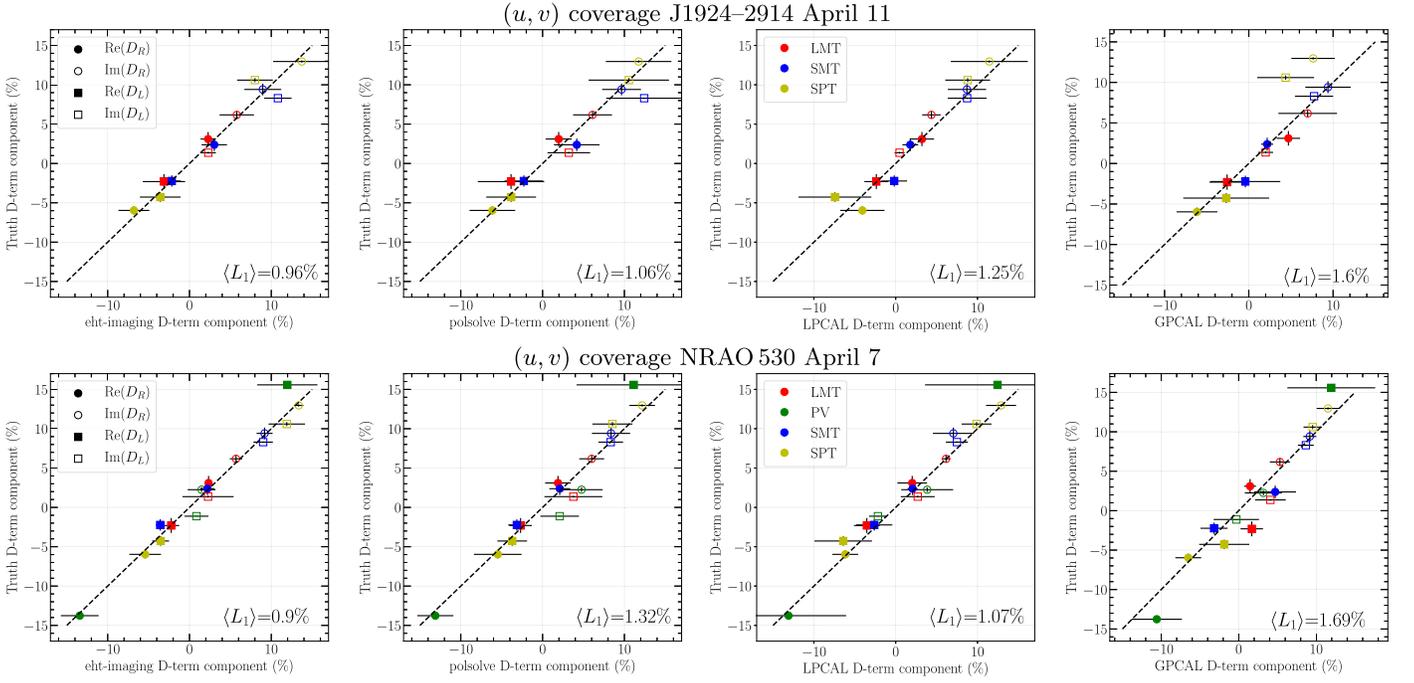

**Figure 33.** D-terms for LMT, SMT, PV, and SPT derived from synthetic data sets. A comparison of estimates to ground-truth values is shown per software (eht-imaging, polsolve, LPCAL, and GPCAL results are shown in first through fourth columns, respectively) and per $(u, v)$ coverage of the real observations (results for $(u, v)$ coverage of J1924–2914 on 2017 April 11 and the $(u, v)$ coverage of NRAO 530 on 2017 April 7 are shown in the top and bottom rows, respectively). Each data point represents the mean and standard deviation for each D-term estimate derived from synthetic data sets 1–7. The norm $L_1 \equiv |D - D_{\rm Truth}|$ is averaged over left, right, real, and imaginary components of the D-terms and over the four EHT stations shown.

J1924–2914 and NRAO 530 exhibit low polarization fractions on most baselines (Figure 30) and have negligible Stokes $\mathcal{V}$ as measured by interferometric-ALMA (Goddi et al. 2021), making them ideal for D-term calibration and polarimetric imaging. Their total-intensity structure, however, is more uncertain and more complex than M87. Both sources are blazars with prominent extended jets (e.g., Wills & Wills 1981; Preston et al. 1989; Shen et al. 1997; Bower & Backer 1998; Healey et al. 2008), and imaging with their current EHT coverage may not capture the complexity of the extended jets in these sources. Nevertheless, their weak polarization allows for better D-term estimates despite uncertainty in modeling their structure.

Following the same methodology as for M87, we generate synthetic data to optimize imaging and calibration parameters for all methods based on J1924–2914 and NRAO 530 low-band coverage. We use the same six ring-like synthetic models as for M87 (see Section 4.3) and add a seventh model constructed with 10 Gaussian sources of varying total and polarization intensity, with some polarization structure offset from Stokes $\mathcal{I}$. This seventh data set is designed to mimic the basic structure seen in the preliminary polarimetric images of the two calibrators, for which the final images will be presented in forthcoming publications (S. Issaoun et al. 2021, in preparation; S. Jorstad et al. 2021, in preparation). We generate seven synthetic EHT observations for each source using their real EHT $(u, v)$ coverage, 2017 April 11 and 2017 April 7 for J1924–2914 and NRAO 530, respectively. Parameter surveys are carried out for each method probing the same parameter space as for M87, and fiducial sets were selected with the same selection metrics; see Appendix G.

In Figures 31 and 32, we present the set of fiducial images from synthetic reconstructions using J1924–2914 and NRAO 530 best-day low-band coverage, respectively. In each panel, the correlations between the ground truth and reconstructed Stokes $\mathcal{I}$ and linear polarization $\mathscr{P}$ images are provided. Consistently with the results with M87 coverage, the Stokes $\mathcal{I}$ correlations are high for all models regardless of method and coverage, and $\mathscr{P}$ correlations seem to worsen for models with complex polarization structure or high polarization.

In Figure 33, we compare the recovered leakage D-terms to the ground-truth D-terms for the synthetic data sets with coverage from J1914–2914 (top row) and NRAO 530 (bottom row) and each method. Similarly to the M87 results, PV and SPT have the largest standard deviations for all methods. Their large deviations stem from all methods having difficulty recovering D-terms for models with no strong polarization substructure due to them being isolated stations with only long baselines. Overall, deviations of the D-terms measured via the $L_1$ norm (and its standard deviation) for the calibrators are comparable to those for M87 for all methods, but the standard deviation on each D-term estimate is noticeably wider for all stations, indicating that while overall image recovery is similar, the coverage differences between the M87 and the calibrator synthetic data do add uncertainty in the D-term recovery for the calibrators.

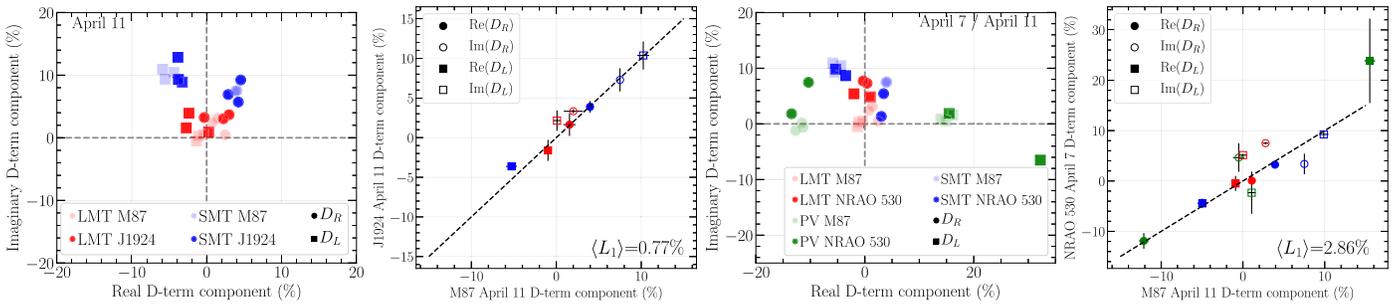

**Figure 34.** Comparison of fiducial D-terms for the telescopes LMT, SMT, and PV estimated from M87 (2017 April 11), J1924–2914 (2017 April 11) and NRAO 530 (2017 April 7) low-band data sets using the `eht-imaging`, `polsolve`, and `GPCAL` pipelines. In the first and third panel from the left the M87 D-terms are depicted with lighter symbols, while heavier symbols mark the calibrator D-terms. In the correlation plots shown in the second and fourth panels from the left, the D-terms for M87 and J1924–2914/NRAO 530 are averaged over different methods. LMT and SMT D-terms derived from J1924–2914 are found to be highly consistent with those from M87. The D-terms derived from NRAO 530 imaging on average show larger deviation from M87 the D-terms; in particular, the PV D-terms estimated by `eht-imaging` show the largest deviation from all other estimates.

Finally, we estimate LMT, SMT, and PV D-terms via polarimetric imaging of the J1924–2914 and NRAO 530 EHT data. The polarimetric images of these two calibrators will be presented in forthcoming publications (S. Issaoun et al. 2021, in preparation; S. Jorstad et al. 2021, in preparation). Here, in Figure 34, we show that D-terms of LMT, SMT, and PV estimated by imaging the calibrators roughly agree with those of M87. We note that a better agreement is obtained between the M87 and J1924–2914 D-terms compared to between M87 and NRAO 530. The calibrators have sparser $(u, v)$ coverage (fewer scans), a narrower field rotation range, and more complex Stokes $\mathcal{I}$ (extended structure and higher noise level) and polarimetric images compared to M87, which all impact the quality of our D-term estimation. Given these additional complexities, we argue that the calibrator D-terms are consistent with those of M87 (the D-term consistency within 2%–3% is expected for the calibrators; see also Appendix K) and that M87 itself is the best source for polarimetric leakage calibration.

Furthermore, while imaging calibrators we found that the quality of the Stokes $\mathcal{I}$ image is critical for calibration. Both NRAO 530 and J1924–2914, as blazar sources, have complex jet structure that is not fully recovered with the current EHT coverage, and thus our Stokes $\mathcal{I}$ reconstructions have larger uncertainties and noise levels that those of M87, due to unconstrained flux density on large scales not sampled by our array configuration. Assumptions about the Stokes $\mathcal{I}$ image affect the results of the polarimetric imaging and calibration methods, for example in the self-similarity assumption employed for CLEAN reconstructions in our sub-component methods (see Appendix K).

# Appendix K
# Validation of the Similarity Approximation in CLEAN Algorithms

The D-term estimates using the M87 data with `polsolve` and `LPCAL` reported in Section 4.2 are based on the similarity approximation. In this approach, the Stokes $\mathcal{I}$ CLEAN models are divided into many sub-models to provide many degrees of freedom for modeling the source's linear polarization structure. Nevertheless, the complex linear polarization structure of M87 (Figure 6) may not be perfectly modeled with this approximation. This could be a source of uncertainty in the D-term estimation.

We investigate the effect of the similarity approximation by using the instrumental polarization self-calibration mode in `GPCAL`, which iterates (i) imaging of the source's linear polarization structures and (ii) solving for the D-terms using the images (Park et al. 2021). We ran `GPCAL` on the M87 data on 2017 April 11. The Stokes $\mathcal{I}$ CLEAN components are divided into 15 sub-models for initial D-term estimation using the similarity approximation. The D-terms of ALMA, APEX, and SMA are fixed to be zero for fitting, as they were already calibrated using the intra-site baselines (Section 4.2). The intra-site baselines are flagged, as the limited field of view of the EHT does not allow us to properly model the source structure observed on large scales. Instrumental polarization self-calibration was then performed with 10 iterations by employing `Difmap` for producing the Stokes $\mathcal{Q}$ and Stokes $\mathcal{U}$ images with CLEAN.

The left panel of Figure 35 compares the D-terms obtained by `GPCAL` with the average of the fiducial imaging pipeline results on the same day (Figure 2). Both results using (i) the similarity approximation only (open squares) and (ii) the similarity approximation followed by 10 iterations of instrumental polarization self-calibration (filled circles) are shown. Both show a good consistency with the fiducial D-terms with the $L_1$ norms of $\approx 1.0\%$–$1.2\%$, consistent with the deviations between values of different pipelines seen in Figure 2. This result indicates that the D-terms obtained by `polsolve` and `LPCAL` using the M87 data are robust against the similarity approximation.

However, this may not be the case for the calibrators. We ran `GPCAL` on the J1924–2914 data on 2017 April 11 and NRAO 530 data on 2017 April 7. Similar parameters to the M87 data analysis are used. The results are shown in the middle and right panels of Figure 35. The D-terms obtained with instrumental polarization self-calibration are more consistent with the fiducial D-terms than those obtained with the similarity approximation only. The $L_1$ norms improve from 9.2% to 1.4% and 2.9% to 2.0% for J1924−2914 and NRAO 530, respectively, with the instrumental polarization self-calibration. This result indicates that the linearly polarized structures of the calibrators are complex and cannot be easily modeled with the similarity approximation.

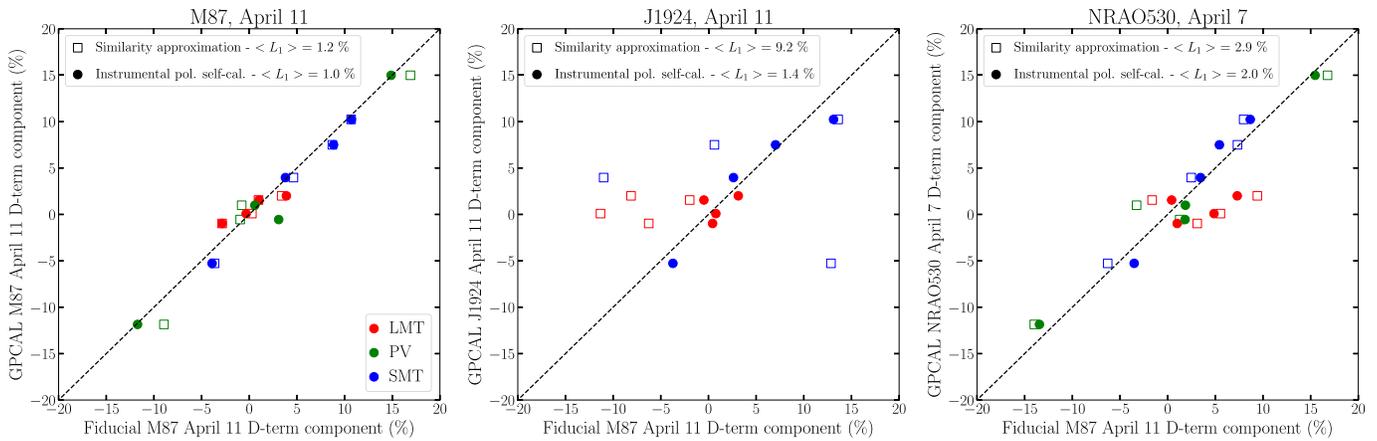

**Figure 35.** Comparison of D-terms estimated with GPCAL with and without instrumental polarization self-calibration. The D-term averages estimated with the imaging pipelines on the 2017 April 11 M87 data are shown on the $x$-axis. The GPCAL results with (filled circles) and without (open squares) instrumental polarization self-calibration are shown on the $y$ − axis (see the text for more details). The left, middle, and right panels show the results for M87 on 2017 April 11, J1924−2914 on 2017 April 11, and NRAO 530 on 2017 April 7, respectively. The $L_1$ norms do not change much with instrumental polarization self-calibration for M87, while they are significantly improved for the calibrators, especially for J1924−2914. This indicates that the similarity approximation employed by polsolve and LPCAL for the D-term estimation from M87 (Section 4.2) is reasonable. The calibrators may have complex linear polarization structure and D-term estimation from those sources can be improved with instrumental polarization self-calibration.

### ORCID iDs

Kazunori Akiyama ⓘ https://orcid.org/0000-0002-9475-4254
Juan Carlos Algaba ⓘ https://orcid.org/0000-0001-6993-1696
Antxon Alberdi ⓘ https://orcid.org/0000-0002-9371-1033
Richard Anantua ⓘ https://orcid.org/0000-0003-3457-7660
Rebecca Azulay ⓘ https://orcid.org/0000-0002-2200-5393
Anne-Kathrin Baczko ⓘ https://orcid.org/0000-0003-3090-3975
Mislav Baloković ⓘ https://orcid.org/0000-0003-0476-6647
John Barrett ⓘ https://orcid.org/0000-0002-9290-0764
Bradford A. Benson ⓘ https://orcid.org/0000-0002-5108-6823
Lindy Blackburn ⓘ https://orcid.org/0000-0002-9030-642X
Raymond Blundell ⓘ https://orcid.org/0000-0002-5929-5857
Katherine L. Bouman ⓘ https://orcid.org/0000-0003-0077-4367
Geoffrey C. Bower ⓘ https://orcid.org/0000-0003-4056-9982
Hope Boyce ⓘ https://orcid.org/0000-0002-6530-5783
Christiaan D. Brinkerink ⓘ https://orcid.org/0000-0002-2322-0749
Roger Brissenden ⓘ https://orcid.org/0000-0002-2556-0894
Silke Britzen ⓘ https://orcid.org/0000-0001-9240-6734
Avery E. Broderick ⓘ https://orcid.org/0000-0002-3351-760X
Do-Young Byun ⓘ https://orcid.org/0000-0003-1157-4109
Andrew Chael ⓘ https://orcid.org/0000-0003-2966-6220
Chi-kwan Chan ⓘ https://orcid.org/0000-0001-6337-6126
Shami Chatterjee ⓘ https://orcid.org/0000-0002-2878-1502
Koushik Chatterjee ⓘ https://orcid.org/0000-0002-2825-3590
Paul M. Chesler ⓘ https://orcid.org/0000-0001-6327-8462
Ilje Cho ⓘ https://orcid.org/0000-0001-6083-7521
Pierre Christian ⓘ https://orcid.org/0000-0001-6820-9941
John E. Conway ⓘ https://orcid.org/0000-0003-2448-9181
James M. Cordes ⓘ https://orcid.org/0000-0002-4049-1882
Thomas M. Crawford ⓘ https://orcid.org/0000-0002-9000-5013

Geoffrey B. Crew ⓘ https://orcid.org/0000-0002-2079-3189
Alejandro Cruz-Osorio ⓘ https://orcid.org/0000-0002-3945-6342
Yuzhu Cui ⓘ https://orcid.org/0000-0001-6311-4345
Jordy Davelaar ⓘ https://orcid.org/0000-0002-2685-2434
Mariafelicia De Laurentis ⓘ https://orcid.org/0000-0002-9945-682X
Roger Deane ⓘ https://orcid.org/0000-0003-1027-5043
Jessica Dempsey ⓘ https://orcid.org/0000-0003-1269-9667
Gregory Desvignes ⓘ https://orcid.org/0000-0003-3922-4055
Jason Dexter ⓘ https://orcid.org/0000-0003-3903-0373
Sheperd S. Doeleman ⓘ https://orcid.org/0000-0002-9031-0904
Ralph P. Eatough ⓘ https://orcid.org/0000-0001-6196-4135
Heino Falcke ⓘ https://orcid.org/0000-0002-2526-6724
Joseph Farah ⓘ https://orcid.org/0000-0003-4914-5625
Vincent L. Fish ⓘ https://orcid.org/0000-0002-7128-9345
Ed Fomalont ⓘ https://orcid.org/0000-0002-9036-2747
H. Alyson Ford ⓘ https://orcid.org/0000-0002-9797-0972
Raquel Fraga-Encinas ⓘ https://orcid.org/0000-0002-5222-1361
Per Friberg ⓘ https://orcid.org/0000-0002-8010-8454
Antonio Fuentes ⓘ https://orcid.org/0000-0002-8773-4933
Peter Galison ⓘ https://orcid.org/0000-0002-6429-3872
Charles F. Gammie ⓘ https://orcid.org/0000-0001-7451-8935
Roberto García ⓘ https://orcid.org/0000-0002-6584-7443
Boris Georgiev ⓘ https://orcid.org/0000-0002-3586-6424
Ciriaco Goddi ⓘ https://orcid.org/0000-0002-2542-7743
Roman Gold ⓘ https://orcid.org/0000-0003-2492-1966
José L. Gómez ⓘ https://orcid.org/0000-0003-4190-7613
Arturo I. Gómez-Ruiz ⓘ https://orcid.org/0000-0001-9395-1670
Minfeng Gu (顾敏峰) ⓘ https://orcid.org/0000-0002-4455-6946
Mark Gurwell ⓘ https://orcid.org/0000-0003-0685-3621
Kazuhiro Hada ⓘ https://orcid.org/0000-0001-6906-772X
Daryl Haggard ⓘ https://orcid.org/0000-0001-6803-2138
Ronald Hesper ⓘ https://orcid.org/0000-0003-1918-6098
Luis C. Ho (何子山) ⓘ https://orcid.org/0000-0001-6947-5846
Mareki Honma ⓘ https://orcid.org/0000-0003-4058-9000
Chih-Wei L. Huang ⓘ https://orcid.org/0000-0001-5641-3953

Shiro Ikeda  https://orcid.org/0000-0002-2462-1448
Sara Issaoun  https://orcid.org/0000-0002-5297-921X
David J. James  https://orcid.org/0000-0001-5160-4486
Michael Janssen  https://orcid.org/0000-0001-8685-6544
Britton Jeter  https://orcid.org/0000-0003-2847-1712
Wu Jiang (江悟)  https://orcid.org/0000-0001-7369-3539
Michael D. Johnson  https://orcid.org/0000-0002-4120-3029
Svetlana Jorstad  https://orcid.org/0000-0001-6158-1708
Taehyun Jung  https://orcid.org/0000-0001-7003-8643
Mansour Karami  https://orcid.org/0000-0001-7387-9333
Ramesh Karuppusamy  https://orcid.org/0000-0002-5307-2919
Tomohisa Kawashima  https://orcid.org/0000-0001-8527-0496
Garrett K. Keating  https://orcid.org/0000-0002-3490-146X
Mark Kettenis  https://orcid.org/0000-0002-6156-5617
Dong-Jin Kim  https://orcid.org/0000-0002-7038-2118
Jae-Young Kim  https://orcid.org/0000-0001-8229-7183
Jongsoo Kim  https://orcid.org/0000-0002-1229-0426
Junhan Kim  https://orcid.org/0000-0002-4274-9373
Motoki Kino  https://orcid.org/0000-0002-2709-7338
Jun Yi Koay  https://orcid.org/0000-0002-7029-6658
Patrick M. Koch  https://orcid.org/0000-0003-2777-5861
Shoko Koyama  https://orcid.org/0000-0002-3723-3372
Michael Kramer  https://orcid.org/0000-0002-4175-2271
Carsten Kramer  https://orcid.org/0000-0002-4908-4925
Thomas P. Krichbaum  https://orcid.org/0000-0002-4892-9586
Cheng-Yu Kuo  https://orcid.org/0000-0001-6211-5581
Tod R. Lauer  https://orcid.org/0000-0003-3234-7247
Sang-Sung Lee  https://orcid.org/0000-0002-6269-594X
Aviad Levis  https://orcid.org/0000-0001-7307-632X
Yan-Rong Li (李彦荣)  https://orcid.org/0000-0001-5841-9179
Zhiyuan Li (李志远)  https://orcid.org/0000-0003-0355-6437
Michael Lindqvist  https://orcid.org/0000-0002-3669-0715
Rocco Lico  https://orcid.org/0000-0001-7361-2460
Greg Lindahl  https://orcid.org/0000-0002-6100-4772
Jun Liu (刘俊)  https://orcid.org/0000-0002-7615-7499
Kuo Liu  https://orcid.org/0000-0002-2953-7376
Elisabetta Liuzzo  https://orcid.org/0000-0003-0995-5201
Laurent Loinard  https://orcid.org/0000-0002-5635-3345
Ru-Sen Lu (路如森)  https://orcid.org/0000-0002-7692-7967
Nicholas R. MacDonald  https://orcid.org/0000-0002-6684-8691
Jirong Mao (毛基荣)  https://orcid.org/0000-0002-7077-7195
Nicola Marchili  https://orcid.org/0000-0002-5523-7588
Sera Markoff  https://orcid.org/0000-0001-9564-0876
Daniel P. Marrone  https://orcid.org/0000-0002-2367-1080
Alan P. Marscher  https://orcid.org/0000-0001-7396-3332
Iván Martí-Vidal  https://orcid.org/0000-0003-3708-9611
Satoki Matsushita  https://orcid.org/0000-0002-2127-7880
Lynn D. Matthews  https://orcid.org/0000-0002-3728-8082
Lia Medeiros  https://orcid.org/0000-0003-2342-6728
Karl M. Menten  https://orcid.org/0000-0001-6459-0669
Izumi Mizuno  https://orcid.org/0000-0002-7210-6264
Yosuke Mizuno  https://orcid.org/0000-0002-8131-6730
James M. Moran  https://orcid.org/0000-0002-3882-4414
Kotaro Moriyama  https://orcid.org/0000-0003-1364-3761

Monika Moscibrodzka  https://orcid.org/0000-0002-4661-6332
Cornelia Müller  https://orcid.org/0000-0002-2739-2994
Gibwa Musoke  https://orcid.org/0000-0003-1984-189X
Alejandro Mus Mejías  https://orcid.org/0000-0003-0329-6874
Daniel Michalik  https://orcid.org/0000-0002-7618-6556
Andrew Nadolski  https://orcid.org/0000-0001-9479-9957
Hiroshi Nagai  https://orcid.org/0000-0003-0292-3645
Neil M. Nagar  https://orcid.org/0000-0001-6920-662X
Masanori Nakamura  https://orcid.org/0000-0001-6081-2420
Ramesh Narayan  https://orcid.org/0000-0002-1919-2730
Iniyan Natarajan  https://orcid.org/0000-0001-8242-4373
Joey Neilsen  https://orcid.org/0000-0002-8247-786X
Roberto Neri  https://orcid.org/0000-0002-7176-4046
Chunchong Ni  https://orcid.org/0000-0003-1361-5699
Aristeidis Noutsos  https://orcid.org/0000-0002-4151-3860
Michael A. Nowak  https://orcid.org/0000-0001-6923-1315
Héctor Olivares  https://orcid.org/0000-0001-6833-7580
Gisela N. Ortiz-León  https://orcid.org/0000-0002-2863-676X
Daniel C. M. Palumbo  https://orcid.org/0000-0002-7179-3816
Jongho Park  https://orcid.org/0000-0001-6558-9053
Ue-Li Pen  https://orcid.org/0000-0003-2155-9578
Dominic W. Pesce  https://orcid.org/0000-0002-5278-9221
Richard Plambeck  https://orcid.org/0000-0001-6765-9609
Oliver Porth  https://orcid.org/0000-0002-4584-2557
Felix M. Pötzl  https://orcid.org/0000-0002-6579-8311
Ben Prather  https://orcid.org/0000-0002-0393-7734
Jorge A. Preciado-López  https://orcid.org/0000-0002-4146-0113
Hung-Yi Pu  https://orcid.org/0000-0001-9270-8812
Venkatessh Ramakrishnan  https://orcid.org/0000-0002-9248-086X
Ramprasad Rao  https://orcid.org/0000-0002-1407-7944
Mark G. Rawlings  https://orcid.org/0000-0002-6529-202X
Alexander W. Raymond  https://orcid.org/0000-0002-5779-4767
Luciano Rezzolla  https://orcid.org/0000-0002-1330-7103
Angelo Ricarte  https://orcid.org/0000-0001-5287-0452
Bart Ripperda  https://orcid.org/0000-0002-7301-3908
Freek Roelofs  https://orcid.org/0000-0001-5461-3687
Eduardo Ros  https://orcid.org/0000-0001-9503-4892
Mel Rose  https://orcid.org/0000-0002-2016-8746
Alan L. Roy  https://orcid.org/0000-0002-1931-0135
Chet Ruszczyk  https://orcid.org/0000-0001-7278-9707
Kazi L. J. Rygl  https://orcid.org/0000-0003-4146-9043
David Sánchez-Arguelles  https://orcid.org/0000-0002-7344-9920
Mahito Sasada  https://orcid.org/0000-0001-5946-9960
Tuomas Savolainen  https://orcid.org/0000-0001-6214-1085
Karl-Friedrich Schuster  https://orcid.org/0000-0003-2890-9454
Lijing Shao  https://orcid.org/0000-0002-1334-8853
Zhiqiang Shen (沈志强)  https://orcid.org/0000-0003-3540-8746
Des Small  https://orcid.org/0000-0003-3723-5404
Bong Won Sohn  https://orcid.org/0000-0002-4148-8378
Jason SooHoo  https://orcid.org/0000-0003-1938-0720
He Sun (孙赫)  https://orcid.org/0000-0003-1526-6787

Fumie Tazaki https://orcid.org/0000-0003-0236-0600
Alexandra J. Tetarenko https://orcid.org/0000-0003-3906-4354
Paul Tiede https://orcid.org/0000-0003-3826-5648
Remo P. J. Tilanus https://orcid.org/0000-0002-6514-553X
Michael Titus https://orcid.org/0000-0003-2423-4505
Kenji Toma https://orcid.org/0000-0002-7114-6010
Pablo Torne https://orcid.org/0000-0001-8700-6058
Efthalia Traianou https://orcid.org/0000-0002-1209-6500
Sascha Trippe https://orcid.org/0000-0003-0465-1559
Ilse van Bemmel https://orcid.org/0000-0001-5473-2950
Huib Jan van Langevelde https://orcid.org/0000-0002-0230-5946
Daniel R. van Rossum https://orcid.org/0000-0001-7772-6131
Jan Wagner https://orcid.org/0000-0003-1105-6109
Derek Ward-Thompson https://orcid.org/0000-0003-1140-2761
John Wardle https://orcid.org/0000-0002-8960-2942
Jonathan Weintroub https://orcid.org/0000-0002-4603-5204
Norbert Wex https://orcid.org/0000-0003-4058-2837
Robert Wharton https://orcid.org/0000-0002-7416-5209
Maciek Wielgus https://orcid.org/0000-0002-8635-4242
George N. Wong https://orcid.org/0000-0001-6952-2147
Qingwen Wu (吴庆文) https://orcid.org/0000-0003-4773-4987
Doosoo Yoon https://orcid.org/0000-0001-8694-8166
André Young https://orcid.org/0000-0003-0000-2682
Ken Young https://orcid.org/0000-0002-3666-4920
Ziri Younsi https://orcid.org/0000-0001-9283-1191
Feng Yuan (袁峰) https://orcid.org/0000-0003-3564-6437
Ye-Fei Yuan (袁业飞) https://orcid.org/0000-0002-7330-4756
J. Anton Zensus https://orcid.org/0000-0001-7470-3321
Guang-Yao Zhao https://orcid.org/0000-0002-4417-1659
Shan-Shan Zhao https://orcid.org/0000-0002-9774-3606

Kazunori Akiyama[1,2,3] , Juan Carlos Algaba[4] , Antxon Alberdi[5] , Walter Alef[6], Richard Antantua[3,7,8] , Keiichi Asada[9], Rebecca Azulay[6,10,11] , Anne-Kathrin Baczko[6] , David Ball[12], Mislav Baloković[13,14] , John Barrett[1] , Bradford A. Benson[15,16] , Dan Bintley[17], Lindy Blackburn[3,7] , Raymond Blundell[7] , Wilfred Boland[18], Katherine L. Bouman[3,7,19] , Geoffrey C. Bower[20] , Hope Boyce[21,22] , Michael Bremer[23], Christiaan D. Brinkerink[24] , Roger Brissenden[3,7] , Silke Britzen[6] , Avery E. Broderick[25,26,27] , Dominique Broguiere[23], Thomas Bronzwaer[24], Do-Young Byun[28,29] , John E. Carlstrom[16,30,31,32], Andrew Chael[33,127] , Chi-kwan Chan[12,34] , Shami Chatterjee[35] , Koushik Chatterjee[36] , Ming-Tang Chen[20], Yongjun Chen (陈永军)[37,38], Paul M. Chesler[3] , Ilje Cho[28,29] , Pierre Christian[39] , John E. Conway[40] , James M. Cordes[35] , Thomas M. Crawford[16,30] , Geoffrey B. Crew[1] , Alejandro Cruz-Osorio[41] , Yuzhu Cui[42,43] , Jordy Davelaar[8,24,44] , Mariafelicia De Laurentis[41,45,46] , Roger Deane[47,48,49] , Jessica Dempsey[17] , Gregory Desvignes[50] , Jason Dexter[51] , Sheperd S. Doeleman[3,7] , Ralph P. Eatough[6,52] , Heino Falcke[24] , Joseph Farah[3,7,53] , Vincent L. Fish[1] , Ed Fomalont[54] , H. Alyson Ford[12] , Raquel Fraga-Encinas[24] , William T. Freeman[55,56], Per Friberg[17] , Christian M. Fromm[3,7,41] , Antonio Fuentes[5] , Peter Galison[3,57,58] , Charles F. Gammie[59,60] , Roberto García[23] , Olivier Gentaz[23] , Boris Georgiev[26,27] , Ciriaco Goddi[24,61] , Roman Gold[25,62] , José L. Gómez[5] , Arturo I. Gómez-Ruiz[63,64] , Minfeng Gu (顾敏峰)[37,65] , Mark Gurwell[7] , Kazuhiro Hada[42,43] , Daryl Haggard[21,22] , Michael H. Hecht[1], Ronald Hesper[66] , Luis C. Ho (何子山)[67,68] , Paul Ho[9], Mareki Honma[42,43,69] , Chih-Wei L. Huang[9] , Lei Huang (黄磊)[37,65] , David H. Hughes[63], Shiro Ikeda[2,70,71,72] , Makoto Inoue[9], Sara Issaoun[24] , David J. James[73] , Buell T. Jannuzi[12], Michael Janssen[6] , Britton Jeter[26,27] , Wu Jiang (江悟)[37] , Alejandra Jimenez-Rosales[24] , Michael D. Johnson[3,7] , Svetlana Jorstad[74,75] , Taehyun Jung[28,29] , Mansour Karami[25,26] , Ramesh Karuppusamy[6] , Tomohisa Kawashima[76] , Garrett K. Keating[7] , Mark Kettenis[77] , Dong-Jin Kim[6] , Jae-Young Kim[6,28] , Jongsoo Kim[28] , Junhan Kim[12,19] , Motoki Kino[2,78] , Jun Yi Koay[9] , Yutaro Kofuji[42,69], Patrick M. Koch[9] , Shoko Koyama[9] , Michael Kramer[6] , Carsten Kramer[23] , Thomas P. Krichbaum[6] , Cheng-Yu Kuo[79,9] , Tod R. Lauer[80] , Sang-Sung Lee[28] , Aviad Levis[19] , Yan-Rong Li (李彦荣)[81] , Zhiyuan Li (李志远)[82,83] , Michael Lindqvist[40] , Rocco Lico[5,6] , Greg Lindahl[7] , Jun Liu (刘俊)[6] , Kuo Liu[6] , Elisabetta Liuzzo[84] , Wen-Ping Lo[9,85], Andrei P. Lobanov[6], Laurent Loinard[86,87] , Colin Lonsdale[1], Ru-Sen Lu (路如森)[37,38,6] , Nicholas R. MacDonald[6] , Jirong Mao (毛基荣)[88,89,90] , Nicola Marchili[6,84] , Sera Markoff[36,91] , Daniel P. Marrone[12] , Alan P. Marscher[74] , Iván Martí-Vidal[10,11] , Satoki Matsushita[9] , Lynn D. Matthews[1] , Lia Medeiros[12,92] , Karl M. Menten[6] , Izumi Mizuno[17] , Yosuke Mizuno[41,93] , James M. Moran[3,7] , Kotaro Moriyama[1,42] , Monika Moscibrodzka[24] , Cornelia Müller[6,24] , Gibwa Musoke[24,36] , Alejandro Mus Mejías[10,11] , Daniel Michalik[94,95] , Andrew Nadolski[60] , Hiroshi Nagai[2,43] , Neil M. Nagar[96] , Masanori Nakamura[9,97] , Ramesh Narayan[3,7] , Gopal Narayanan[98], Iniyan Natarajan[47,49,99] , Antonios Nathanail[41,100] , Joey Neilsen[101] , Roberto Neri[23] , Chunchong Ni[26,27] , Aristeidis Noutsos[6] , Hiroki Okino[42,69], Héctor Olivares[24] , Gisela N. Ortiz-León[6] , Tomoaki Oyama[42], Feryal Özel[12], Daniel C. M. Palumbo[3,7] , Jongho Park[9,128] , Nimesh Patel[7], Ue-Li Pen[25,103,104,105] , Dominic W. Pesce[3,7] , Vincent Piétu[23], Richard Plambeck[106] , Aleksandar PopStefanija[98], Oliver Porth[36,41] , Felix M. Pötzl[6] , Ben Prather[58] , Jorge A. Preciado-López[25] , Dimitrios Psaltis[12], Hung-Yi Pu[9,25,107] , Venkatessh Ramakrishnan[96] , Ramprasad Rao[20] , Mark G. Rawlings[17], Alexander W. Raymond[3,7] , Luciano Rezzolla[41,108,109] , Angelo Ricarte[3,7] , Bart Ripperda[8,110] , Freek Roelofs[24] , Alan Rogers[1], Eduardo Ros[6] , Mel Rose[12] , Arash Roshanineshat[12], Helge Rottmann[6], Alan L. Roy[6] , Chet Ruszczyk[1] , Kazi L. J. Rygl[84] , Salvador Sánchez[111], David Sánchez-Arguelles[63,64] , Mahito Sasada[42,112] , Tuomas Savolainen[6,113,114] , F. Peter Schloerb[98], Karl-Friedrich Schuster[23] , Lijing Shao[6,68] , Zhiqiang Shen (沈志强)[37,38] , Des Small[77] , Bong Won Sohn[28,29,115] , Jason SooHoo[1] , He Sun (孙赫)[19] , Fumie Tazaki[42] , Alexandra J. Tetarenko[17] , Paul Tiede[26,27] , Remo P. J. Tilanus[12,24,61,116] , Michael Titus[1] , Kenji Toma[117,118] , Pablo Torne[6,111] , Tyler Trent[12], Efthalia Traianou[6] , Sascha Trippe[119] , Ilse van Bemmel[77] , Huib Jan van Langevelde[77,120] , Daniel R. van Rossum[24] , Jan Wagner[6] , Derek Ward-Thompson[121] , John Wardle[122] , Jonathan Weintroub[3,7] , Norbert Wex[6] , Robert Wharton[6] , Maciek Wielgus[3,7] , George N. Wong[59] , Qingwen Wu (吴庆文)[123] , Doosoo Yoon[36] , André Young[24] , Ken Young[7] ,

Ziri Younsi[41,124,129] , Feng Yuan (袁峰)[37,65,125] , Ye-Fei Yuan (袁业飞)[126] , J. Anton Zensus[6] , Guang-Yao Zhao[5] , and
Shan-Shan Zhao[37]

The Event Horizon Telescope Collaboration

[1] Massachusetts Institute of Technology Haystack Observatory, 99 Millstone Road, Westford, MA 01886, USA
[2] National Astronomical Observatory of Japan, 2-21-1 Osawa, Mitaka, Tokyo 181-8588, Japan
[3] Black Hole Initiative at Harvard University, 20 Garden Street, Cambridge, MA 02138, USA
[4] Department of Physics, Faculty of Science, University of Malaya, 50603 Kuala Lumpur, Malaysia
[5] Instituto de Astrofísica de Andalucía-CSIC, Glorieta de la Astronomía s/n, E-18008 Granada, Spain
[6] Max-Planck-Institut für Radioastronomie, Auf dem Hügel 69, D-53111 Bonn, Germany
[7] Center for Astrophysics | Harvard & Smithsonian, 60 Garden Street, Cambridge, MA 02138, USA
[8] Center for Computational Astrophysics, Flatiron Institute, 162 Fifth Avenue, New York, NY 10010, USA
[9] Institute of Astronomy and Astrophysics, Academia Sinica, 11F of Astronomy-Mathematics Building, AS/NTU No. 1, Sec. 4, Roosevelt Rd, Taipei 10617, Taiwan, R.O.C.
[10] Departament d'Astronomia i Astrofísica, Universitat de València, C. Dr. Moliner 50, E-46100 Burjassot, València, Spain
[11] Observatori Astronòmic, Universitat de València, C. Catedrático José Beltrán 2, E-46980 Paterna, València, Spain
[12] Steward Observatory and Department of Astronomy, University of Arizona, 933 North Cherry Avenue, Tucson, AZ 85721, USA
[13] Yale Center for Astronomy & Astrophysics, 52 Hillhouse Avenue, New Haven, CT 06511, USA
[14] Department of Physics, Yale University, P.O. Box 2018120, New Haven, CT 06520, USA
[15] Fermi National Accelerator Laboratory, MS209, P.O. Box 500, Batavia, IL, 60510, USA
[16] Department of Astronomy and Astrophysics, University of Chicago, 5640 South Ellis Avenue, Chicago, IL 60637, USA
[17] East Asian Observatory, 660 North A'ohoku Place, Hilo, HI 96720, USA
[18] Nederlandse Onderzoekschool voor Astronomie (NOVA), PO Box 9513, 2300 RA Leiden, The Netherlands
[19] California Institute of Technology, 1200 East California Boulevard, Pasadena, CA 91125, USA
[20] Institute of Astronomy and Astrophysics, Academia Sinica, 645 North A'ohoku Place, Hilo, HI 96720, USA
[21] Department of Physics, McGill University, 3600 rue University, Montréal, QC H3A 2T8, Canada
[22] McGill Space Institute, McGill University, 3550 rue University, Montréal, QC H3A 2A7, Canada
[23] Institut de Radioastronomie Millimétrique, 300 rue de la Piscine, F-38406 Saint Martin d'Hères, France
[24] Department of Astrophysics, Institute for Mathematics, Astrophysics and Particle Physics (IMAPP), Radboud University, P.O. Box 9010, 6500 GL Nijmegen, The Netherlands
[25] Perimeter Institute for Theoretical Physics, 31 Caroline Street North, Waterloo, ON, N2L 2Y5, Canada
[26] Department of Physics and Astronomy, University of Waterloo, 200 University Avenue West, Waterloo, ON, N2L 3G1, Canada
[27] Waterloo Centre for Astrophysics, University of Waterloo, Waterloo, ON, N2L 3G1, Canada
[28] Korea Astronomy and Space Science Institute, Daedeok-daero 776, Yuseong-gu, Daejeon 34055, Republic of Korea
[29] University of Science and Technology, Gajeong-ro 217, Yuseong-gu, Daejeon 34113, Republic of Korea
[30] Kavli Institute for Cosmological Physics, University of Chicago, 5640 South Ellis Avenue, Chicago, IL 60637, USA
[31] Department of Physics, University of Chicago, 5720 South Ellis Avenue, Chicago, IL 60637, USA
[32] Enrico Fermi Institute, University of Chicago, 5640 South Ellis Avenue, Chicago, IL 60637, USA
[33] Princeton Center for Theoretical Science, Jadwin Hall, Princeton University, Princeton, NJ 08544, USA
[34] Data Science Institute, University of Arizona, 1230 North Cherry Avenue, Tucson, AZ 85721, USA
[35] Cornell Center for Astrophysics and Planetary Science, Cornell University, Ithaca, NY 14853, USA
[36] Anton Pannekoek Institute for Astronomy, University of Amsterdam, Science Park 904, 1098 XH, Amsterdam, The Netherlands
[37] Shanghai Astronomical Observatory, Chinese Academy of Sciences, 80 Nandan Road, Shanghai 200030, People's Republic of China
[38] Key Laboratory of Radio Astronomy, Chinese Academy of Sciences, Nanjing 210008, People's Republic of China
[39] Physics Department, Fairfield University, 1073 North Benson Road, Fairfield, CT 06824, USA
[40] Department of Space, Earth and Environment, Chalmers University of Technology, Onsala Space Observatory, SE-43992 Onsala, Sweden
[41] Institut für Theoretische Physik, Goethe-Universität Frankfurt, Max-von-Laue-Straße 1, D-60438 Frankfurt am Main, Germany
[42] Mizusawa VLBI Observatory, National Astronomical Observatory of Japan, 2-12 Hoshigaoka, Mizusawa, Oshu, Iwate 023-0861, Japan
[43] Department of Astronomical Science, The Graduate University for Advanced Studies (SOKENDAI), 2-21-1 Osawa, Mitaka, Tokyo 181-8588, Japan
[44] Department of Astronomy and Columbia Astrophysics Laboratory, Columbia University, 550 West 120th Street, New York, NY 10027, USA
[45] Dipartimento di Fisica "E. Pancini", Universitá di Napoli "Federico II", Compl. Univ. di Monte S. Angelo, Edificio G, Via Cinthia, I-80126, Napoli, Italy
[46] INFN Sez. di Napoli, Compl. Univ. di Monte S. Angelo, Edificio G, Via Cinthia, I-80126, Napoli, Italy
[47] Wits Centre for Astrophysics, University of the Witwatersrand, 1 Jan Smuts Avenue, Braamfontein, Johannesburg 2050, South Africa
[48] Department of Physics, University of Pretoria, Hatfield, Pretoria 0028, South Africa
[49] Centre for Radio Astronomy Techniques and Technologies, Department of Physics and Electronics, Rhodes University, Makhanda 6140, South Africa
[50] LESIA, Observatoire de Paris, Université PSL, CNRS, Sorbonne Université, Université de Paris, 5 place Jules Janssen, F-92195 Meudon, France
[51] JILA and Department of Astrophysical and Planetary Sciences, University of Colorado, Boulder, CO 80309, USA
[52] National Astronomical Observatories, Chinese Academy of Sciences, 20A Datun Road, Chaoyang District, Beijing 100101, People's Republic of China
[53] University of Massachusetts Boston, 100 William T. Morrissey Boulevard, Boston, MA 02125, USA
[54] National Radio Astronomy Observatory, 520 Edgemont Road, Charlottesville, VA 22903, USA
[55] Department of Electrical Engineering and Computer Science, Massachusetts Institute of Technology, 32-D476, 77 Massachusetts Avenue, Cambridge, MA 02142, USA
[56] Google Research, 355 Main Street, Cambridge, MA 02142, USA
[57] Department of History of Science, Harvard University, Cambridge, MA 02138, USA
[58] Department of Physics, Harvard University, Cambridge, MA 02138, USA
[59] Department of Physics, University of Illinois, 1110 West Green Street, Urbana, IL 61801, USA
[60] Department of Astronomy, University of Illinois at Urbana-Champaign, 1002 West Green Street, Urbana, IL 61801, USA
[61] Leiden Observatory—Allegro, Leiden University, P.O. Box 9513, 2300 RA Leiden, The Netherlands
[62] CP3-Origins, University of Southern Denmark, Campusvej 55, DK-5230 Odense M, Denmark
[63] Instituto Nacional de Astrofísica, Óptica y Electrónica, Apartado Postal 51 y 216, 72000. Puebla Pue., México
[64] Consejo Nacional de Ciencia y Tecnología, Av. Insurgentes Sur 1582, 03940, Ciudad de México, México
[65] Key Laboratory for Research in Galaxies and Cosmology, Chinese Academy of Sciences, Shanghai 200030, People's Republic of China
[66] NOVA Sub-mm Instrumentation Group, Kapteyn Astronomical Institute, University of Groningen, Landleven 12, 9747 AD Groningen, The Netherlands
[67] Department of Astronomy, School of Physics, Peking University, Beijing 100871, People's Republic of China
[68] Kavli Institute for Astronomy and Astrophysics, Peking University, Beijing 100871, People's Republic of China

[69] Department of Astronomy, Graduate School of Science, The University of Tokyo, 7-3-1 Hongo, Bunkyo-ku, Tokyo 113-0033, Japan

[70] The Institute of Statistical Mathematics, 10-3 Midori-cho, Tachikawa, Tokyo, 190-8562, Japan

[71] Department of Statistical Science, The Graduate University for Advanced Studies (SOKENDAI), 10-3 Midori-cho, Tachikawa, Tokyo 190-8562, Japan

[72] Kavli Institute for the Physics and Mathematics of the Universe, The University of Tokyo, 5-1-5 Kashiwanoha, Kashiwa, 277-8583, Japan

[73] ASTRAVEO, LLC, PO Box 1668,Gloucester, MA 01931, USA

[74] Institute for Astrophysical Research, Boston University, 725 Commonwealth Avenue, Boston, MA 02215, USA

[75] Astronomical Institute, St. Petersburg University, Universitetskij pr., 28, Petrodvorets,198504 St.Petersburg, Russia

[76] Institute for Cosmic Ray Research, The University of Tokyo, 5-1-5 Kashiwanoha, Kashiwa, Chiba 277-8582, Japan

[77] Joint Institute for VLBI ERIC (JIVE), Oude Hoogeveensedijk 4, 7991 PD Dwingeloo, The Netherlands

[78] Kogakuin University of Technology & Engineering, Academic Support Center, 2665-1 Nakano, Hachioji, Tokyo 192-0015, Japan

[79] Physics Department, National Sun Yat-Sen University, No. 70, Lien-Hai Road, Kaosiung City 80424, Taiwan, R.O.C.

[80] National Optical Astronomy Observatory, 950 North Cherry Avenue, Tucson, AZ 85719, USA

[81] Key Laboratory for Particle Astrophysics, Institute of High Energy Physics, Chinese Academy of Sciences, 19B Yuquan Road, Shijingshan District, Beijing, People's Republic of China

[82] School of Astronomy and Space Science, Nanjing University, Nanjing 210023, People's Republic of China

[83] Key Laboratory of Modern Astronomy and Astrophysics, Nanjing University, Nanjing 210023, People's Republic of China

[84] Italian ALMA Regional Centre, INAF-Istituto di Radioastronomia, Via P. Gobetti 101, I-40129 Bologna, Italy

[85] Department of Physics, National Taiwan University, No. 1, Sect. 4, Roosevelt Road, Taipei 10617, Taiwan, R.O.C.

[86] Instituto de Radioastronomía y Astrofísica, Universidad Nacional Autónoma de México, Morelia 58089, México

[87] Instituto de Astronomía, Universidad Nacional Autónoma de México, CdMx 04510, México

[88] Yunnan Observatories, Chinese Academy of Sciences, 650011 Kunming, Yunnan Province, People's Republic of China

[89] Center for Astronomical Mega-Science, Chinese Academy of Sciences, 20A Datun Road, Chaoyang District, Beijing, 100012, People's Republic of China

[90] Key Laboratory for the Structure and Evolution of Celestial Objects, Chinese Academy of Sciences, 650011 Kunming, People's Republic of China

[91] Gravitation Astroparticle Physics Amsterdam (GRAPPA) Institute, University of Amsterdam, Science Park 904, 1098 XH Amsterdam, The Netherlands

[92] School of Natural Sciences, Institute for Advanced Study, 1 Einstein Drive, Princeton, NJ 08540, USA

[93] Tsung-Dao Lee Institute and School of Physics and Astronomy, Shanghai Jiao Tong University, 800 Dongchuan Road, Shanghai, 200240, People's Republic of China

[94] Science Support Office, Directorate of Science, European Space Research and Technology Centre (ESA/ESTEC), Keplerlaan 1, 2201 AZ Noordwijk, The Netherlands

[95] University of Chicago, 5640 South Ellis Avenue, Chicago, IL 60637, USA

[96] Astronomy Department, Universidad de Concepción, Casilla 160-C, Concepción, Chile

[97] National Institute of Technology, Hachinohe College, 16-1 Uwanotai, Tamonoki, Hachinohe City, Aomori 039-1192, Japan

[98] Department of Astronomy, University of Massachusetts, 01003, Amherst, MA, USA

[99] South African Radio Astronomy Observatory, Observatory 7925, Cape Town, South Africa

[100] Department of Physics, National and Kapodistrian University of Athens, Panepistimiopolis, GR 15783 Zografos, Greece

[101] Villanova University, Mendel Science Center Rm. 263B, 800 East Lancaster Avenue, Villanova PA 19085, USA

[102] Physics Department, Washington University CB 1105, St Louis, MO 63130, USA

[103] Canadian Institute for Theoretical Astrophysics, University of Toronto, 60 St. George Street, Toronto, ON M5S 3H8, Canada

[104] Dunlap Institute for Astronomy and Astrophysics, University of Toronto, 50 St. George Street, Toronto, ON M5S 3H4, Canada

[105] Canadian Institute for Advanced Research, 180 Dundas Street West, Toronto, ON M5G 1Z8, Canada

[106] Radio Astronomy Laboratory, University of California, Berkeley, CA 94720, USA

[107] Department of Physics, National Taiwan Normal University, No. 88, Sec. 4, Tingzhou Rd., Taipei 116, Taiwan, R.O.C.

[108] Frankfurt Institute for Advanced Studies, Ruth-Moufang-Strasse 1, 60438 Frankfurt, Germany

[109] School of Mathematics, Trinity College, Dublin 2, Ireland

[110] Department of Astrophysical Sciences, Peyton Hall, Princeton University, Princeton, NJ 08544, USA

[111] Instituto de Radioastronomía Milimétrica, IRAM, Avenida Divina Pastora 7, Local 20, E-18012, Granada, Spain

[112] Hiroshima Astrophysical Science Center, Hiroshima University, 1-3-1 Kagamiyama, Higashi-Hiroshima, Hiroshima 739-8526, Japan

[113] Aalto University Department of Electronics and Nanoengineering, PL 15500, FI-00076 Aalto, Finland

[114] Aalto University Metsähovi Radio Observatory, Metsähovintie 114, FI-02540 Kylmälä, Finland

[115] Department of Astronomy, Yonsei University, Yonsei-ro 50, Seodaemun-gu, 03722 Seoul, Republic of Korea

[116] Netherlands Organisation for Scientific Research (NWO), Postbus 93138, 2509 AC Den Haag, The Netherlands

[117] Frontier Research Institute for Interdisciplinary Sciences, Tohoku University, Sendai 980-8578, Japan

[118] Astronomical Institute, Tohoku University, Sendai 980-8578, Japan

[119] Department of Physics and Astronomy, Seoul National University, Gwanak-gu, Seoul 08826, Republic of Korea

[120] Leiden Observatory, Leiden University, Postbus 2300, 9513 RA Leiden, The Netherlands

[121] Jeremiah Horrocks Institute, University of Central Lancashire, Preston PR1 2HE, UK

[122] Physics Department, Brandeis University, 415 South Street, Waltham, MA 02453, USA

[123] School of Physics, Huazhong University of Science and Technology, Wuhan, Hubei, 430074, People's Republic of China

[124] Mullard Space Science Laboratory, University College London, Holmbury St. Mary, Dorking, Surrey, RH5 6NT, UK

[125] School of Astronomy and Space Sciences, University of Chinese Academy of Sciences, No. 19A Yuquan Road, Beijing 100049, People's Republic of China

[126] Astronomy Department, University of Science and Technology of China, Hefei 230026, People's Republic of China